\documentclass[]{jfm}

\usepackage{graphicx}
\usepackage{newtxtext}
\usepackage{comment}
\usepackage{newtxmath}
\usepackage{natbib}
\usepackage{bm}
\usepackage{hyperref}
\hypersetup{
	colorlinks = true,
	urlcolor   = blue,
	citecolor  = blue,
}
\usepackage{xcolor}
\usepackage{algorithm}
\usepackage{algpseudocode}
\usepackage{float}
\usepackage{caption}
\usepackage{subcaption}

\newcommand{\RomanNumeralCaps}[1]
\linenumbers

\title{Simulating the swimming motion of a flagellated bacterium in a microstructured bio-fluid}
\author{Arjun Sharma\aff{1}, Sabarish V. Narayanan\aff{2}, Sarah Hormozi\aff{2}, \and Donald L. Koch\aff{2}\corresp{\email{dlk15@cornell.edu}}}

\affiliation{\aff{1}Computational Science Research Institute, Sandia National Laboratories, Albuquerque, NM, 87185, USA
	\aff{2}Smith School of Chemical and Biomolecular Engineering, Cornell University, Ithaca, NY, 14853, USA}

\begin{document}
	\maketitle
			
\begin{abstract}
	We develop a numerical framework to simulate the locomotion of a flagellated bacterium with a spheroidal head (such as Escherichia coli) in biological fluids like mucus, which are entangled polymer solutions exhibiting elasto-viscoplastic (EVP) rheology and porous microstructure. To account for the scale disparity between the large bacterial head and the slender flagellar bundle, whose thickness is comparable to the pore size, we employ a two-fluid model in which the bundle directly drives the solvent and exchanges momentum with the polymer phase via drag proportional to their relative velocity. The numerical implementation combines a finite-difference discretization of the two-fluid equations with a slender-body theory (SBT) to model flagellar forcing. A key observation is that the coupled mass and momentum equations for these inertialess flows, together with SBT, are linear in the pressure and velocity fields and in the force distribution along the flagellar bundle. By treating the polymer stress as a body force, we decompose the flow field and hydrodynamic moments into three additive contributions: kinematic (motion), flagellar forcing, and polymer stress. This decomposition allows several components of the flow to be precomputed and enables the determination of swimming velocity via a resistivity formulation driven by polymer-induced forces, which greatly improves computational efficiency during transient calculations of the polymer stress and the resulting flow. We validate the method and use it to analyze how polymer microstructure and its interactions with the bacterial head and tail affect motility in complex biofluids.
\end{abstract}
	
	
	
\section{Introduction}\label{sec:Intro}
Flagellated bacteria such as \textit{Escherichia coli}, a common and sometimes pathogenic member of the human microbiome, inhabit complex biological environments such as mucus, a protective hydrogel coating all wet epithelial surfaces \citep{Ribbeck,tenaillon2010population,kaper2004pathogenic,MobleyNatRev}. Mucus exhibits elastoviscoplastic (EVP) rheology and contains a porous microstructure formed by entangled polymers \citep{Kirch12}. During locomotion, \textit{E.~coli} furls up its individual flagellar filaments into a rotating helical bundle that propels the organism by overcoming drag on its spheroidal head \citep{chattopadhyay2006swimming,kaya2009characterization}. The head and flagellar bundle differ in radius by nearly two orders of magnitude ($1$--$2\,\mu$m vs.\ $\sim$30--40\,nm~\citep{LaugaARFM}), resulting in qualitatively different interactions with the mucosal medium: the head senses a continuum, while the flagellar bundle probes the microstructure. This scale disparity, combined with length-scale dependent spatial heterogeneity in mucus and swimmer geometry, gives rise to diverse swimming behaviors. Computational modeling serves as a valuable complement to experiments by enabling controlled isolation of geometrical and rheological effects. However, simulating both flagellar nanoscale geometry and mucus microstructure remains computationally prohibitive despite a few recent attempts using first principles approach\,\cite{Yeomans}. To address this, we develop a multiscale framework that models mucus as a two-fluid interpenetrating continua\,\citep{Sabarish}, and develop a novel computational framework coupling two-fluid slender-body theory for the flagellar bundle with a finite-difference solver for the flow of polymeric fluid\,\citep{Sharma23}. This approach captures the distinct interactions of the bacterial head and the flagellar bundle with the surrounding fluid and enables simulations in fluids with microstructure and non-Newtonian (EVP) rheology.

The Reynolds number, defined as the ratio of inertial to viscous forces, at the scale of an individual \textit{E.coli} bacterium is on the order of $10^{-4}$ to $10^{-5}$ \citep{purcell2014life}, indicating effectively inertialess dynamics. These bacteria swim via a characteristic \textit{run-and-tumble} motion, consisting of straight, ballistic runs interrupted by tumbling events that randomly reorient their direction, enabled by the transient unfurling of the flagellar bundle. Experiments in polymer solutions show that \textit{E.~coli} exhibits altered motility compared to Newtonian fluids of the same viscosity: it follows straighter trajectories \citep{Allison15}, tumbles less frequently \citep{Qu20}, bundles its flagella more quickly \citep{Qu18}, and swims faster \citep{Poon14,Allison15,Qu20}. One proposed explanation is extreme shear-thinning near the flagellar bundle, which reduces local viscosity and drag \citep{Man15,Poon14,Qu20}, though this is not supported by bulk rheology. An alternative hypothesis attributes the enhanced speed to reduced wobbling of the cell body caused by elastic stresses \citep{Allison15}. More recently, \cite{Xiang22} reported qualitative similarities in motility between dilute polymeric and colloidal solutions, suggesting that polymer microstructure, rather than viscoelasticity, may dominate swimming speed in dilute and semi-dilute regimes. Viscoelasticity may instead influence secondary behaviors such as reduced tumbling and faster bundling, which indirectly enhance motility. A detailed review of how polymer elasticity and microstructure affect bacterial swimming is provided in \cite{Sabarish}.

In addition to elastic effects, the yield-stress behavior of EVP biofluids has attracted growing interest. EVP fluids exhibit a critical yield stress - below which they behave as elastic solids and above which they flow like viscoelastic liquids. This creates spatial regions around a swimmer that may be either yielded or unyielded, depending on the local stress field. Yield stress plays a biologically significant role in microbial motility. For instance, \textit{Helicobacter pylori} overcomes the yield stress of the stomach’s mucosal lining by acidifying its local environment, fluidizing the mucus and allowing penetration of the protective barrier \citep{Celli09}. Similarly, \cite{Figueroa-Morales19} showed that \textit{E.~coli} can penetrate mucus by mechanically shearing it. Motivated by such findings, several theoretical and experimental studies have explored the motion of simplified swimmers in yield-stress fluids, including waving sheets \citep{Balmforth10,Hewitt17}, flexible filaments \citep{Hewitt22}, rigid helices \citep{Hewitt18}, squirmers \citep{Eastham22}, and synthetic swimmers \citep{Nazari23}.

These studies demonstrate that yield stress can strongly influence swimming behavior, sometimes even enhancing swimming speed \citep{Hewitt17,Hewitt22}, and that performance depends sensitively on swimmer geometry \citep{Nazari23}. A commonly proposed mechanism is that the swimmer yields the surrounding material in a narrow region, creating a lubricated tunnel that allows it to slip through the medium. In particular, the experiments of \cite{Nazari23} show that if the swimmer can exceed the yield strain to rotate its helical flagella, and if the yield stress is not too large compared to its thrust, then a sufficiently wide yielded region can form to enable locomotion. The pitch angle of the helix is critical: low angles are hindered by yield stress, while higher angles can lead to enhanced propulsion. However, the mechanistic role of the hydrodynamic interactions between the flagellar bundle and the influence of the bacterial head remain poorly understood. These open questions motivate the need for numerical simulations. Moreover, no existing theoretical or computational model captures a bacterium swimming in a yield-stress fluid while accounting for microscale heterogeneity in the surrounding medium.

Our simulations isolate specific behaviors such as tumbling and provide detailed access to fields including polymer stretch, polymer stress, and local velocity perturbations. This enables mechanistic insight into how bacteria exploit the non-Newtonian rheology and microstructure of mucus to enhance motility compared to Newtonian fluids. To improve model accuracy, we incorporate a complete rheological description including yield stress and account for non-continuum effects. These features are essential for accurately modeling biofluids such as mucus, which behave as concentrated polymer solutions with microstructural scales comparable to flagellar bundle thickness\,\citep{LiJNNFM,SpagnolieARFM}.

To resolve fluid heterogeneity at the flagellar scale, we employ a two-fluid model that allows the solvent and polymer phases to respond independently to flagellar forcing. Originally introduced by \cite{Doi}, two-fluid models describe the relative motion between polymer and solvent phases due to concentration gradients in inhomogeneous flows. In our formulation, the two interpenetrating phases consist of a Newtonian solvent and an elastoviscoplastic polymer network. Prior applications of this model to microorganisms or active probes typically assume a purely elastic polymer phase and restrict attention to simple geometries such as spheres \citep{Fu08,Diamant15,Lehshansky23}, cylinders \citep{Lehshansky22}, or filaments \citep{Keaveny24}. A recent exception is the work of \cite{Moradi22}, which generalizes the two-fluid formulation to spherical particles in viscoelastic media.

Around \textit{E.~coli}, the flagellar bundle, whose diameter is comparable to the microstructural scale, exerts different forces on the polymer and solvent phases, thereby driving relative motion. The pore size of mucus varies with physiological state; for example, bacterial infection can increase pore size while reducing viscosity and yield stress \citep{Ribbeck}. In this work, we consider cases where the pore size is much larger than the flagellar diameter, so the bundle does not directly contact the polymer network. In such cases, the flagellar bundle exerts force only on the solvent, which then couples indirectly to the polymer. When the bundle and pore sizes are comparable, the flagellar bundle interacts with both phases directly. This scenario can be modeled by enforcing slip at the polymer phase and no-slip at the solvent. Both cases were explored by \citet{Sabarish} using a Newtonian two-fluid model, revealing that swimming behavior depends sensitively on how the bundle couples to the two phases. Here, we extend that approach to include a non-Newtonian polymer phase with yield stress and incorporate hydrodynamic interactions with the full bacterial geometry.

The numerical method builds on the finite-difference scheme of \cite{Sharma23}, formulated in prolate spheroidal coordinates, and extends it to the two-fluid setting. This coordinate system enables exact representation of the particle (bacterial head) shape. A uniform grid in spheroidal coordinates concentrates grid points near the particle surface in Euclidean space, improving resolution where flow gradients are strongest. It also allows practical approximation of an infinite domain with a finite number of points. The helical flagellar bundle is modeled as an external body force using slender-body theory. In contrast to previous approaches for modeling flagellar forcing \citep{Cortez01,Peskin12,Montenegro12,Montenegro13,Griffith24}, our method avoids the need for regularization.

The remainder of the paper is organized as follows. Section~\ref{sec:MathForm} introduces the governing equations for particle motion, flagellar forcing, background flow, and polymer stress. These equations are decomposed exactly into kinematic, non-Newtonian, and flagellar components, as described in section~\ref{sec:LinDecomp}. The numerical method, presented in section~\ref{sec:NumericalMethod}, leverages this decomposition to precompute basis flows that can be combined online to construct full solutions. This modular approach also avoids iterative enforcement of integral constraints. In section~\ref{sec:ResultsMicrostructure}, we apply the method to investigate the role of polymer microstructure in swimming through a quiescent medium. Section~\ref{sec:Conclusions} summarizes our findings and provides directions for future research.

\section{Problem formulation} \label{sec:MathForm}
\begin{figure}
	\centering
	\includegraphics[width=0.6\textwidth]{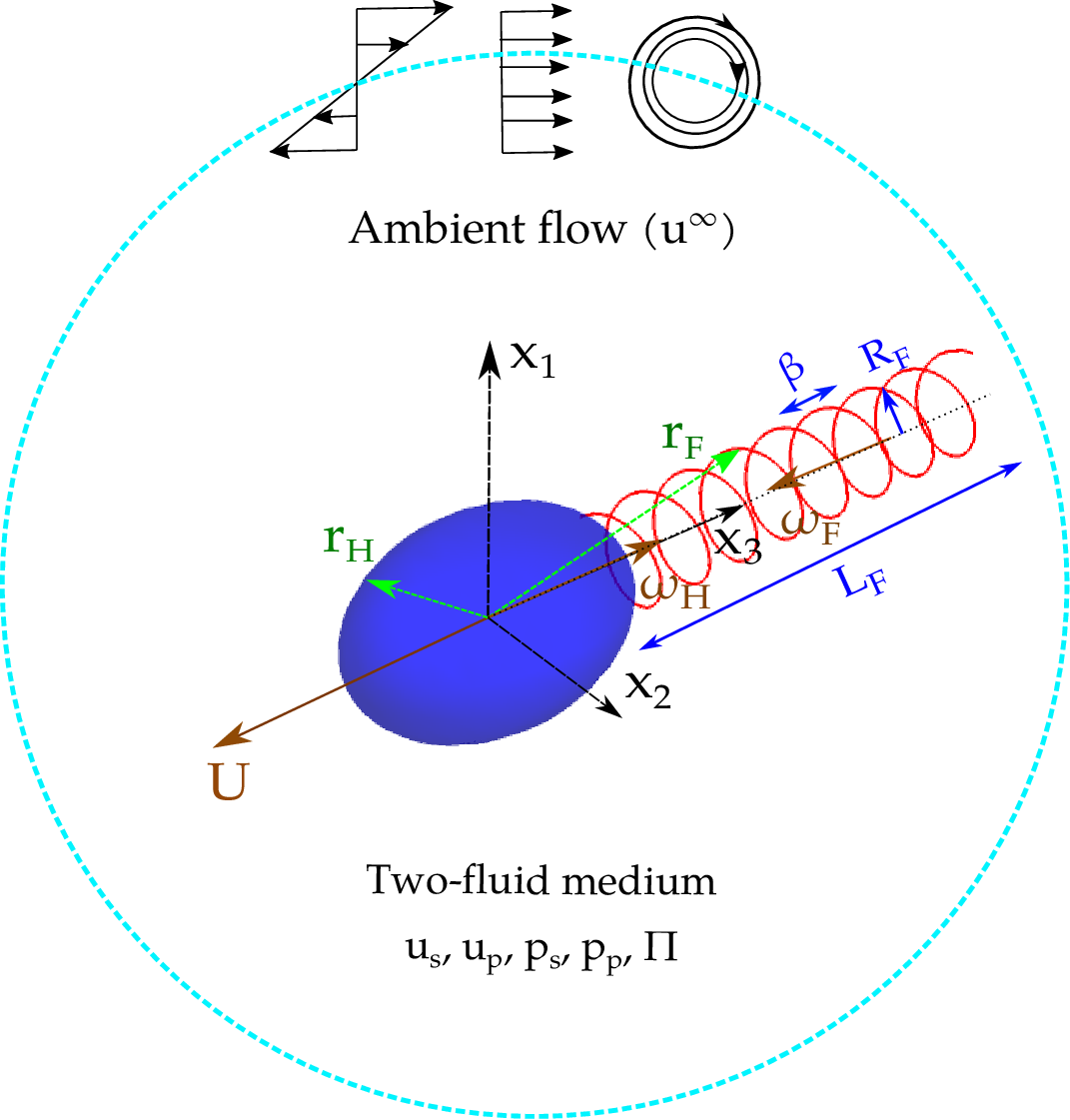}
	\caption{Schematic of a bacterium with a spheroidal head and helical flagellar bundle, showing helix geometry and key variables.}
	\label{fig:CompDomainSchematic}
\end{figure}

We consider a flagellated bacterium, swimming in an unbounded fluid (figure~\ref{fig:CompDomainSchematic}). The organism consists of a prolate spheroidal head (aspect ratio $\chi$) and a helical flagellar bundle. The surrounding medium is a two-phase fluid comprising a Newtonian solvent and a polymeric phase.

In the low Reynolds number regime, the incompressible mass and momentum conservation equations for the solvent and polymer phases are
\begin{align}
	&\nabla \cdot \bm{u}_s = 0, \quad \nabla \cdot \bm{u}_p = 0, \label{EQ1} \\
	&\nabla \cdot \bm{\sigma}_s - \frac{\mu_s}{L_B^2}(\bm{u}_s - \bm{u}_p) = \int \text{d}s\, \bm{f}(\bm{r}_F(s)) \delta (\bm{r}-\bm{r}_F(s)), \label{EQ2}\\
	&\nabla \cdot \bm{\sigma}_p + \frac{\mu_s}{L_B^2}(\bm{u}_s - \bm{u}_p) = 0. \label{EQ3}
\end{align}
Here, $\bm{u}_i$, $p_i$, and $\mu_i$ denote the velocity, pressure, and viscosity of each phase ($i = s$ for solvent and $i = p$ for polymer), and $\bm{f}$ is the force per unit length along the flagellar centerline. The stresses in the solvent and polymer phase are
\begin{align}
	\bm{\sigma}_s &= -p_s \bm{I} + \mu_s (\nabla \bm{u}_s + \nabla \bm{u}_s^T), \\
	\bm{\sigma}_p &= -p_p \bm{I} + \mu_p (\nabla \bm{u}_p + \nabla \bm{u}_p^T) + \bm{\Pi}. \label{eq:PolyStress}
\end{align}

The flagellar bundle, being slender and having a size comparable to pore size, directly forces only the solvent. This results in a relative velocity $\bm{u}_s - \bm{u}_p$, which decays over a screening length $L_B$, associated with the microstructural pore size. The far-field velocity is prescribed as:
\begin{align}
	\bm{u}_s, \bm{u}_p \rightarrow \bm{u}^{\infty} = \langle\bm{u}\rangle + \bm{r} \cdot \bm{\Gamma}, \quad \text{as } r \rightarrow \infty. \label{BCfar}
\end{align}
Boundary conditions on the bacterial head depend on the ratio of pore size to head size. For small pores, both phases satisfy no-slip:
\begin{equation}
	\bm{u}_s, \bm{u}_p = \bm{U} + \bm{\omega}_H \times \bm{r}, \quad \text{on head}. \label{BCnoslip}
\end{equation}
For pore sizes comparable to the head, polymers can slip past the surface \citep{BergTurner}, leading to free-slip boundary conditions on the head:
\begin{align}
	\bm{u}_s &= \bm{U} + \bm{\omega}_H \times \bm{r}, \label{BC2B} \\
	\bm{u}_p \cdot \bm{n} &= (\bm{U} + \bm{\omega}_H \times \bm{r}) \cdot \bm{n}, \label{BC2C} \\
	(\bm{I} - \bm{n n}) \cdot (\bm{\sigma}_p \cdot \bm{n}) &= 0. \label{BC2D}
\end{align}

The flagella's helical centerline is parameterized by arc length $s$ as
\begin{align}
	\bm{r}_F(s) = R_F \begin{bmatrix}
		\cos \left( \tfrac{2\pi s}{\beta} \cos\theta \right) &
		\sin \left( \tfrac{2\pi s}{\beta} \cos\theta \right) &
		\tfrac{2\pi s}{\beta} \cos\theta \cot\theta
	\end{bmatrix}, \label{eq:FlagShape}
\end{align}
where $R_F$, $\theta$, and $\beta$ are the flagellar bundle’s radius, pitch angle, and pitch length, respectively.

The remainder of this section discusses (i) polymer stress modeling, (ii) computation of $\bm{f}$ via slender body theory, (iii) force and torque constraints, and (iv) the coupling between these components.

\subsection{Polymer stress and constitutive modeling}
Polymer elasticity enters the momentum equation via the stress tensor $\bm{\Pi}$, modeled using an ensemble of elastic dumbbells \citep{bird1987dynamics}. Each dumbbell consists of a pair of spherical beads connected by a spring. The polymer stress, generated by stretching of these springs, is related to the configuration tensor $\bm{\Lambda}$ as
\begin{align}
	\bm{\Pi} = \frac{c \mu_p}{\tau} \mathcal{F}(\bm{\Lambda}), \label{eq:PiDef}
\end{align}
where $c$ is the polymer concentration, $\tau$ is the relaxation time, and $\mathcal{F}$ depends on the chosen constitutive model (see table~\ref{tab:Constitutive}).
\begin{table}
	\centering
	\begin{tabular}{ | c | c | c | c | c | }
		\hline
		{\bf Model} & Oldroyd-B  & FENE-P  & FENE-CR & Giesekus\\
		\hline			
		$\mathcal{F}(\bm{\Lambda})$ & $\bm{\Lambda} - \bm{I}$ & $\xi \bm{\Lambda} - \hat{\chi} \bm{I}$ & $\xi (\bm{\Lambda} - \bm{I})$ & $\bm{\Lambda} - \bm{I}$\\ 
		$\mathcal{R}(\bm{\Lambda})$ & $\bm{\Lambda} - \bm{I}$ & $\xi \bm{\Lambda} - \hat{\chi} \bm{I}$ & $\xi (\bm{\Lambda} - \bm{I})$ & $(\bm{\Lambda} - \bm{I}) - \hat{\alpha} (\bm{\Lambda} - \bm{I})^2 $\\
		\hline
	\end{tabular}
	\caption{Stress and relaxation functions $\mathcal{F}$ and $\mathcal{R}$ for common models. In FENE models, $\xi = 1/(1 - \text{tr}(\bm{\Lambda})/L^2)$ and $\hat{\chi} = 1/(1 - \text{tr}(\bm{I})/L^2)$, with $L$ being the ratio of the maximum extensibility to the radius of gyration of the polymers. The Giesekus model introduces a mobility factor $\hat{\alpha}$ that implicitly restricts the maximum polymer stretch.}
	\label{tab:Constitutive}
\end{table}

The configuration tensor $\bm{\Lambda} = \langle \bm{r}_\text{poly} \bm{r}_\text{poly} \rangle$ (defined as the outer product of the polymer end-to-end vector, $\bm{r}_\text{poly}$ averaged over spring orientations) evolves as
\begin{align}
	\frac{\partial \bm{\Lambda}}{\partial t} + \bm{u}_p \cdot \nabla \bm{\Lambda} = (\nabla \bm{u}_p)^T \cdot \bm{\Lambda} + \bm{\Lambda} \cdot (\nabla \bm{u}_p) - \Gamma_Y \frac{1}{\tau} \mathcal{R}(\bm{\Lambda}), \label{ConstEq}
\end{align}
where $\Gamma_Y$, first proposed by \cite{Saramito09}, is a yielding limiter:
\begin{align}
	\Gamma_Y = \max \left(0, \frac{|\bm{\Pi}_d| - \sigma_Y}{|\bm{\Pi}_d|} \right), \quad 
	|\bm{\Pi}_d| = \sqrt{\tfrac{1}{2} \bm{\Pi}_d : \bm{\Pi}_d},\quad \bm{\Pi}_d =\bm{\Pi} - \frac{1}{3} \text{tr}(\bm{\Pi}) \bm{I}.\label{eq:yield}
\end{align}
The deviatoric stress norm $|\bm{\Pi}_d|$ compares the polymer stress against the yield stress $\sigma_Y$. When $|\bm{\Pi}_d| < \sigma_Y$, the polymer behaves as a viscoelastic solid with viscosity $\mu_p$ and elasticity $c \mu_p / \tau$. Once yielding occurs, the material flows as a viscoelastic liquid. Here, $\mathcal{R}(\bm{\Lambda})$ is the relaxation function and varies with the choice of constitutive model (table \ref{tab:Constitutive}). The Oldroyd-B model assumes linear springs and is widely used for its simplicity, though it predicts unbounded stretch in extensional flows. The FENE-P and FENE-CR models introduce finite extensibility and are more suitable for dilute polymer solutions \citep{anna2008effect}. For more concentrated systems like melts, the Giesekus model offers a better fit \citep{bird1987dynamics}. We next describe how to compute the flagellar forcing using slender body theory, and how it couples with the fluid and polymer fields.

\subsection{Slender-body theory and hydrodynamic coupling}
The force distribution $\bm{f}$ along the flagellar bundle is computed using slender body theory (SBT), which approximates the flagellar bundle as a distribution of point forces along its centerline. Matched asymptotic expansions yield the following integral equation \citep{Sabarish} for $\bm{f}$ at each point $s$ along the flagellar bundle:
\begin{equation}
	\begin{split}
		&\frac{\bm{f} \cdot(\bm{I} + \bm{p p})}{4 \pi \mu_s} \log(2\gamma) + \frac{\bm{f} \cdot(\bm{I}-3\bm{p p})}{8 \pi \mu_s}  \\&+ \frac{1}{8\pi \mu_s} \int \left[\left( \frac{\bm{I}}{|\bm{r}_{ss'}|} + \frac{\bm{r}_{ss'}\bm{r}_{ss'}}{|\bm{r}_{ss'}|^3} \right)\cdot\bm{f}(s') - \frac{\bm{I} + \bm{p p}}{|s - s'|} \cdot \bm{f}(s) \right] ds' \\&+ \frac{\mu_p}{8 \pi \mu_s(\mu_s + \mu_p)} \int \bm{f}_s(s') \cdot \left( \bm{G}_{Br}(\bm{r}_{ss'}) - \bm{G}_{St}(\bm{r}_{ss'}) \right)\, ds' \\&= \bm{U} + \bm{\omega}_F \times \bm{r}_F(s) - \bm{u}_s^{\infty}(\bm{r}_F(s)) + \bm{u}_{s,\text{Flag}}(\bm{r}_F(s))+\langle\bm{u}\rangle + \bm{r} \cdot \bm{\Gamma},
	\end{split}
	\label{SBTEQ}
\end{equation}
where $\bm{r}_{ss'} = \bm{r}_F(s) - \bm{r}_F(s')$ is the vector between two points on the flagellar centerline, and $\gamma = L_F / (2a)$ is the slenderness ratio. We assume that the uncoiled flagellar bundle resembles a prolate spheroid with major and minor axes $L_F/2$ and $a$, respectively. If the geometry deviates significantly, corrections are required to account for the blunt ends.

The background velocity $\bm{u}_s^{\infty}$ includes contributions from external flow, the head-induced disturbance, and polymer stresses. The final term on the left-hand side of equation~\eqref{SBTEQ} accounts for corrections due to polymer–solvent interactions via Brinkman and Stokes Green's functions:
\begin{align}
	\bm{G}_{St}(\tilde{\bm{r}}) &= \frac{1}{8 \pi} \left( \frac{\bm{I}}{\tilde{r}} + \frac{\tilde{\bm{r}} \tilde{\bm{r}}}{\tilde{r}^3} \right), \label{Gst}\\
	\bm{G}_{Br}(\tilde{\bm{r}}) &= \frac{1}{8 \pi} \left[ (\bm{I} - \tilde{\bm{r}} \tilde{\bm{r}}) \phi_1(\tilde{r}) + \tilde{\bm{r}} \tilde{\bm{r}} \phi_2(\tilde{r}) \right], \label{Gbr}
\end{align}
where the screening parameter is $\alpha = \frac{1}{L_B} \sqrt{(1+\lambda)/\lambda}$, and the scalar functions are given by
\begin{equation}
	\phi_1 = \frac{2}{\alpha^2 \tilde{r}^3} \left[ (1 + \alpha \tilde{r} + \alpha^2 \tilde{r}^2)\exp(-\alpha \tilde{r}) - 1 \right], \quad
	\phi_2 = \frac{4}{\alpha^2 \tilde{r}^3} \left[ 1 - (1 + \alpha \tilde{r})\exp(-\alpha \tilde{r}) \right].
\end{equation}

The velocity field $\bm{u}_{s,\text{Flag}}$ represents the solvent flow induced by $\bm{f}$ in a Newtonian two-fluid system:
\begin{align}
	\nabla \cdot \bm{\sigma}_{s,\text{Flag}} - \frac{\mu_s}{L_B^2} (\bm{u}_{s,\text{Flag}} - \bm{u}_{p,\text{Flag}}) &=  \int \text{d}s\, \bm{f}(\bm{r}_F(s)) \delta (\bm{r}-\bm{r}_F(s)), \label{uflagEQ2}\\
	\nabla \cdot \bm{\sigma}_{p,\text{Flag}}  + \frac{\mu_s}{L_B^2} (\bm{u}_{s,\text{Flag}} - \bm{u}_{p,\text{Flag}}) &=0 , \label{uflagEQ3}
\end{align}
subject to incompressibility and far-field conditions:
\begin{align}
	\nabla \cdot \bm{u}_{s,\text{Flag}} = \nabla \cdot \bm{u}_{p,\text{Flag}} = 0, \quad \bm{u}_{s,\text{Flag}}, \bm{u}_{p,\text{Flag}} \rightarrow 0 \quad \text{as } \bm{r} \to \infty. \label{uflagBC}
\end{align}
These equations resemble the full flow equations~\eqref{EQ1}–\eqref{EQ3}, but exclude the bacterial head and polymer stress $\bm{\Pi}$, and replace the line force with a single point force source term. Here,
\begin{align}
	\bm{\sigma}_{s,\text{Flag}} &= -p_{s,\text{Flag}}\bm{I} + \mu_s (\nabla \bm{u}_{s,\text{Flag}} + \nabla \bm{u}_{s,\text{Flag}}^T), \\
	\bm{\sigma}_{p,\text{Flag}} &= -p_{p,\text{Flag}} \bm{I} + \mu_p (\nabla \bm{u}_{p,\text{Flag}} + \nabla \bm{u}_{p,\text{Flag}}^T).
\end{align}

Since the flagellar bundle is too slender to directly interact with the polymer network, the polymer's influence on $\bm{f}$ arises indirectly through the background flow $\bm{u}_s^{\infty}$. The regular disturbance from the head is captured by the difference $\bm{u}_s^{\infty}(\bm{r}_F(s)) - \bm{u}_{s,\text{Flag}}(\bm{r}_F(s))$. These hydrodynamic interactions inform the computation of $\bm{f}$. To complete the determination of bacterial swimming, we now impose global force and torque balance conditions on the bacterium.

\subsection{Force and torque free constraints}
\begin{figure}
	\centering
	\includegraphics[width=0.6\textwidth]{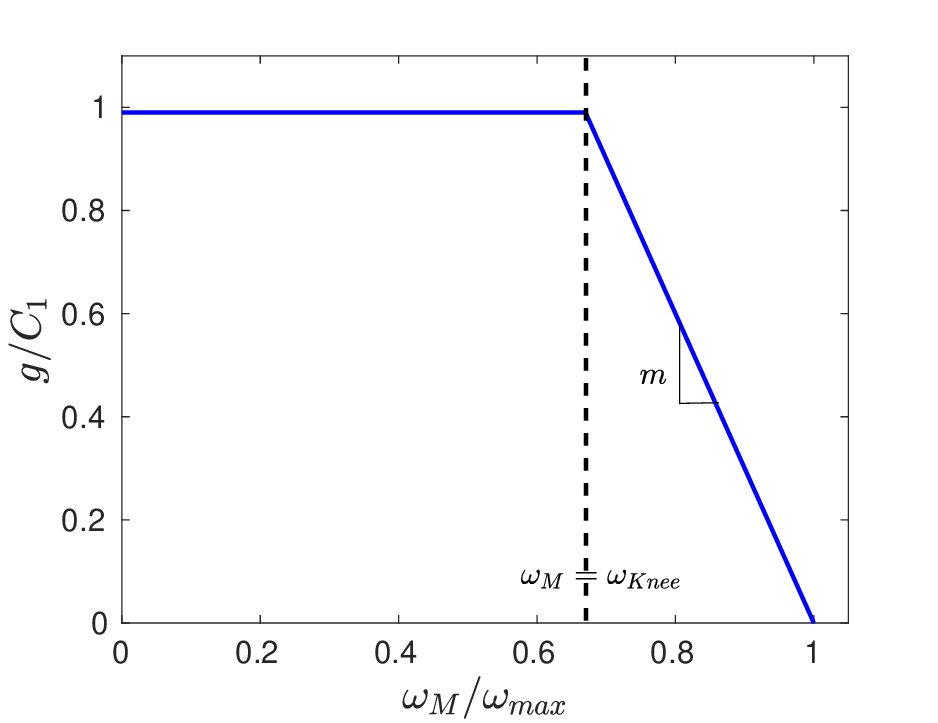}
	\caption{Plot of the motor torque $g$ as a function of motor angular velocity $\omega_M$. See table~\ref{tab:EcoliBergPoon} for typical values of the constants.}
	\label{fig:MotorTorque}
\end{figure}

To close the governing system of equations introduced above, the bacterial translational velocity, $\bm{U}$, and the rotational velocity of the head, $\bm{\omega}_H$, must be determined. These quantities are determined by enforcing global force and torque balance on the organism:
\begin{align}
	&\bm{F}^{\infty} + \bm{F}_\text{Head} + \bm{F}_\text{Flag} = 0, \label{FF} \\
	&\bm{L}^{\infty} + \bm{L}_\text{Head} + \bm{L}_\text{Flag} = 0, \label{TF} \\
	&\bm{L}_\text{Flag} = -\frac{\bm{L}_M}{2}. \label{TM}
\end{align}
The motor applies a torque $\bm{L}_M$ that drives the flagellar bundle to rotate at angular velocity $\bm{\omega}_F$. To maintain torque balance, the head counter-rotates at $\bm{\omega}_H$. The motor torque depends on their relative rotation, $\bm{\omega}_M$:
\begin{equation}
	\bm{L}_M = g(\bm{\omega}_M),\quad \bm{\omega}_M=\bm{\omega}_H - \bm{\omega}_F
\end{equation}
where $g$ characterizes the torque–speed relationship. Based on experimental measurements \citep{Berg}, $g$ takes the piecewise form:
\begin{equation}
	g = \begin{cases}
		C_1 & \text{if } \bm{\omega}_M < \bm{\omega}_\text{Knee}, \\
		C_2 + m \omega_M & \text{if } \bm{\omega}_\text{Knee} < \bm{\omega}_M < \bm{\omega}_\text{max},
	\end{cases} \label{eq:MotorParams}
\end{equation}
with constants defined in figure~\ref{fig:MotorTorque}) and table~\ref{tab:EcoliBergPoon}.

The total stress on the head is the sum of solvent and polymer contributions, $\bm{\sigma} = \bm{\sigma}_s + \bm{\sigma}_p$. The resulting hydrodynamic force and torque are:
\begin{equation}
	\bm{F}_\text{Head} = \int d\bm{r} \, \bm{\sigma} \cdot \bm{n}, \quad
	\bm{L}_\text{Head} = \int d\bm{r} \, (\bm{\sigma} \cdot \bm{n}) \times \bm{r}. \label{eq:FandLHead}
\end{equation}
The flagellar force and torque are computed from the force per unit length $\bm{f}(s)$:
\begin{equation}
	\bm{F}_\text{Flag} = -\int ds \, \bm{f}(s), \quad
	\bm{L}_\text{Flag} = -\int ds \, \bm{r}_F(s) \times \bm{f}(s).
\end{equation}

In the following part, we summarize the role of each component of our model described above and provide an overview of how everything fits together.

\subsection{Operator formulation and model overview}
The objective is to determine the swimming velocities $\bm{U}$, $\bm{\omega}_F$, and $\bm{\omega}_H$ of a flagellated bacterium given an internal motor torque $\bm{L}_M$. This torque drives the flagellar bundle to rotate at an angular velocity $\bm{\omega}_F$, generating a hydrodynamic torque $\bm{L}_\text{Flag}$ (equation~\eqref{TM}). To satisfy torque balance (equation~\eqref{TF}), the head counter-rotates at $\bm{\omega}_H$, producing a resisting torque $\bm{L}_\text{Head}$. Meanwhile, the rotating flagellar bundle also induces a distributed force $\bm{f}$ along its length, which generates a net thrust $\bm{F}_\text{Flag}$ that is balanced by the hydrodynamic drag on the head, $\bm{F}_\text{Head}$, resulting in a net swimming velocity $\bm{U}$ (equation~\eqref{FF}).

The force distribution $\bm{f}$ is computed from slender-body theory (equation~\eqref{SBTEQ}), and depends on the swimming kinematics, flagellar geometry, fluid viscosities, $\mu_s$ and $\mu_p$, the screening length $L_B$, and an effective background velocity $\bm{u}_s^{\infty}$. This background includes contributions from the imposed flow, head–tail hydrodynamic interactions, and the polymeric stress field. In turn, $\bm{f}$ drives the fluid velocity field $\bm{u}_s$ via the solvent momentum equation~\eqref{EQ2}, which is coupled to the velocity of polymer phase $\bm{u}_p$ through $L_B$-dependent drag (equations~\eqref{EQ2}–\eqref{EQ3}) and to the stress field $\bm{\Pi}$ through polymer momentum balance~\eqref{EQ3}. The polymer stress $\bm{\Pi}$ evolves according to the configuration tensor $\bm{\Lambda}$ via the constitutive relation~\eqref{eq:PiDef} and~\eqref{ConstEq}, completing a closed nonlinear system involving bacterial motion, fluid mechanics, and polymer microstructure. This modular formulation enables efficient, non-iterative evaluation by treating each subsystem separately and exploiting linearity in the inputs.

As a starting point, we temporarily treat the polymer stress $\bm{\Pi}$ as a known input. With this assumption, the remaining system, comprising the hydrodynamic equations and the slender-body formulation, is linear in its unknowns. This structure is best expressed in operator form, which we now present.

\vspace{1ex}
\noindent\textbf{Main fluid problem (MFP).}
The viscoelastic flow around the bacterium, governed by equations~\eqref{EQ1}–\eqref{EQ3} and boundary conditions~\eqref{BCfar}–\eqref{BC2D}, can be compactly written as:
\begin{equation}
	\hat{\mathcal{P}}_\text{MFP}: \hat{\mathcal{S}}_\text{Mass-Momentum} \cdot \bm{v}_\text{MFP} = \hat{\bm{b}}_\text{MFP}(\bm{f}, \bm{U} + \bm{\omega}_H \times \bm{r}, \langle\bm{u}\rangle + \bm{r} \cdot \bm{\Gamma}, \bm{\Pi}), \label{eq:FullOper}
\end{equation}
with the interior and boundary parts:
\begin{align}
	\mathcal{P}_\text{MFP}:&\quad \mathcal{S}_\text{Mass-Momentum} \cdot \bm{v}_\text{MFP} = \bm{b}_\text{MFP}(\bm{f}, \bm{\Pi}), \label{eq:IntMFP}\\
	\mathcal{B}_\text{MFP}^{\text{BC}}:&\quad \mathcal{S}_\text{MFP}^{\text{BC}} \cdot \bm{v}_\text{MFP} = \bm{b}_\text{MFP}^{\text{BC}}(\bm{U} + \bm{\omega}_H \times \bm{r}, \langle\bm{u}\rangle + \bm{r} \cdot \bm{\Gamma}, \bm{\Pi}). \label{eq:BCMFP}
\end{align}

\vspace{1ex}
\noindent\textbf{Flagellum-only flow.}
The velocity field generated by the flagellar bundle in isolation (equations~\eqref{uflagEQ2}–\eqref{uflagBC}) follows a similar structure:
\begin{equation}
	\hat{\mathcal{P}}_\text{Flag}: {\mathcal{S}}_\text{Mass-Momentum} \cdot \bm{v}_\text{Flag} = {\bm{b}}_\text{Flag}(\bm{f}).\label{eq:FlagProb}
\end{equation}

\vspace{1ex}
\noindent\textbf{Slender-body theory (SBT).}
The force distribution $\bm{f}$ and swimming kinematics are determined from:
\begin{equation}
	\mathcal{P}_\text{SBT}: \mathcal{S}_\text{SBT} \cdot \bm{v}_\text{SBT} = \bm{b}_\text{SBT}(\bm{u}_s, \bm{u}_p, p_s, p_p, \bm{u}_{s,\text{Flag}}, \bm{u}_s^\infty, \langle\bm{u}\rangle,  \bm{\Pi}, \bm{\Gamma}, g \bm{\omega}_F/2). \label{SBTOperator}
\end{equation}

The unknown vectors are:
\begin{equation}
	\bm{v}_\text{MFP} = \begin{bmatrix} \bm{u}_s \\ \bm{u}_p \\ p_s \\ p_p \end{bmatrix}, \quad
	\bm{v}_\text{Flag} = \begin{bmatrix} \bm{u}_{s,\text{Flag}} \\ \bm{u}_{p,\text{Flag}} \\ p_{s,\text{Flag}} \\ p_{p,\text{Flag}} \end{bmatrix}, \quad
	\bm{v}_\text{SBT} = \begin{bmatrix} \bm{f} \\ \bm{U} \\ \bm{\omega}_H \\ \bm{\omega}_F \end{bmatrix}. \label{eq:Unknowns}
\end{equation}
The interior source vectors are
\begin{equation}
	\bm{b}_\text{MFP} = \begin{bmatrix} \mathcal{L} \cdot \bm{f} \\ -\nabla \cdot \bm{\Pi} \\ 0 \\ 0 \end{bmatrix}, \quad
	\bm{b}_\text{Flag} = \begin{bmatrix} \mathcal{L} \cdot \bm{f} \\ 0 \\ 0 \\ 0 \end{bmatrix}, \quad
	\bm{b}_\text{SBT} = \begin{bmatrix}
		\bm{u}_{s,\text{Flag}} - \bm{u}_s^\infty + \langle\bm{u}\rangle + \bm{r} \cdot \bm{\Gamma} \\
		-\bm{F}_\text{Head} - \bm{F}^\infty \\
		-\bm{L}_\text{Head} - \bm{L}^\infty \\
		-\tilde{C}/2
	\end{bmatrix}.\label{eq:InternalSourceVectors}
\end{equation}
Here, $\mathcal{L} \cdot \bm{f} = \int ds\, \bm{f}(\bm{r}_F(s)) \delta(\bm{r} - \bm{r}_F(s))$. The quantities $\bm{F}_\text{Head}$ and $\bm{L}_\text{Head}$ depend linearly on the fluid stresses and polymer contribution. The shared mass–momentum operator for the main fluid problem and  flagellum-only flow system is:
\begin{equation}
	\mathcal{S}_\text{Mass-Momentum} = \begin{bmatrix}
		\mu_s(\nabla^2 - 1/L_B^2) & \mu_s/L_B^2 & -\nabla & 0 \\
		\mu_s/L_B^2 & \mu_p \nabla^2 - \mu_s/L_B^2 & 0 & -\nabla \\
		\nabla \cdot & 0 & 0 & 0 \\
		0 & \nabla \cdot & 0 & 0
	\end{bmatrix},\label{eq:MFMOper}
\end{equation}
and that for the SBT system is:
\begin{align}
	\mathcal{S}_\text{SBT} &= \begin{bmatrix}
		\mathcal{S}_\text{SBT}^f \circ & -\bm{I} & 0 & -\bm{r} \times \\
		\int_{\bm{r}_F(s)} ds & 0 & 0 & 0 \\
		-\int_{\bm{r}_F(s)} ds\, \bm{r}_F(s) \times & 0 & 0 & 0 \\
		-\int_{\bm{r}_F(s)} ds\, \bm{r}_F(s) \times & 0 & \tilde{m}/2 & -\tilde{m}/2
	\end{bmatrix},\label{eq:FlagOper}
\end{align}
where the operator $\mathcal{S}_\text{SBT}^f \circ$ acts only on $\bm{f}$ such that $\mathcal{S}_\text{SBT}^f \circ\bm{f}$ represents all the terms on the left hand side of equation \eqref{SBTEQ}. Parameters $\tilde{C}$ and $\tilde{m}$ (either equal to $C_1$ and 0 or $C_2$ and $m$) account for piecewise motor torque, following equation~\eqref{eq:MotorParams} and figure~\ref{fig:MotorTorque}). 

\begin{figure}
	\centering
	\includegraphics[width=0.7\textwidth]{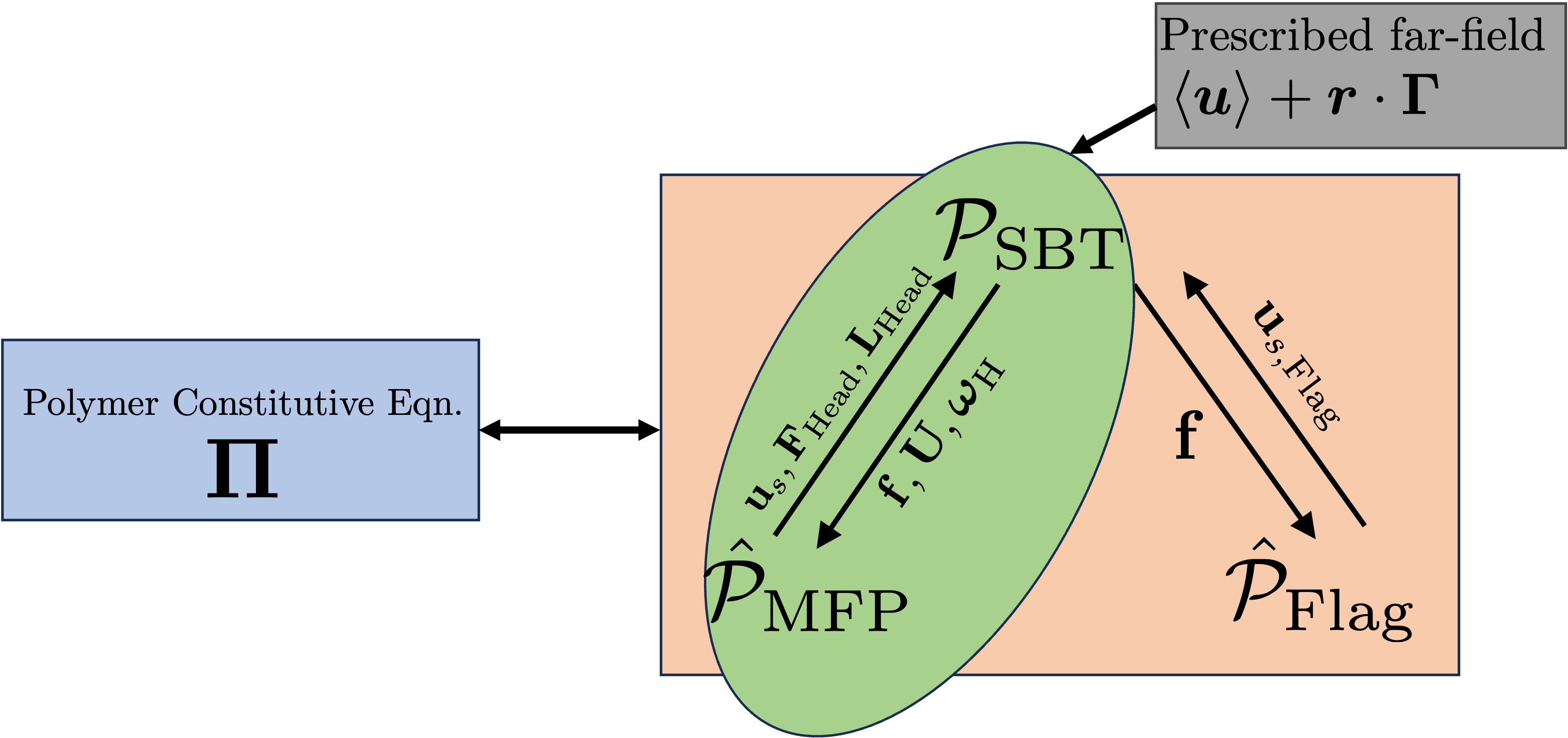}
	\caption{Schematic showing the coupling between the polymer stress, $\bm{\Pi}$ and three subproblems involving unknowns in equation \eqref{eq:Unknowns}. The three subproblems are the main fluid–polymer system, the flagellum-only flow, and the slender-body theory solver.}
	\label{fig:ModelSchematic}
\end{figure}

Figure~\ref{fig:ModelSchematic} shows the directional dependencies between subsystems. The main fluid problem $\hat{\mathcal{P}}_\text{MFP}$ depends on $\bm{f}$, $\bm{U}$, and $\bm{\omega}_H$ provided by $\mathcal{P}_\text{SBT}$, while the flagellum-only flow $\hat{\mathcal{P}}_\text{Flag}$ uses the same $\bm{f}$. In turn, the SBT system $\mathcal{P}_\text{SBT}$ requires $\bm{u}_s$, $\bm{u}_p$, $p_s$, and $p_p$ from $\hat{\mathcal{P}}_\text{MFP}$ and $\bm{u}_{s,\text{Flag}}$ from $\hat{\mathcal{P}}_\text{Flag}$. 

\vspace{1ex}
\noindent\textbf{Nonlinear coupling and decomposition.}
In this sub-section, so far, we assumed the polymer stress, $\bm{\Pi}$, was known. In reality, $\bm{\Pi}$ is tightly coupled to the flow field (figure~\ref{fig:ModelSchematic}): it modifies the velocity and pressure fields through momentum balance, which in turn alter $\bm{f}$, $\bm{F}_\text{Head}$, and $\bm{L}_\text{Head}$. The $\bm{F}_\text{Head}$, and $\bm{L}_\text{Head}$ then determine the appropriate values of $\bm{U}$, $\bm{\omega}_H$, and $\bm{\omega}_F$ to maintain a force and torque free motion. These $\bm{U}$ and $\bm{\omega}_H$ are required as boundary conditions in the momentum balance. Meanwhile, $\bm{\Pi}$ evolves in response to $\bm{u}_p$. This creates a feedback loop that, if approached naively, would require iterative solution strategies such as a secant method~\citep{padhy2013simulations}.

Instead, we exploit a key property: the linearity of the hydrodynamic subproblems. The entire system can be exactly decomposed into a Newtonian (motion-induced) component and a non-Newtonian (polymer stress-driven) component. The swimming velocities $\bm{U}$, $\bm{\omega}_F$, and $\bm{\omega}_H$ influence only the Newtonian part. The polymer stress $\bm{\Pi}$ contributes additively through a non-Newtonian subproblem. This resistivity formulation allows us to isolate the effect of $\bm{\Pi}$ and solve for bacterial motion without iterations. A similar decomposition, valid for elastoviscoplastic media, was introduced by~\citet{Sharma23} for inertialess flow of non-Newtonian liquids past passive particles. In the next section, we extend that approach to include active forcing from a flagellar bundle and the additional coupling via slender-body theory and the flagellum-only flow.

\section{Exact linear decomposition of flow fields and hydrodynamic moments} \label{sec:LinDecomp}
We begin by considering the main fluid problem, $\hat{\mathcal{P}}_\text{MFP}$. As illustrated in figure~\ref{fig:ModelSchematic}, its inputs include the head velocity $\bm{U} + \bm{\omega}_H \times \bm{r}$, the background flow $\langle\bm{u}\rangle + \bm{r} \cdot \bm{\Gamma}$, the flagellar force distribution $\bm{f}(\bm{r}_F(s))$, and the polymer stress $\bm{\Pi}$. These, whether specified or arising from coupled components, are assembled into the composite source vector $\hat{\bm{b}}_\text{MFP}$ in equation~\eqref{eq:FullOper}.

As seen in equations~\eqref{eq:FullOper} and \eqref{eq:InternalSourceVectors}, $\hat{\bm{b}}_\text{MFP}$ depends linearly on its inputs and may therefore be decomposed as
\begin{equation}
	\hat{\bm{b}}_\text{MFP} = \hat{\bm{b}}_\text{MFP}^{(M)} + \hat{\bm{b}}_\text{MFP}^{(P)} + \hat{\bm{b}}_\text{MFP}^{(\text{Flag})},
\end{equation}
where the superscripts $(M)$, $(P)$, and $(\text{Flag})$ denote contributions from motion (bacterial head motion and background flow), polymer stress, and flagellar forcing, respectively.

This defines three sub-problems:
\begin{equation}
	\hat{\mathcal{P}}_\text{MFP} = \hat{\mathcal{P}}_\text{MFP}^{(M)} + \hat{\mathcal{P}}_\text{MFP}^{(P)} + \hat{\mathcal{P}}_\text{MFP}^{(\text{Flag})},\label{eq:MFPDecomp}
\end{equation}
each obtained by substituting the appropriate source term and unknowns into~\eqref{eq:FullOper}. For example the polymer stress induced problem is:
\begin{equation}
	\hat{\mathcal{P}}_\text{MFP}^{(P)}:\quad \mathcal{S}_\text{Mass-Momentum} \cdot \bm{v}_\text{MFP}^{(P)} = \bm{b}_\text{MFP}^{(P)}, \quad \bm{b}_\text{MFP}^{(P)}=\begin{bmatrix} 0 & -\nabla \cdot \bm{\Pi} & 0 & 0 \end{bmatrix}^T.\label{eq:IntMFPPoly}
\end{equation} such that the field $\bm{v}_\text{MFP}^{(P)}$  is driven only by $\nabla \cdot \bm{\Pi}$ in a quiescent fluid with a stationary head.
The motion-driven field $\bm{v}_\text{MFP}^{(M)}$ corresponds to a translating–rotating bacterial head (without flagella or polymer stress) in a Newtonian two-fluid medium with background flow $\langle\bm{u}\rangle + \bm{r} \cdot \bm{\Gamma}$. The field $\bm{v}_\text{MFP}^{(\text{Flag})}$ represents the flow induced by $\int ds\, \bm{f}(\bm{r}_F(s)) \delta(\bm{r} - \bm{r}_F(s))$ in a quiescent fluid with a stationary head. Only $\bm{v}_\text{MFP}^{(M)}$ satisfies the full boundary conditions~\eqref{BCfar}–\eqref{BC2D}, with replacements such as $\bm{u}_s \rightarrow \bm{u}_s^{(M)}$, etc. Sub-problems $\hat{\mathcal{P}}_\text{MFP}^{(P)}$ and $\hat{\mathcal{P}}_\text{MFP}^{(\text{Flag})}$ satisfy the same boundary conditions with homogeneous (zero) right-hand sides and are purely source-driven.

The flagellar sub-problem, $\hat{\mathcal{P}}_\text{MFP}^{(\text{Flag})}$, corresponds to the flow generated by a line distribution of two-fluidlets adjacent to a stationary head. These two-fluidlets, analogous to Stokeslets in Newtonian media, are the singularity solutions of the two-fluid Newtonian equations~\citep{Sabarish}. The modular decomposition of $\hat{\mathcal{P}}_\text{MFP}$, particularly the separation of motion- and flagella-induced components, enables precomputation of flow fields associated with $\hat{\mathcal{P}}_\text{MFP}^{(M)}$ and $\hat{\mathcal{P}}_\text{MFP}^{(\text{Flag})}$ across a broad range of fluid parameters. This significantly improves computational efficiency by avoiding repeated solutions of these sub-problems during the time integration of the polymer constitutive equations.

The hydrodynamic force and torque on the head can also be decomposed into corresponding components:
\begin{align}
	\bm{F}_\text{Head} &= \bm{F}_\text{Head}^{(M)} + \bm{F}_\text{Head}^{(P)} + \bm{F}_\text{Head}^{(\text{Flag})}, \\
	\bm{L}_\text{Head} &= \bm{L}_\text{Head}^{(M)} + \bm{L}_\text{Head}^{(P)} + \bm{L}_\text{Head}^{(\text{Flag})}. \label{eq:MomentMFM}
\end{align}
The components $\bm{F}_\text{Head}^{(M)}$ and $\bm{L}_\text{Head}^{(M)}$ are due to the motion of an isolated head in a quiescent two-fluid and can be obtained from the resistivity problem:
\begin{align}
	\bm{F}_\text{Head}^{(M)} &= \bm{R}_u \cdot \bm{U} + \bm{R}_\omega \cdot \bm{\omega}_H, \\
	\bm{L}_\text{Head}^{(M)} &= \bm{Q}_u \cdot \bm{U} + \bm{Q}_\omega \cdot \bm{\omega}_H.
\end{align}
Here, $\bm{R}_u$, $\bm{R}_\omega$, $\bm{Q}_u$, and $\bm{Q}_\omega$ are resistance tensors that may be computed analytically for a sphere or numerically for a spheroid in a two-fluid medium. These calculations can be performed offline prior to the transient simulation. These calculations need to be performed only once and they remain valid across a wide range of flow parameters. In principle, the spheroidal harmonics formulation of~\cite{dabade2015effects,dabade2016effect} could also be extended to Newtonian two-fluid systems to enable analytical evaluation for a spheroid. This is possible since the general solutions for flows past spheroids are available for both Stokes and Brinkman flow, and the two-fluid equations can be readily re-written as Stokes and Brinkman equations for the combined (mixture) and difference fields~\cite{Sabarish}. However, in our current implementation, we stick with numerical calculations for a spheroid. The resistance tensors depend on head shape, screening length, and the viscosity ratio between the solvent and polymer phases. The components $\bm{F}_\text{Head}^{(P)}$ and $\bm{L}_\text{Head}^{(P)}$ account for contributions from polymer-induced velocity, pressure, and stress fields:
\begin{equation}
	\bm{F}_\text{Head}^{(P)} = \int d\bm{r} \, \bm{\sigma}^{(P)} \cdot \bm{n}, \quad
	\bm{L}_\text{Head}^{(P)} = \int d\bm{r} \, (\bm{\sigma}^{(P)} \cdot \bm{n}) \times \bm{r},\label{eq:ForceP}
\end{equation}
where
\begin{align}
	\bm{\sigma}^{(P)} = [-p_s^{(P)} \bm{I} + \mu_s (\nabla \bm{u}_s^{(P)} + (\nabla \bm{u}_s^{(P)})^T)] 
	+ [-p_p^{(P)} \bm{I} + \mu_p (\nabla \bm{u}_p^{(P)} + (\nabla \bm{u}_p^{(P)})^T) + \bm{\Pi}].
\end{align}
Similarly, the remaining components $\bm{F}_\text{Head}^{(\text{Flag})}$ and $\bm{L}_\text{Head}^{(\text{Flag})}$, due to flagellar forcing, can be expressed in a matrix–vector form, with the vector being $\bm{f}$ and the matrix precomputed offline. Critically, expressing the dependence of hydrodynamic forces and torques on $\bm{f}$, $\bm{U}$, and $\bm{\omega}_H$ in this matrix–vector form allows us to avoid iterative solutions for these quantities, thereby enhancing computational efficiency. In the remaining part of this section we will consider further details on these precomputations and reformulation to the equations presented in the previous section to enable iteration free numerical evaluations.

\subsection{Precomputations for flow due to kinematic motion ($\hat{\mathcal{P}}_\text{MFP}^{(M)}$)}
The bacterial head undergoes translational and rotational motion, while the imposed background flow consists of uniform and linear components. The relative far-field flow experienced by the head can therefore be represented as a linear combination of translational and linear basis flows:
\begin{align}
	\mathbf{u}_\infty^\text{relative} &= \sum_{i=1}^3 a_i \hat{\mathbf{u}}^{(i)} + \mathbf{r} \cdot \sum_{i=1}^8 b_i \bm{S}^{(i)}, \label{eq:BasisFlowsDecomp}
\end{align}
where,
\begin{align}
\begin{split}
		&\hat{\mathbf{u}}^{(1)} = \begin{bmatrix}1 & 0 & 0\end{bmatrix}^T, 
		\hat{\mathbf{u}}^{(2)} = \begin{bmatrix}0 & 1 & 0\end{bmatrix}^T, 
		\hat{\mathbf{u}}^{(3)} = \begin{bmatrix}0 & 0 & 1\end{bmatrix}^T, \\
		&\bm{S}^{(1)} = \begin{bmatrix}1 & 0 & 0\\ 0 & -1 & 0\\ 0 & 0 & 0\end{bmatrix}, 
		\bm{S}^{(2)} = \begin{bmatrix}-\frac{1}{2} & 0 & 0\\ 0 & -\frac{1}{2} & 0\\ 0 & 0 & 1\end{bmatrix}, 
		\bm{S}^{(3)} = \begin{bmatrix}0 & 1 & 0\\ 1 & 0 & 0\\ 0 & 0 & 0\end{bmatrix}, 
		\bm{S}^{(4)} = \begin{bmatrix}0 & 0 & 1\\ 0 & 0 & 0\\ 1 & 0 & 0\end{bmatrix}, \\
		&\bm{S}^{(5)} = \begin{bmatrix}0 & 0 & 0\\ 0 & 0 & 1\\ 0 & 1 & 0\end{bmatrix}, 
		\bm{S}^{(6)} = \begin{bmatrix}0 & 1 & 0\\ -1 & 0 & 0\\ 0 & 0 & 0\end{bmatrix}, 
		\bm{S}^{(7)} = \begin{bmatrix}0 & 0 & -1\\ 0 & 0 & 0\\ 1 & 0 & 0\end{bmatrix}, 
		\bm{S}^{(8)} = \begin{bmatrix}0 & 0 & 0\\ 0 & 0 & 1\\ 0 & -1 & 0\end{bmatrix}.
	\end{split}
\end{align}
The coefficients $a_i$ and $b_i$ are determined by the bacterial velocity and background flow through an appropriate change of reference frame. Accordingly, the vector $\hat{\bm{b}}_\text{MFP}^{(M)}$ can be constructed as a linear combination of these eleven basis flows, and the resulting motion-induced flow field is given by
\begin{align}
	\bm{v}_\text{MFP}^{(M)} = \sum_{i=1}^3 a_i \bm{v}^{(M, U_i)} + \sum_{i=1}^8 b_i \bm{v}^{(M, S_i)}, \label{eq:BasisFlowFieldsAddition}
\end{align}
where each basis flow $\bm{v}^{(M, U_i)}$ or $\bm{v}^{(M, S_i)}$ represents the response of an isolated bacterial head (without flagella) to a unit translational or linear flow in a Newtonian two-fluid medium.

These basis fields can be computed either analytically using spheroidal harmonics~\citep{dabade2015effects,dabade2016effect} or numerically using finite difference methods~\citep{Sharma23}. A similar basis decomposition strategy was used in~\cite{sharma2025sedimentation} to analyze spheroid dynamics in viscosity-stratified flows. Using the precomputed basis fields, the velocity components $\bm{u}_s^{(M)}$ and $\bm{u}_p^{(M)}$ as well as the corresponding pressures $p_s^{(M)}$ and $p_p^{(M)}$ can be expressed as linear combinations of stored field solutions, scaled by the translational and rotational velocities ($\bm{U}$, $\bm{\omega}_H$) and background flow parameters ($\langle\bm{u}\rangle$, $\bm{\Gamma}$). The fields proportional to $\bm{U}$ and $\bm{\omega}_H$ are denoted as $\bm{u}_s^{(M,U)}$ and $\bm{u}_s^{(M,\bm{\omega})}$  for later use.

\subsection{Pre-computations for flows due to flagellar forcing ($\hat{\mathcal{P}}_\text{MFP}^{(\text{Flag})}$ and $\hat{\mathcal{P}}_\text{Flag}$)}\label{sec:FlagFlow}
\begin{figure}
	\centering
	\includegraphics[width=0.6\textwidth]{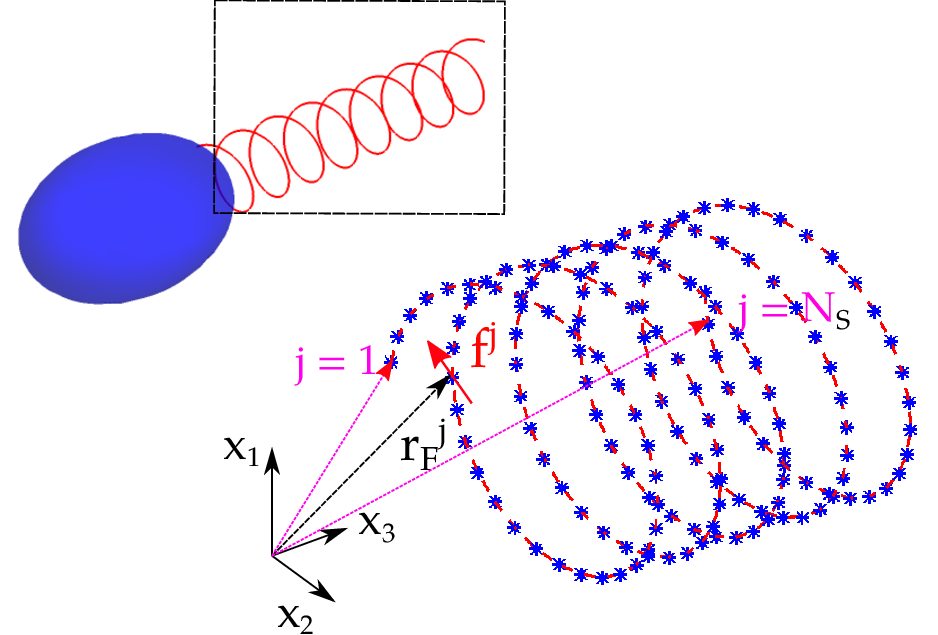}
	\caption{Discretization of the helical line of singularities into $N_S$ points. Each point on the helix (blue stars) carries a two-fluidlet, a fundamental singularity of the two-fluid medium.}
	\label{fig:HelixSchematic}
\end{figure}

We discretize the helical centerline of the flagellar bundle into $N_S$ points, whose position vectors are given by $\bm{r}_F^j$ for $j \in [1, N_S]$. At each point, we place a point force, thereby replacing the continuous force integral with a discrete sum (see figure~\ref{fig:HelixSchematic}). 
This discretization strategy, standard in slender-body theories, reduces the original integral formulation to a matrix–vector system~\citep{Swinney,Sabarish}. The force distribution $\bm{f}$ is obtained by solving $\mathcal{P}_\text{SBT}$ and is treated as unknown for now. The source term in $\hat{\mathcal{P}}_\text{MFP}^{(\text{Flag})}$ is approximated via a Riemann sum:
\begin{equation}
	\int \! \mathrm{d}s\, \bm{f}(\bm{r}_F(s)) \delta(\bm{r} - \bm{r}_F(s))
	\approx \sum_{i=1}^3 \sum_{j=1}^{N_S} f_i^j\, \mathcal{F}_{\text{TF-let},i}(\bm{r}_F^j)\, \Delta s. \label{PTFlet}
\end{equation}
Here, $\mathcal{F}_{\text{TF-let},i}(\bm{r}_F^j)$ denotes the field induced by a unit point force in the $i$-th direction placed at $\bm{r}_F^j$, adjacent to a stationary spheroidal head in a quiescent two-fluid medium.

The problem $\hat{\mathcal{P}}_\text{Flag}$, defined in equation~\eqref{eq:FlagProb}, is structurally identical to $\hat{\mathcal{P}}_\text{MFP}^{(\text{Flag})}$ but assumes that the point forces act in isolation, i.e., without the bacterial head. The resulting flow from each singularity is denoted $\bm{v}_{\text{TF-let}}^{(i)}(\bm{r}_F^j)$.
Using equations~\eqref{eq:Unknowns} and \eqref{eq:MFMOper}, the two-fluidlet vector is written as:
\begin{equation}
	\bm{v}_{\text{TF-let}}^{(i)}(\bm{r}_F^j)=\begin{bmatrix} \bm{u}_{s,\text{TF-let}}^{ij} & \bm{u}_{p,\text{TF-let}}^{ij} & p_{s,\text{TF-let}}^{ij} & p_{p,\text{TF-let}}^{ij} \end{bmatrix}^T,
\end{equation}
governed by:
\begin{equation}
	\mathcal{S}_\text{Mass-Momentum}\cdot \bm{v}_{\text{TF-let}}^{(i)}(\bm{r}_F^j) = \bm{b}_{\text{TF-let}}^{ij},
\end{equation}
with source term $\bm{b}_{\text{TF-let}}^{ij}=\begin{bmatrix}
	0&\bm{e}_i \delta&0&0
\end{bmatrix}$ and boundary conditions
\begin{align}
	\bm{u}_{s,\text{TF-let}}^{ij}= \bm{u}_{p,\text{TF-let}}^{ij} \rightarrow 0 \quad \text{as } \bm{r} \rightarrow \infty.
\end{align} 
The flow due to a single two-fluidlet can be computed analytically using canonical Stokeslet and Brinkmanlet~\citep{Sabarish}, and the total flagellar bundle-induced flow is:
\begin{equation}
	\bm{v}_\text{Flag}
	= \sum_{i=1}^3 \sum_{j=1}^{N_S} f_i^j\, \bm{v}_{\text{TF-let}}^{(i)}(\bm{r}_F^j). \label{eq:PTFletFlow}
\end{equation}

To account for the bacterial head, we compute the modified fields $\bm{v}_{\text{TF-let}}^{\text{Head},(i)}(\bm{r}_F^j)$ numerically. These are decomposed as:
\begin{equation}
	\bm{v}_{\text{TF-let}}^{\text{Head},(i)}(\bm{r}_F^j) = \tilde{\bm{v}}_{\text{TF-let}}^{\text{Head},(i)}(\bm{r}_F^j) + \bm{v}_{\text{TF-let}}^{(i)}(\bm{r}_F^j), \label{eq:SIngFlowDecomp}
\end{equation}
where $\tilde{\bm{v}}_{\text{TF-let}}^{\text{Head},(i)}$ satisfies
\begin{equation}
	\mathcal{S}_\text{Mass-Momentum}\cdot \tilde{\bm{v}}_{\text{TF-let}}^{\text{Head},(i)}(\bm{r}_F^j) = 0,\label{eq:UtildeEqn}
\end{equation}
with $\tilde{\bm{v}}_{\text{TF-let}}^{\text{Head},(i)} \rightarrow 0$ as $\bm{r} \rightarrow \infty$, and boundary conditions at the head surface:
\begin{align}
	\tilde{\bm{u}}_{s,\text{TF-let}}^{\text{Head},ij} = -{\bm{u}}_{s,\text{TF-let}}^{ij}, \quad \tilde{\bm{u}}_{p,\text{TF-let}}^{\text{Head},ij} = -{\bm{u}}_{p,\text{TF-let}}^{ij}\label{eq:UtildeBC1}
\end{align}
for no slip condition on the polymer phase, and
\begin{align}
	\tilde{\bm{u}}_{s,\text{TF-let}}^{\text{Head},ij} &= -{\bm{u}}_{s,\text{TF-let}}^{ij},\label{eq:UtildeBC2a} \\
	\tilde{\bm{u}}_{p,\text{TF-let}}^{\text{Head},ij} \cdot \bm{n} &= -{\bm{u}}_{p,\text{TF-let}}^{ij} \cdot \bm{n},\label{eq:UtildeBC2b} \\
	(\bm{I} - \bm{n n}) \cdot (\tilde{\bm{\sigma}}_{p,\text{TF-let}}^{ij} \cdot \bm{n}) &= (\bm{I} - \bm{n n}) \cdot ({\bm{\sigma}}_{p,\text{TF-let}}^{ij} \cdot \bm{n})\label{eq:UtildeBC2c}
\end{align}
for a perfectly slipping polymer phase.

Here, ${\bm{\sigma}}_{p,\text{TF-let}}^{ij} = -p_p \bm{I} + \mu_p (\nabla {\bm{u}}_{p,\text{TF-let}}^{ij} + \nabla {\bm{u}}_{p,\text{TF-let}}^{ij}{}^T)$, and similarly for $\tilde{\bm{\sigma}}_{p,\text{TF-let}}^{\text{Head},ij}$. Solving for $\tilde{\bm{v}}_{\text{TF-let}}^{\text{Head},(i)}$ avoids directly resolving singularities, as the flow at the head surface is smooth and numerically tractable.

After computing $\tilde{\bm{v}}_{\text{TF-let}}^{\text{Head},(i)}$, we recover $\bm{v}_{\text{TF-let}}^{\text{Head},(i)}$ and obtain the total flow:
\begin{equation}
	\bm{v}_\text{MFP}^{(\text{Flag})}
	= \sum_{i=1}^3 \sum_{j=1}^{N_S} f_i^j\, \bm{v}_{\text{TF-let}}^{\text{Head},(i)}(\bm{r}_F^j). \label{eq:PTFletFlowWithHead}
\end{equation}
This procedure also yields the hydrodynamic force and torque on the head:
\begin{equation}
	\bm{F}_\text{Head}^{(\text{Flag})} = \bm{A}_1 \cdot \bm{f}, \quad
	\bm{L}_\text{Head}^{(\text{Flag})} = \bm{A}_2 \cdot \bm{f}, \quad \bm{A}_1, \bm{A}_2 \in \mathbb{R}^{3 \times 3N_S}.
\end{equation}
Here, columns $3j{-}2$ to $3j$ of $\bm{A}_1$ and $\bm{A}_2$ represent the head's force and torque from unit forces in directions 1 to 3 at $\bm{r}_F^j$.

Finally, since the solvent velocity from flagellar forcing is the first component of $\bm{v}_\text{Flag}$ in equation~\eqref{eq:Unknowns}, we write:
\begin{equation}
	\bm{u}_{s,\text{Flag}} = \bm{A}_3^\text{flag} \cdot \bm{f},
\end{equation}
where $\bm{A}_3^\text{flag}$ is assembled from the solvent velocity fields generated by individual point forces.

\subsection{Reformulating the SBT problem ($\hat{\mathcal{P}}_\text{SBT}$)}\label{sec:ReformSBT}
Based on the previous two subsections, several components in the SBT forcing term (equation~\eqref{eq:InternalSourceVectors}) can be expressed as matrix–vector products involving the unknown vector $\bm{v}_\text{SBT}$ (defined in equation~\eqref{eq:Unknowns}). We therefore reformulate the SBT problem in operator form as:
\begin{equation}
	\hat{\mathcal{P}}_\text{SBT}: \quad \hat{\mathcal{S}}_\text{SBT} \cdot \bm{v}_\text{SBT} = \hat{\bm{b}}_\text{SBT}, \label{SBTOperator2}
\end{equation}
where
\begin{eqnarray}
	\hat{\mathcal{S}}_\text{SBT}=&
	\begin{bmatrix}
		&-\bm{A}_3^\text{flag}+\mathcal{S}_\text{SBT}^f\circ & -\bm{I}+\bm{u}_s^{(M,U)}  & \bm{u}_s^{(M,\omega)}  &-\bm{r}_F(s) \times\\
		&\bm{A}_1+\int_{\bm{r}_F(s)} ds &\bm{R}_u &\bm{R}_{\omega}  &0\\
		&\bm{A}_2-\int_{\bm{r}_F(s)} ds \,  \bm{r}_F(s) \times &\bm{Q}_u &\bm{Q}_{\omega}  &0\\
		&-\int_{\bm{r}_F(s)} ds \,  \bm{r}_F(s) \times &0 &\frac{m}{2} &-\frac{m}{2}
	\end{bmatrix},\label{eq:SBTTensor}\\
	\hat{\bm{b}}_\text{SBT}=&\begin{bmatrix}
		-\bm{w}&
		-\bm{F}^\infty&
		-\bm{L}^\infty&
		-\tilde{C}/2
	\end{bmatrix}^T+\begin{bmatrix}
		-\bm{u}_s^{(P)} &
		-\bm{F}_\text{Head}^{(P)}&
		-\bm{L}_\text{Head}^{(P)}&
		0
	\end{bmatrix}^T.\label{eq:SBTVector}
\end{eqnarray}
Here, $\tilde{C}$ arises from the prescribed motor torque parameters, and $\bm{w}$ is the disturbance field generated by the head in a Newtonian two-fluid medium subject to the imposed background flow $\langle\bm{u}\rangle + \bm{r} \cdot \bm{\Gamma}$. The terms $\bm{F}^\infty$ and $\bm{L}^\infty$ denote externally imposed force and torque, respectively. All these components can be evaluated prior to the start of the transient simulation. The only components that must be computed online at each time step are those arising from polymer-induced effects, namely, the polymer-induced solvent velocity $\bm{u}_s^{(P)}$ and the polymer-induced force and torque on the head, $\bm{F}_\text{Head}^{(P)}$ and $\bm{L}_\text{Head}^{(P)}$. These depend on the polymer stress field $\bm{\Pi}$ and must be updated dynamically.

From the discussion in this section, it is evident that only the solution to $\hat{\mathcal{P}}_\text{MFP}^{(P)}$ and the polymer constitutive equations must be computed afresh at each time step. All other components, $\hat{\mathcal{P}}_\text{MFP}^{(M)}$, $\hat{\mathcal{P}}_\text{MFP}^{(\text{Flag})}$, and $\hat{\mathcal{P}}_\text{Flag}$, are obtained by superposing precomputed flow fields. Moreover, solving the SBT system $\hat{\mathcal{P}}_\text{SBT}$ involves inverting a fixed matrix $\hat{\mathcal{S}}_\text{SBT} \in \mathbb{R}^{12 \times (3N_S + 9)}$, which is computationally inexpensive relative to evaluating $\hat{\mathcal{P}}_\text{MFP}^{(P)}$ and evolving the polymer model. In the next section, we describe how this formulation is implemented in our numerical algorithm.

\section{Time-stepping algorithm for coupled swimmer–polymer dynamics}\label{sec:NumericalMethod}
Our numerical algorithm based on the formulation presented in the previous two sections is summarized in algorithm~\ref{alg:MainAlg}.
\begin{algorithm}
	\caption{Numerical solution of the two-fluid swimming problem \label{alg:MainAlg}}
	\begin{algorithmic}[1]
		
		\State \textbf{Input:} Head and flagellar geometry, fluid parameters, motor torque characteristics, and imposed forces and torques used to compute $\bm{F}^\infty$, $\bm{L}^\infty$, $\tilde{m}$ and $\tilde{C}$ in equation~\eqref{eq:SBTVector}.
		
		\State \textbf{Precompute:} Basis and singularity-driven flow fields, resistivity tensors for $\hat{\mathcal{P}}_\text{MFP}^{(M)}$ and $\hat{\mathcal{P}}_\text{MFP}^{(\text{Flag})}$, and assemble the tensors in equation~\eqref{eq:SBTTensor} and background flow $\bm{w}$ in equation~\eqref{eq:SBTVector}.
		
		\vspace{1mm}
		\Statex \textbf{Initialization} (at $t_0$, i.e., $t = 0$):
		\State Set $\bm{\Pi} = \bm{0}$ (unstretched polymers), implying $\bm{u}_s^{(P)} = \bm{u}_p^{(P)} = \bm{F}_\text{Head}^{(P)} = \bm{L}_\text{Head}^{(P)} = \bm{0}$
		\State Invert $\hat{\mathcal{S}}_\text{SBT}$ (equation \eqref{SBTOperator2}) to obtain $\bm{f}$, $\bm{U}$, $\bm{\omega}_H$, and $\bm{\omega}_F$
		\State Assemble $\bm{v}_\text{MFP}^{(M)}$, $\bm{v}_\text{MFP}^{(\text{Flag})}$, and $\bm{v}_\text{Flag}$ using precomputed basis flows (see equations~\eqref{eq:BasisFlowFieldsAddition}, \eqref{eq:PTFletFlow}, and \eqref{eq:PTFletFlowWithHead})
		\State Compute $\bm{u}_p(t_0) = \bm{u}_p^{(M)} + \bm{u}_p^{(\text{Flag})}$
		
		\vspace{1mm}
		\For{each time step $t_n$, $n \in [1, N]$}
		
		\State \textbf{Given:} $\bm{u}_p(t_{n-1})$, and $\bm{\Pi}(t_{n-1})$
		
		\State \textbf{(A) Evolve polymer stress:} Solve equations~\eqref{eq:PiDef}–\eqref{eq:yield} to compute $\bm{\Pi}(t_n)$ using $\bm{u}_p(t_n - 1)$ and $\bm{\Pi}(t_{n-1})$
		
		\State \textbf{(B) Compute polymer-induced flow:} Solve $\hat{\mathcal{P}}_\text{MFP}^{(P)}$ (see equation~\eqref{eq:IntMFPPoly}) to obtain $\bm{u}_s^{(P)}(t_n)$, $\bm{u}_p^{(P)}(t_n)$, $p_s^{(P)}(t_n)$, and $p_p^{(P)}(t_n)$, and update $\bm{F}_\text{Head}^{(P)}$ and $\bm{L}_\text{Head}^{(P)}$ from equation~\eqref{eq:ForceP}
		
		\State \textbf{(C) Update swimming kinematics:} Invert $\hat{\mathcal{S}}_\text{SBT}$ (equation \eqref{SBTOperator2}) with current $\bm{u}_s^{(P)}(t_n)$, $\bm{F}_\text{Head}^{(P)}$, and $\bm{L}_\text{Head}^{(P)}$ to recompute $\bm{f}$, $\bm{U}$, $\bm{\omega}_H$, and $\bm{\omega}_F$ (i.e., redo step 4)
		
		\State \textbf{(D) Update polymer phase velocity:} Reassemble $\bm{v}_\text{MFP}^{(M)}$, $\bm{v}_\text{MFP}^{(\text{Flag})}$, and $\bm{v}_\text{Flag}$; then compute $\bm{u}_p(t_n) = \bm{u}_p^{(M)} + \bm{u}_p^{(\text{Flag})} + \bm{u}_p^{(P)}$ (i.e., redo step 5 and the non-Newtonian extension of step 6)
		
		\State Advance to next time step
		
		\EndFor
		
		\vspace{1mm}
		\State \textbf{Output:} Time-evolving fields $\bm{u}_{s,p}(t)$, $\bm{\Pi}(t)$, $\bm{f}(s,t)$, $\bm{U}(t)$, $\bm{\omega}_H(t)$, and $\bm{\omega}_F(t)$
		
	\end{algorithmic}
\end{algorithm}

The algorithm proceeds with prescribed geometric and rheological inputs: head and flagellar aspect ratio and length, flagellar radius, pitch angle and length (equation~\eqref{eq:FlagShape}), solvent and polymer viscosities $\mu_s$ and $\mu_p$, and the screening length $L_B$. The motor parameters $C_1$, $C_2$, and $m$ (equation~\eqref{eq:MotorParams}) are grouped as $\tilde{C}$ and $\tilde{m}$. External forces ($\bm{F}^\infty$) and torques ($\bm{L}^\infty$) are also prescribed, and the computational domain radius $\bm{r}_\infty \gg 1$ is chosen to approximate far-field conditions.

The operator $\hat{\mathcal{S}}_\text{MFP}$ in equation~\eqref{eq:MFMOper} is inverted using a two-fluid extension of the finite-difference solver developed in \cite{Sharma23}, hereafter referred to as the \textit{two fluid solver}. This solver discretizes the continuum operators $\nabla \cdot$, $\nabla^2$, and $\nabla()$ using variable-order central differences, with specialized treatment near poles and internal boundaries in prolate spheroidal coordinates. In the present work, we extend the single-fluid viscoelastic formulation of \cite{Sharma23} to a two-fluid system by appending the governing equations of the solvent and polymer fluids.

For the prescribed geometry and rheology, the basis flows $\bm{v}^{(M, U_i)}$ and $\bm{v}^{(M, S_{ji})}$ from equation~\eqref{eq:BasisFlowFieldsAddition} are obtained using the \textit{two fluid solver} with zero head boundary conditions and either $\hat{\mathbf{u}}^{(i)}$ or $\mathbf{r} \cdot \bm{S}^{(i)}$ imposed at $\bm{r}_\infty$. The zero head boundary conditions refer to equation~\eqref{BCnoslip} or equations~\eqref{BC2B}–\eqref{BC2D} with zero right-hand side, depending on polymer size. Mixed boundary conditions may also be implemented as a weighted combination of these limits. These flows are then used to construct the tensors $\bm{R}_u$, $\bm{R}_\omega$, $\bm{Q}_u$, and $\bm{Q}_\omega$ in $\hat{\mathcal{S}}_\text{SBT}$ (equation~\eqref{eq:SBTTensor}) and the vector $\bm{w}$ in $\hat{\bm{b}}_\text{SBT}$ (equation~\eqref{eq:SBTVector}).

As described in section~\ref{sec:FlagFlow}, we compute the regular field $\tilde{\bm{v}}_{\text{TF-let}}^{\text{Head},(i)}(\bm{r}_F^j)$ by subtracting the analytically known singular solution from the full numerical solution of the singularity-driven flow around the head. This is also achieved using the \textit{two fluid solver}. These fields are used to build the tensors $\bm{A}_1$, $\bm{A}_2$, and $\bm{A}_3^\text{flag}$ in $\hat{\mathcal{S}}_\text{SBT}$. The discrete form of $\hat{\mathcal{S}}_\text{SBT}$ is obtained following the procedure laid down by \cite{Swinney}. Matrix inversion is performed using LU decomposition as implemented in the open-source solver AGMG\,\citep{agmg}, which is also employed within the \textit{two fluid solver}.

Initially, the polymers are assumed to be unstretched, so that $\bm{\Pi} = 0$, and consequently $\bm{u}_s^{(P)} = \bm{F}_\text{Head}^{(P)} = \bm{L}_\text{Head}^{(P)} = \bm{0}$. Under this assumption, we solve for the flagellar force distribution $\bm{f}$, translational velocity $\bm{U}$, and angular velocities $\bm{\omega}_H$ and $\bm{\omega}_F$ by inverting the operator $\hat{\mathcal{S}}_\text{SBT}$ in equation~\eqref{SBTOperator2}. This inversion is performed twice: once using $\tilde{C} = C_1$, $\tilde{m} = 0$, and again using $\tilde{C} = C_2$, $\tilde{m} = m$, as defined in equation~\eqref{eq:MotorParams}. The solution that yields a motor angular velocity $\omega_M$ corresponding to the smaller value of $g$ is selected, consistent with the piecewise motor torque–speed curve shown in figure~\ref{fig:MotorTorque}. This provides the full Newtonian two-fluid solution, incorporating flagellar bundle–head interactions beyond the resistive-force approximation used in \citet{Sabarish}. Results from such an investigation are shown in section~\ref{sec:ResultsMicrostructure}.

The resulting polymer velocity field $\bm{u}_p=\bm{u}_p^{(M)} + \bm{u}_p^{(\text{Flag})}$ perturbs the initially unstretched polymer distribution, inducing nonzero $\bm{\Pi}$. We solve the polymer constitutive equations~\eqref{eq:PiDef}–\eqref{eq:yield} using the same numerical strategy as in \cite{Sharma23}, but now convecting and stretching polymers using $\bm{u}_p$ instead of a single velocity for both polymers and solvent. A forward Euler scheme is used for the first step, followed by a second-order Crank–Nicolson scheme. A high-order upwind discretization is used for convective terms to ensure numerical stability.

After obtaining $\bm{\Pi}$, the polymer-induced flow is computed by inverting equation~\eqref{eq:IntMFPPoly} using the \textit{two fluid solver}. The swimming kinematics are then updated by solving $\hat{\mathcal{S}}_\text{SBT}$, completing steps 10–12 of algorithm~\ref{alg:MainAlg}. The steps 8-12 are repeated for a prescribed number of time steps $N$, or until a steady swimming velocity is achieved. 

The flow velocities required for the SBT calculations, namely $\bm{w}$ and $\bm{u}_p$, are computed on the spheroidal grid. However, within the SBT framework, these velocities are needed at the discrete flagellar points shown in figure~\ref{fig:HelixSchematic}, which do not coincide with the vertices of the spheroidal grid. Rather than interpolating between these two sets of discrete points, which would introduce accuracy concerns, we develop a boundary integral formulation to evaluate the velocities directly at the flagellar locations. This method uses the analytical form of the Stokes and Brinkman Green's functions, along with the numerically computed fluid stress on the head. The full formulation is described in appendix~\ref{sec:SphdCoordToFlagella}.

In the remaining part of this section we provide tests validating various components of this numerical solver.

\subsection{Validation}
The numerical solver designed above is novel in its entirety. While direct validation of the full two-phase EVP swimming problem is not possible due to the absence of prior benchmarks, we validate the solver rigorously by comparing individual sub-problems to known analytical and numerical results. Most of these available results are limited to spherical particles, barring a few exceptions. In our implementation of the solver, we simulate dimensionless versions of the governing equations \eqref{}-\eqref{}, \eqref{}-\eqref{}, \eqref{}, where we use the minor radius of the spheroid ($l_c$) to normalize length, and $\mu_s U_c/l_c$ to normalize stress, with the velocity and time scales set by the problem being solved. For problems where the particle motion is prescribed, the prescribed velocity (or angular velocity) sets $U_c$, with $U_c/l_c$ being the time scale. For solving mobility-type problems, where the input forces and torques are specified (as in the swimming bacterium problem described here), $U_c$ is set by these input forces.

\subsubsection{Handling two fluids}
\begin{figure}
	\centering
   	\includegraphics[scale=0.35]{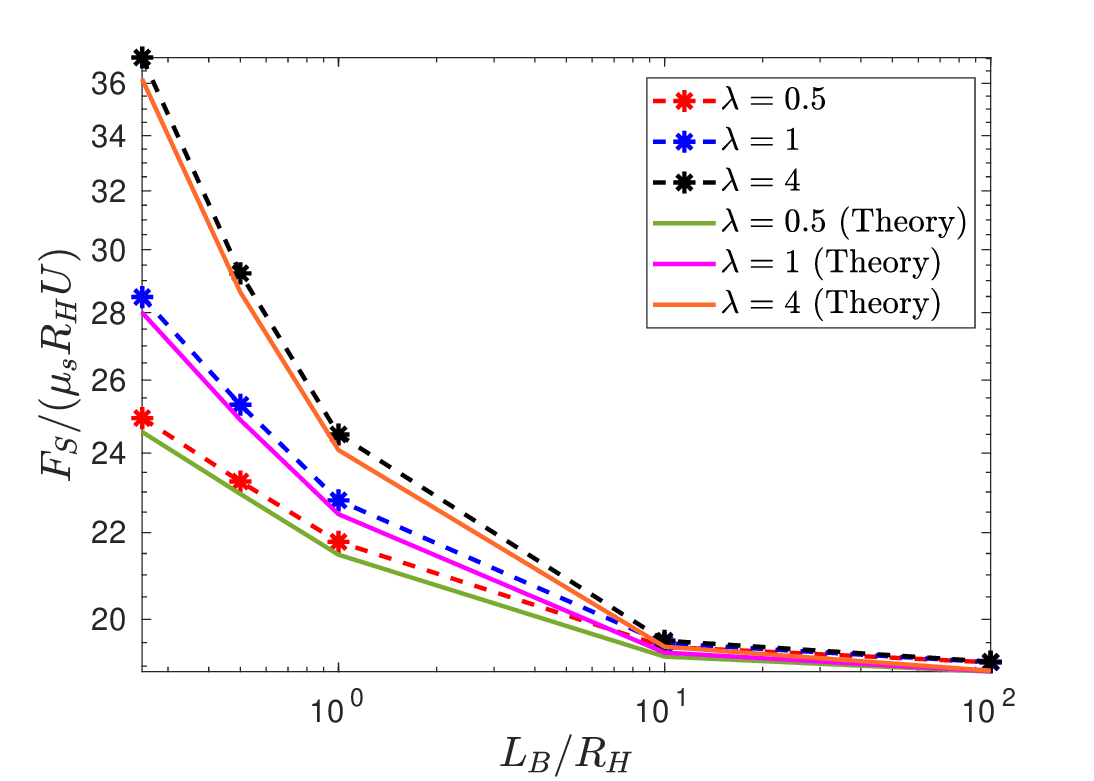}
	\includegraphics[scale=0.35]{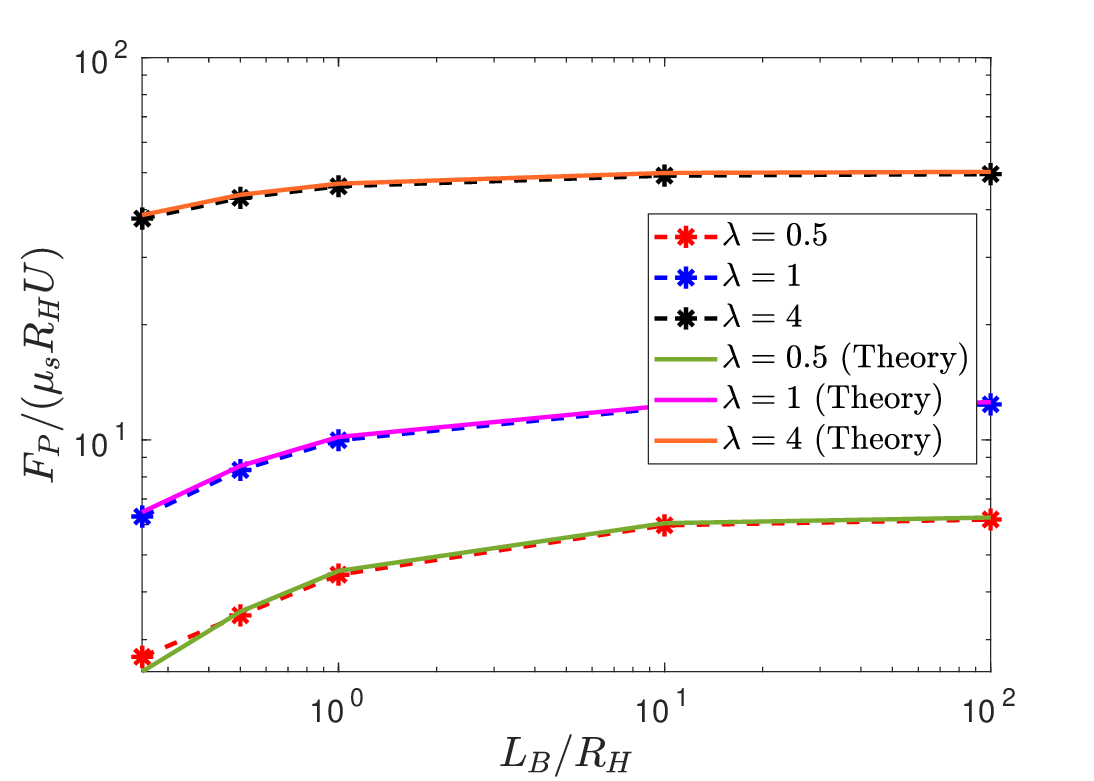}
	\includegraphics[scale=0.35]{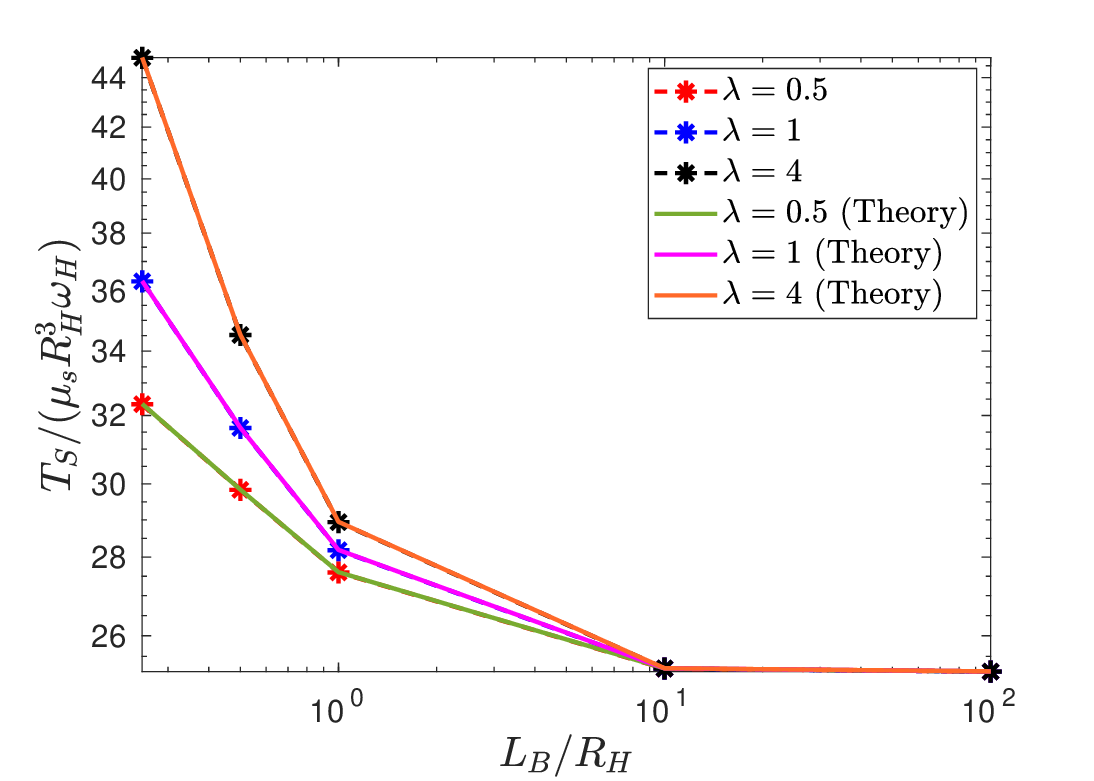}
	\caption{Plot of the dimensionless drag force on a translating sphere due to (a) the solvent and (b) polymer as a function of $L_B/R_H$, for different viscosity ratios, simulated using the two-fluid Newtonian solver and compared with analytical solutions from \citet{Sabarish}. (c) shows the solvent torque on a rotating sphere. The polymer torque is zero due to the slip condition.}
	\label{fig:MotionValidation}
\end{figure}

We first validate the \textit{two fluid solver} for Newtonian (two-fluid) flow (i.e. $\bm{\Pi}=0$), which is directly used in pre-computing $\bm{v}^{(M, U_i)}$ and $\bm{v}^{(M, S_{ji})}$ from equation~\eqref{eq:BasisFlowFieldsAddition}. We compare results for flow around a sphere with theoretical predictions from \cite{Sabarish}. In our spheroidal coordinate framework, a sphere is modeled as a particle with aspect ratio $\chi = 1.0001$, while the normalized radius of the outer boundary of the domain is set to $200$, where the length scale used for normalization is the minor radius of the spheroid, which is almost the same as the head radius $R_H$ for a spherical head. The sphere translates with velocity $U$ and rotates with angular velocity $\omega_H$ in a quiescent fluid, with slip boundary conditions applied on the polymer phase to mimic a large microstructure length scale. The drag and torque on the solvent and polymer phases are plotted against theoretical predictions for a range of $\lambda = \mu_p/\mu_s$ in figure~\ref{fig:MotionValidation}, where $L_B$ is the screening length and $R_H$ is the head radius.

This comparison tests the efficacy of the \textit{two fluid solver} to solve $\hat{\mathcal{P}}_\text{MFP}^{(M)}$, as decomposed in equation~\eqref{eq:MFPDecomp}. We also present evidence of grid-independence in Fig.\ref{fig:MotionValidation}(d), where we plot the force and torque on a translating and a rotating sphere for a medium with $\lambda = 4$ for three different mesh resolutions to show that the numerical results are converged, with finer resolutions leading to closer agreement with theory for $L_B/R_H < 1$. The close agreement (within $3\%$ relative error for the finest mesh resolution) validates the solver for a wide range of $L_B/R_H$, with increasing errors for $L_B/R_H < 0.2$ due to numerical stiffness. Thus, in all simulations we restrict $L_B/R_H \geq 0.2$ and our mesh resolution to $200 \times 151 \times 75$ (the finest resolution in Fig.\ref{fig:MotionValidation}(d)). Good agreement was also observed for the no-slip boundary condition on both solvent and polymer fluids\,(not shown). 

\subsubsection{EVP constitutive law}
\begin{figure}
	\centering
    \includegraphics[scale=0.35]{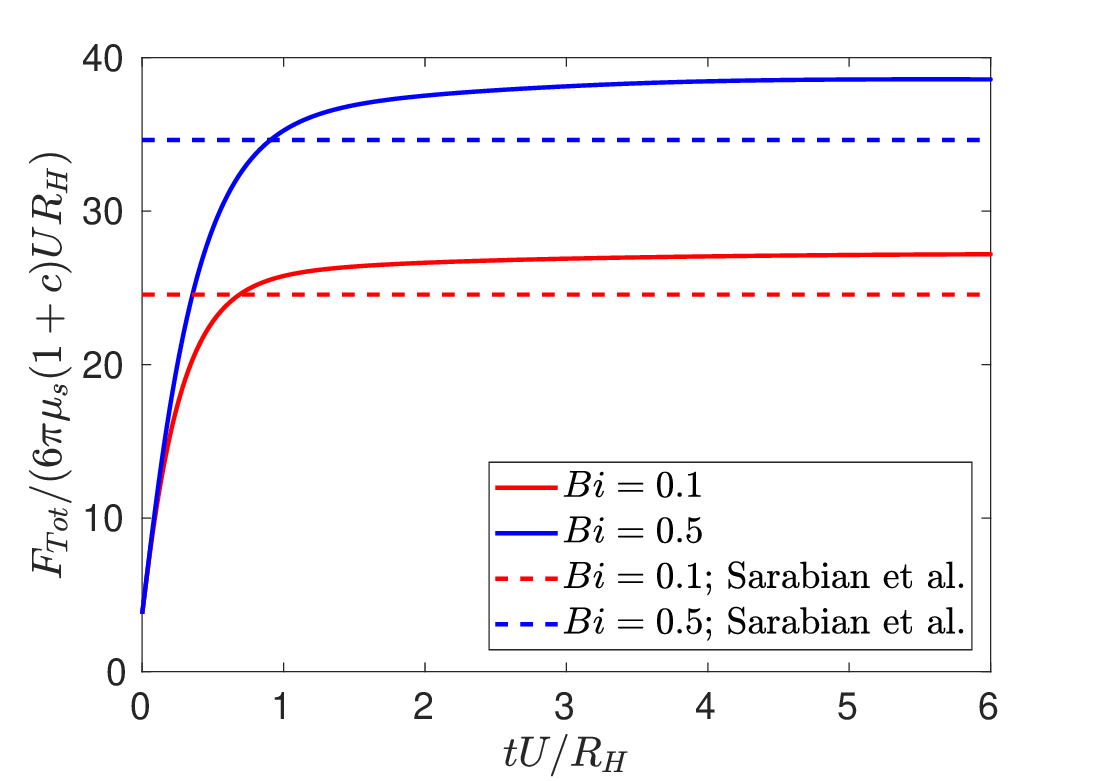}\label{fig:EVPValidA}
	\includegraphics[scale=0.35]{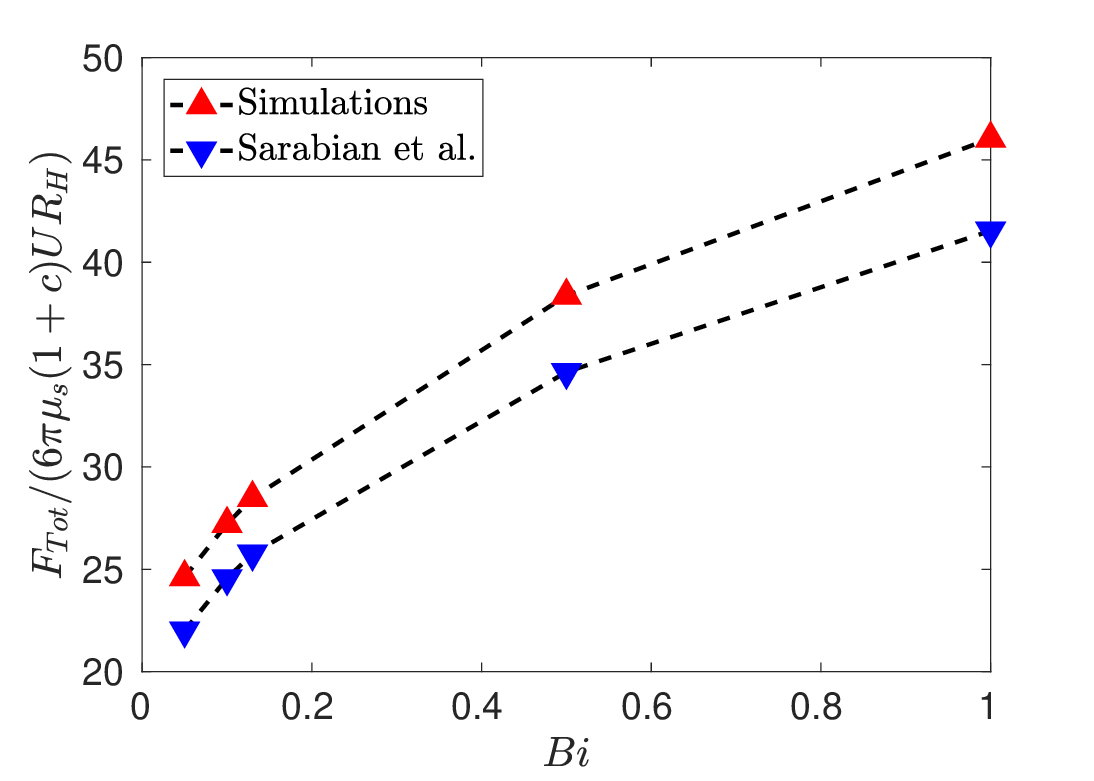}\label{fig:EVPValidB}
	\caption{(a) Drag coefficient on a sphere translating in an EVP fluid as a function of dimensionless time, compared to steady-state value from \citet{Sarabian20}. (b) Steady-state drag versus $Bi$. }
	\label{fig:EVPValid}
\end{figure}

The single-fluid solver of \cite{Sharma23} has previously been validated for various viscoelastic constitutive models but not yet for elastoviscoplastic (EVP) fluids. Also, there is no prior work considered for two-fluid EVP systems. Thus, to validate this aspect, we compare drag on a sphere in a single-fluid EVP liquid with the immersed boundary method simulations of \citet{Sarabian20}. Figure \ref{fig:EVPValid}a shows the evolution of drag on a sphere of radius $R_H$ translating with speed $U$ in an EVP fluid with viscosity $\mu$ and polymer viscosity $c\mu$ for two Bingham numbers $Bi = \sigma_Y R_H / (U\mu)$, where $\sigma_Y$ is the yield stress. Steady-state values from \citet{Sarabian20} are also shown for comparison. Discrepancies are attributed to confinement in \citet{Sarabian20}'s rectangular domain, whereas our simulation is unbounded. Overall, our predictions of steady-state drag (figure~\ref{fig:EVPValid}b) match those from \citet{Sarabian20}.
\begin{figure}
	\centering
	  \includegraphics[scale=0.42]{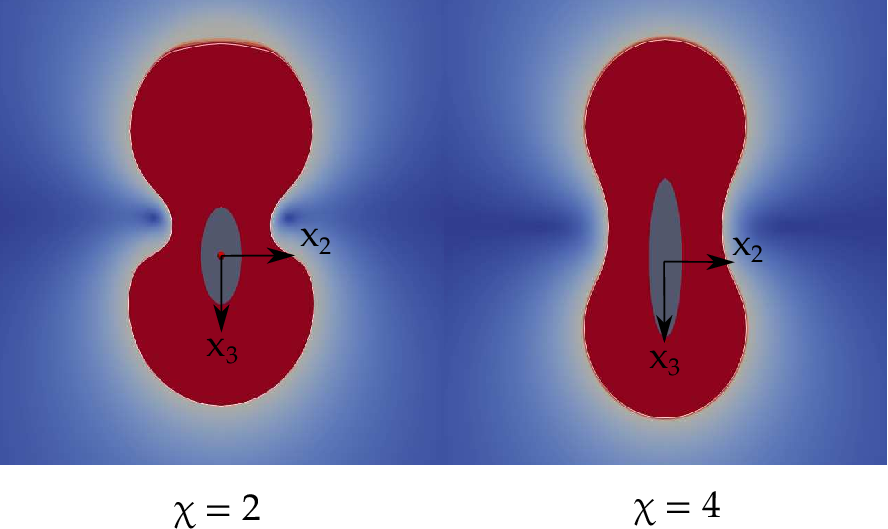}
	 \includegraphics[scale=0.35]{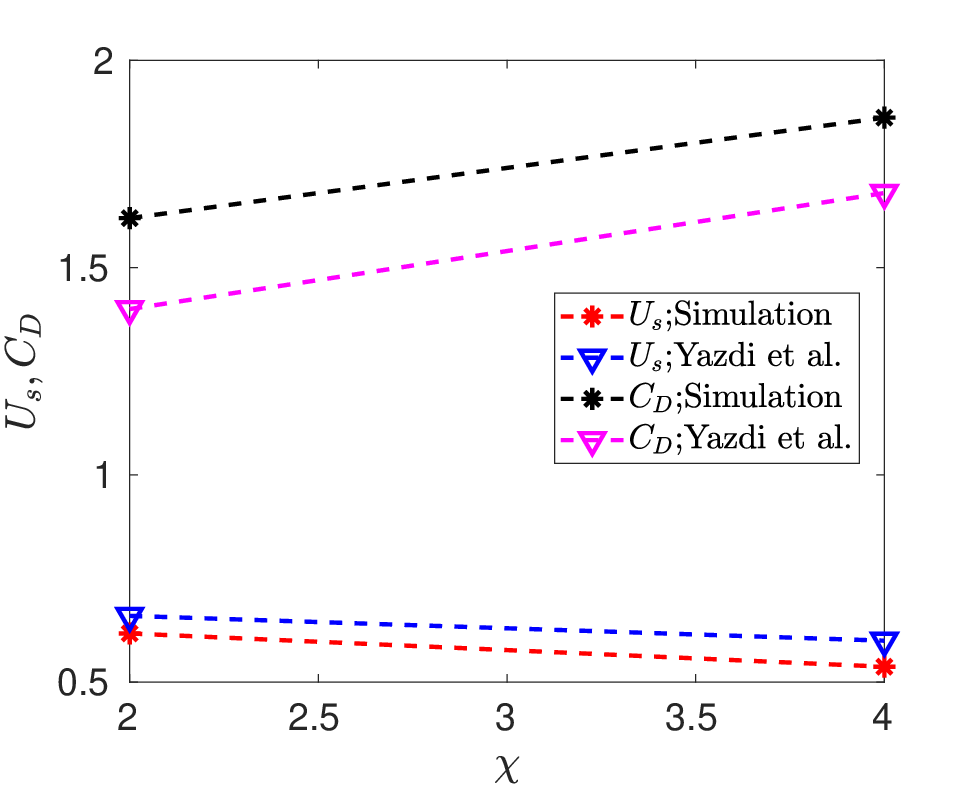}
	\caption{(a) Yielded region (red) around a sedimenting spheroid in a single-fluid EVP medium ($\hat{\alpha} = 0.2$) at $t = 10$ ($t$ - dimensionless simulation time) for $Bi = 0.1$ and $Wi = 1$, where $Bi = \frac{\sigma_Y 6 \pi \mu(1+c) \chi^{2/3}}{F}$ and $Wi = \frac{\tau F}{6 \pi \mu(1+c)\chi^{2/3}}$. The shape of these surfaces agree with those observed by \cite{Avino23} (Fig.5(a),6(a)). (b) Steady-state dimensionless sedimenting velocity ($U_s = 6 \pi U \mu (1+c) \chi^{1/3}/F$) and drag coefficient for the same spheroids as (a) ($C_D = F/(6 \pi \mu (1+c) \chi^{1/3} U_s)$) compared between our work and that of \cite{Avino23}.}
	\label{fig:AvinoEVP}
\end{figure}

Recently, \cite{Avino23} has performed simulations of a sedimenting spheroid in the same EVP media as \cite{Sarabian20}, and has provided results on the drag and yielded domains for spheroids of different aspect ratio. We benchmark our single-fluid solver with this study as well\,(Fig.\ref{fig:AvinoEVP}), where we observe similar yielded regions and a steady-state drag coefficient and sedimentation velocity that is within $~10\%$ of the values provided by \cite{Avino23}. These results once again serve to validate the solver of \cite{Sharma23}, which is now shown to work well even with EVP constitutive laws.

\subsubsection{Flagellar bundle forcing: Stokeslet next to a sphere}
\begin{figure}
	\centering
	\subfloat{\includegraphics[scale=0.36]{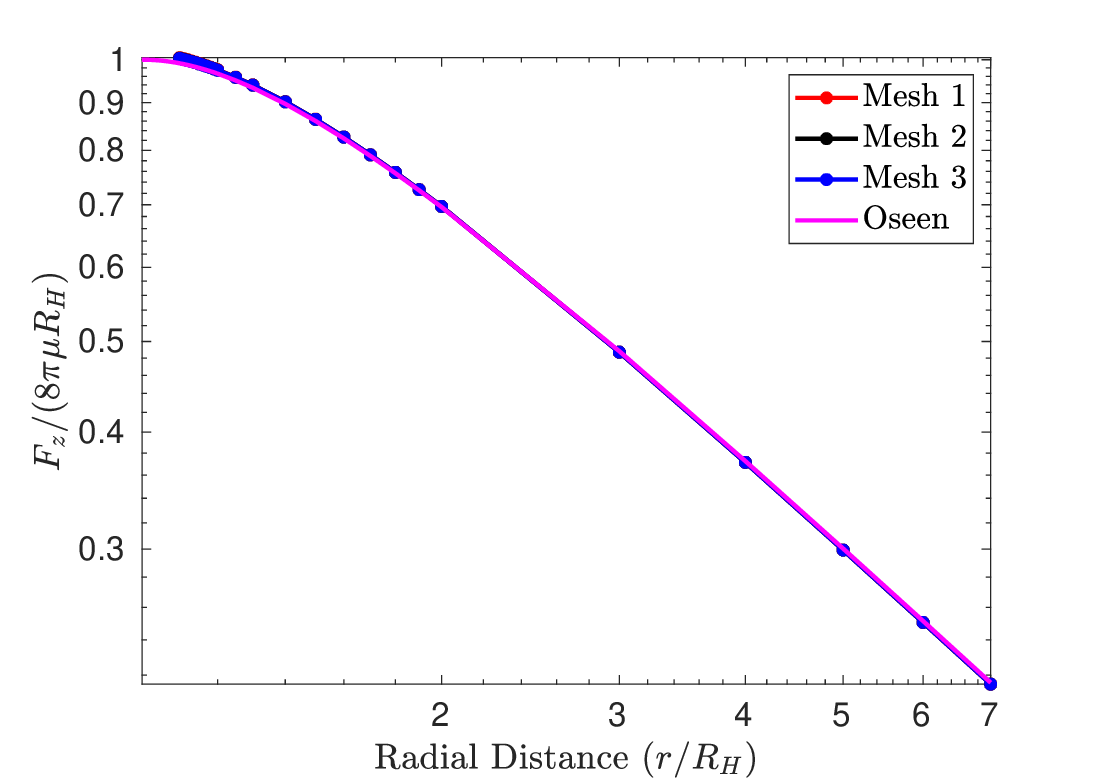}\label{fig:OseenSolsValidA}}
	\subfloat{\includegraphics[scale=0.36]{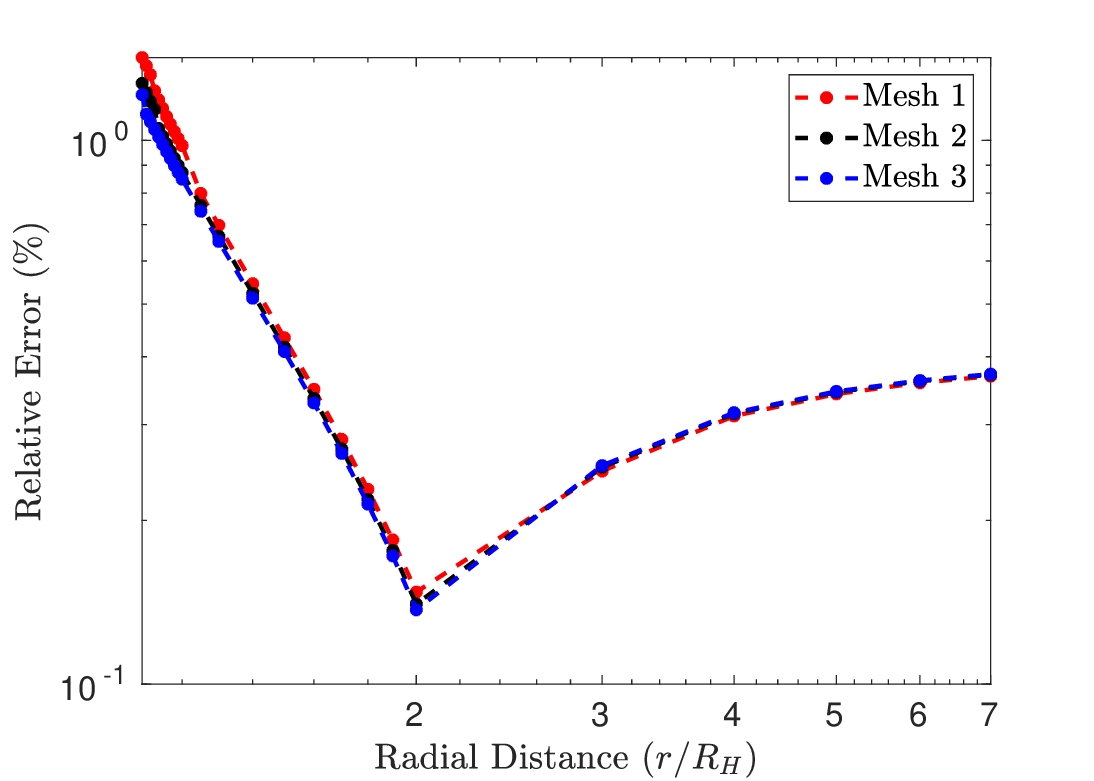}\label{fig:OseenSolsValidB}}
	\caption{(a) Dimensionless force on a sphere due to a Stokeslet along the axial direction, versus radial Stokeslet location, compared with analytical results from \citet{Oseen}. (b) Relative error ($\%$) in the numerical prediction.}
	\label{fig:OseenSolsValid}
\end{figure}

To compute flagellar forcing, we precompute flows due to point forces (two-fluidlets) placed near the bacterial head and assemble them into the tensors $\bm{A}_1$, $\bm{A}_2$, and $\bm{A}_3^\text{flag}$ in $\hat{{S}}_\text{SBT}$. The numerical solution computes only the regular part, $\tilde{\bm{v}}_{\text{TF-let}}^{\text{Head},(i)}$, while the singular part, $\bm{v}_{\text{TF-let}}^{(i)}$, is known analytically. The difference between the two provides boundary conditions on the surface of the head. Since no such (analytical) solution exists for two-fluid medium or spheroids, we validate our method in a single-fluid Newtonian medium by comparing the numerical solution from the solver of \citet{Sharma23} with analytical results from \citet{Oseen}. These correspond to a Stokeslet oriented along the axis of symmetry and placed at varying distances from a spherical particle. Figure~\ref{fig:OseenSolsValid}a shows the computed force on the sphere for three different grid resolutions. The relative error, shown in figure~\ref{fig:OseenSolsValid}b, remains within $\sim1\%$, demonstrating high accuracy.

As the singularity approaches the head, the boundary condition region nears the singular core, reducing accuracy. Similarly, when the singularity is placed far from the head, the local grid resolution is coarser due to spheroidal clustering, again slightly degrading accuracy. Despite these effects, the error remains small. This method avoids artificial regularization of singularities and can be extended to higher-order singularities near spheroidal particles.

In the next section, we utilize the solver to explore novel effects of microstructure on bacteria swimming in a two-fluid medium.

\section{Effect of microstructure on the motion of a swimming bacterium in a bio-fluid} \label{sec:ResultsMicrostructure}
As discussed in section~\ref{sec:MathForm}, the screening length $L_B$ in the two-fluid model parametrizes the pore size in an entangled polymer solution, which constitutes the microstructure length scale of the medium. While the bacterial head is much larger than the pore size and interacts with the polymer solution as a continuum, the slender flagellar bundle has a thickness comparable to the pores of the polymer network, leading to novel effects on the swimming motion. Herein, we simulate a straight swimming flagellated bacterium in this two-fluid medium, and analyze the effect of this length-scale dependent heterogeneity caused by the microstructure of the swimming medium. Although the numerical framework supports a non-Newtonian medium, we restrict our attention here to a Newtonian medium to isolate the role of microstructure. 

In the limit $L_B \rightarrow 0$, the solvent and polymer fluids move together in unison\,($\bm{u}_s = \bm{u}_p$), and the fluid behaves as a standard single-phase Newtonian liquid with viscosity equal to the sum of the two fluids, $\mu_s+\mu_p$. In contrast, for large screening lengths ($L_B/a \gg 1$), the polymer does not experience direct forcing from the flagellar bundle, and the bacterial head imparts independent hydrodynamic stresses on the two fluids. In this section, we explore how the swimming velocity of a flagellated bacterium depends on the screening length $L_B$, as well as on the viscosity ratio $\lambda = \mu_p/\mu_s$ between the solvent and polymer fluids. The geometric parameters of the bacterium and the input motor torque for motion are taken from experimental measurements on {\it E. coli} by \citet{Berg,Poon14}, as summarized in table~\ref{tab:EcoliBergPoon}.
\begin{table}
	\centering
	\resizebox{\textwidth}{!}{
		\begin{tabular}{  c c c c c c c c c c c c  }
			\hline
			$a$ ($\mu$m) & $L_F$ ($\mu$m) & $\beta$ ($\mu$m) & $\theta$ (deg) & $\gamma$ (--) & $R_H$ ($\mu$m) & 
			$\mu_s$ (mPa$\cdot$s) & $C_1$ (pN$\cdot$nm) & $C_2$ (pN$\cdot$nm) & $m$ (pN$\cdot$nm$\cdot$s) & 
			$\omega_{\text{Knee}}$ (rad/s) & $\omega_{\text{max}}$ (rad/s) \\
			
			$0.03$ & $7$ & $2$ & $41$ & $240$ & $1.5$ & $1$ & $1250$ & $3750$ & $-1.91$ & $350\pi$ & $600\pi$ \\
			\hline
		\end{tabular}
	}
	\caption{Values of the various parameters corresponding to {\it E. coli} used in simulations.}
	\label{tab:EcoliBergPoon}
\end{table}

In section \ref{sec:SmallPolymers} we assume the pore size of the polymer network to be much smaller than the bacterial head radius ($R_H$), and thus impose a no-slip condition on the polymer fluid at the head surface (equation~\eqref{BCnoslip}). Then in section \ref{sec:LargePolymers} we consider the pore sizes to be comparable to the bacterial head\,(or equivalently polymer chains to be long enough), such that they slip past the head (equations~\eqref{BC2B}-\eqref{BC2D}). In both these cases the pore size in the polymer network is assumed to be much larger than the flagellar bundle thickness ($L_B/a \gg 1$). These cases highlight the effect of microstructure at the length scale of the flagellar bundle and the entire bacterium respectively. We first describe these results for a bacterium with a spherical head, before describing the effect of the spheroidal shape of the head. 

\subsection{Effects of microstructure at the length scale of flagellar bundle}\label{sec:SmallPolymers}
\begin{figure}
	\centering
	\includegraphics[scale=0.35]{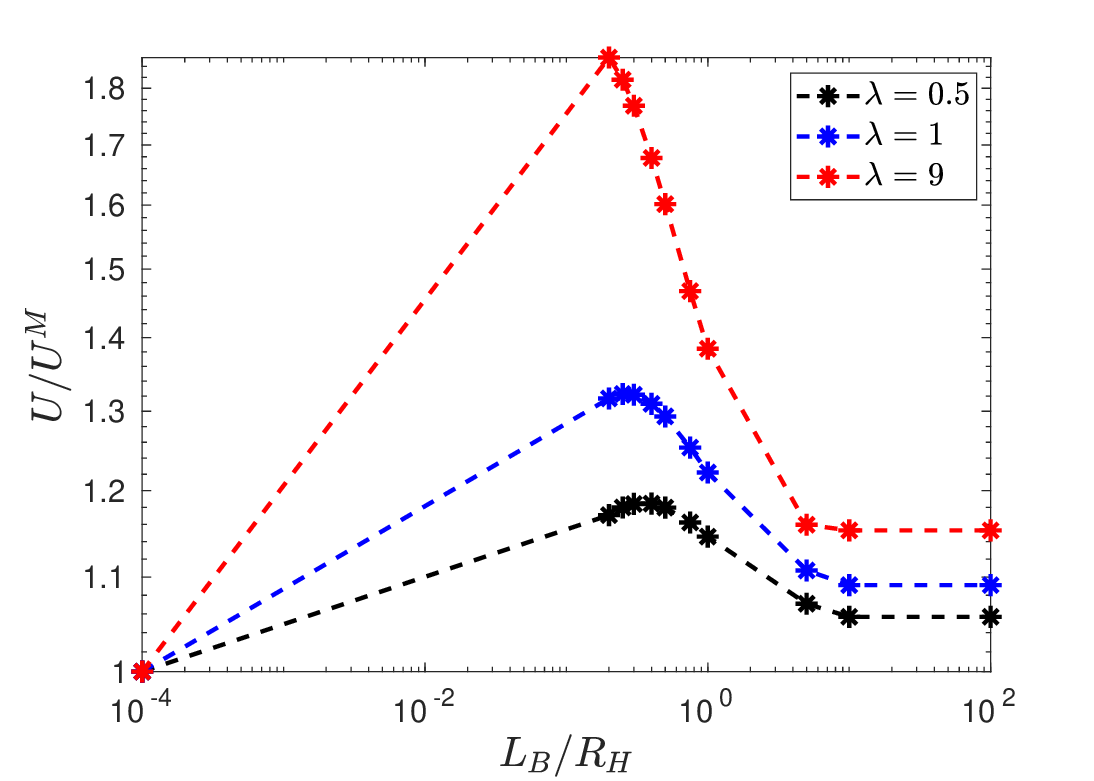}\quad
	\includegraphics[scale=0.35]{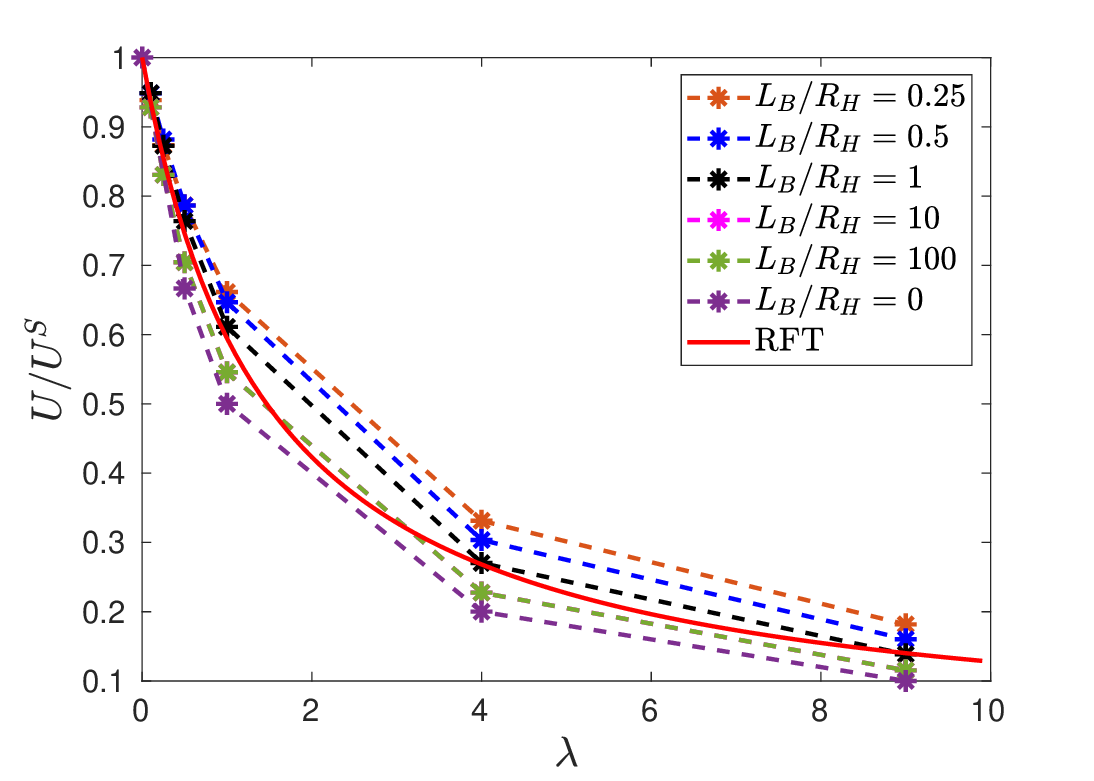}\quad
	\includegraphics[scale=0.35]{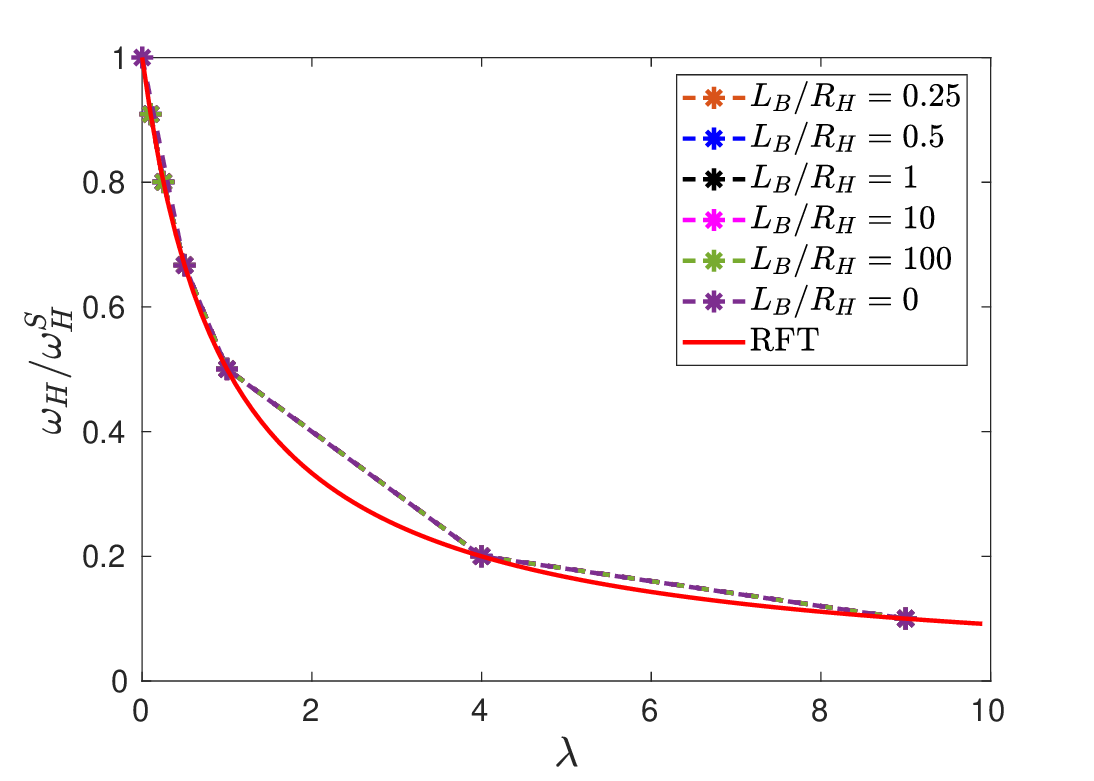}\quad
	\includegraphics[scale=0.35]{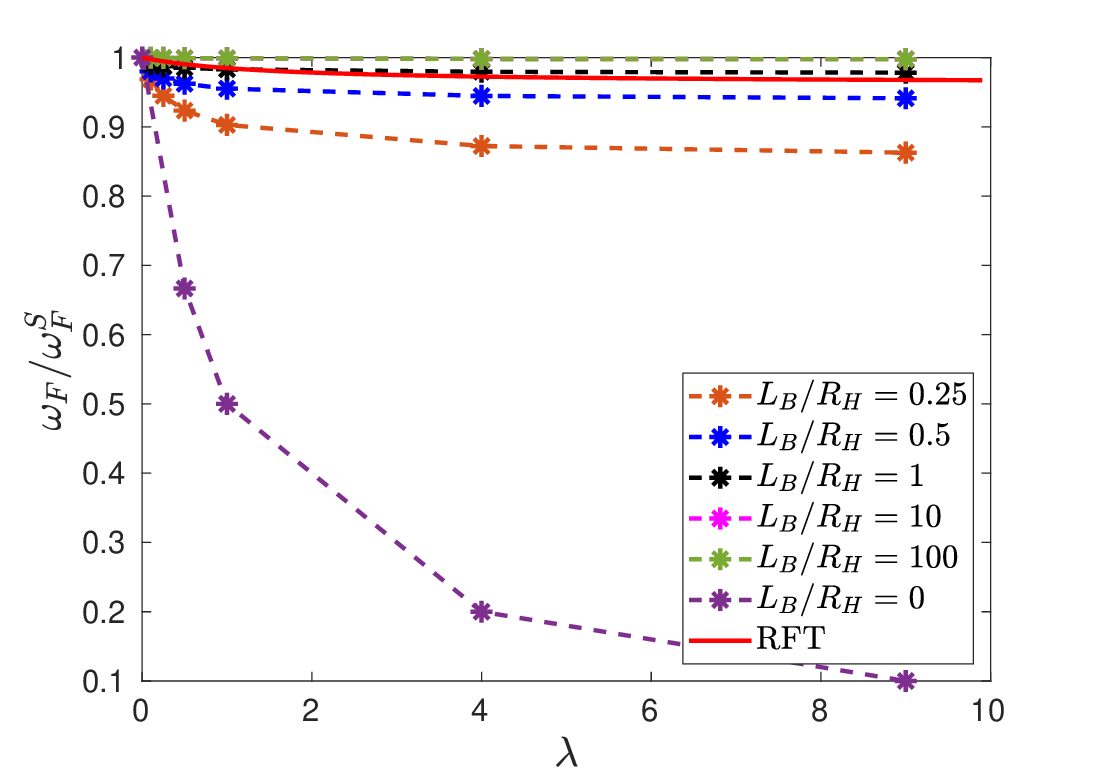}
	\caption{Results with no-slip polymers. (a) Swimming speed normalized by the velocity in a single-fluid Newtonian medium of viscosity $\mu_s(1+\lambda)$ as a function of $L_B/R_H$. (b) Swimming speed normalized by the velocity in a pure solvent ($\lambda = 0$) as a function of $\lambda$. (c) Normalized angular velocity of the head and (d) normalized angular velocity of the flagellar bundle as functions of $\lambda$ for different $L_B/R_H$. Red solid lines denote predictions from resistive force theory \citep{Sabarish}.}
	\label{fig:SwimmingVelocity2FluidNewt}
\end{figure}

Figure~\ref{fig:SwimmingVelocity2FluidNewt}a shows the swimming speed $U$, normalized by $U_M$, the swimming speed in a single-phase Newtonian liquid of viscosity $\mu_s(1+\lambda)$\,(hereafter referred to as 'mixture'), plotted as a function of $L_B/R_H$, where $R_H$ is the radius of the head, for three different values of the viscosity ratio $\lambda$. 

As $L_B$ increases from zero, the swimming speed initially increases, reaching up to 1.8 times the single-phase value for $\lambda = 9$, with a maximum near $L_B/R_F \approx 1$. This enhancement exceeds predictions from resistive force theory (RFT)\,\citep{Sabarish} (dashed horizontal lines in figure \ref{fig:SwimmingVelocity2FluidNewt}a), which neglects hydrodynamic interactions between the head and the flagellar bundle. The degree of enhancement is found to scale with the viscosity ratio $\lambda$. At larger values of $L_B$, the swimming speed gradually decreases but remains above the mixture value within the range of $L_B/R_H$ considered. 

\begin{figure}
	\centering
	\includegraphics[scale=0.35]{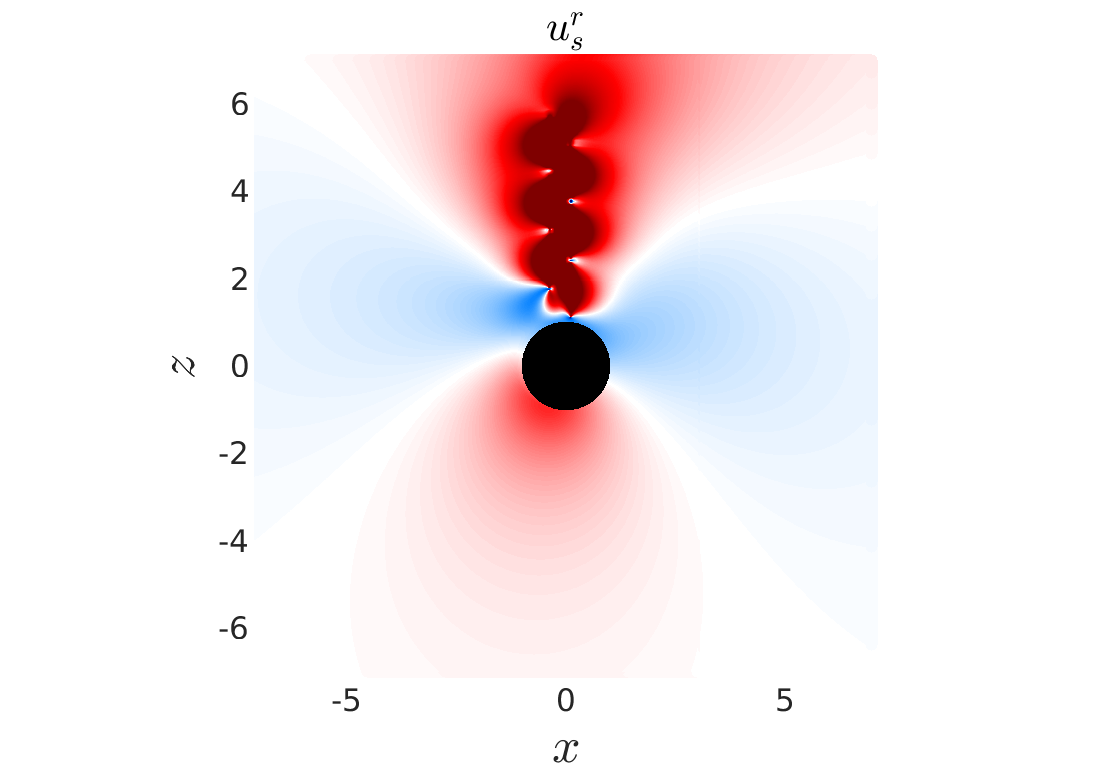}
	\includegraphics[scale=0.35]{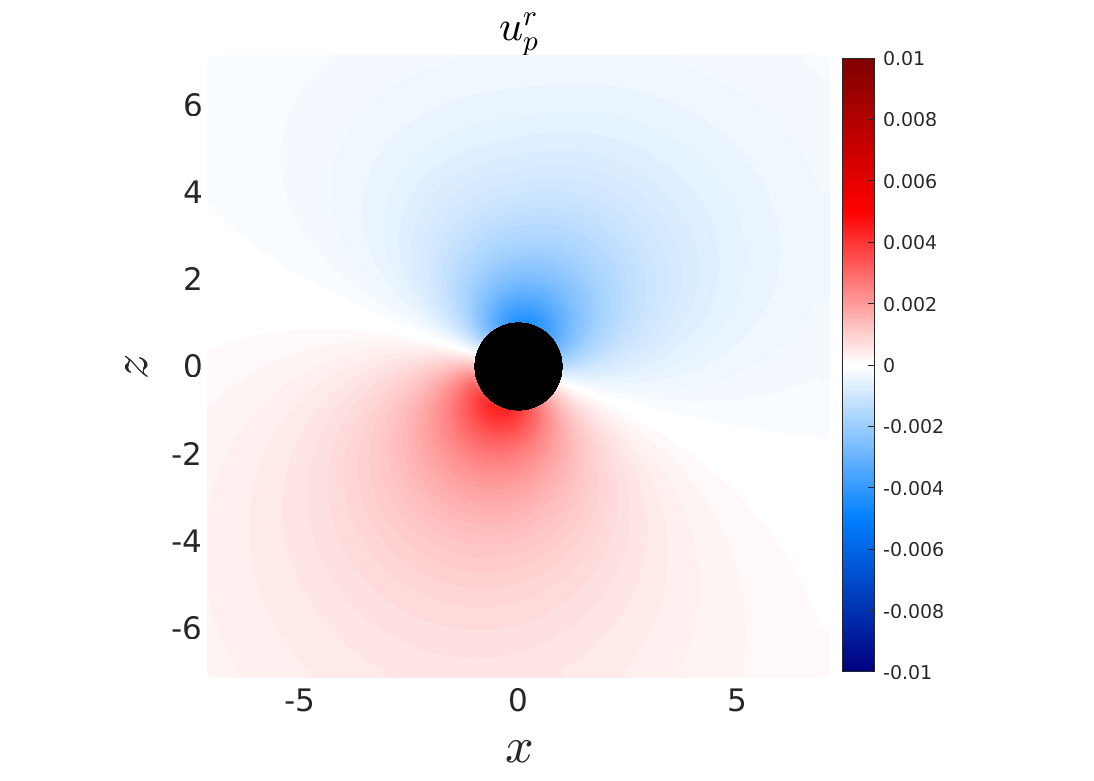}\\
	\includegraphics[scale=0.35]{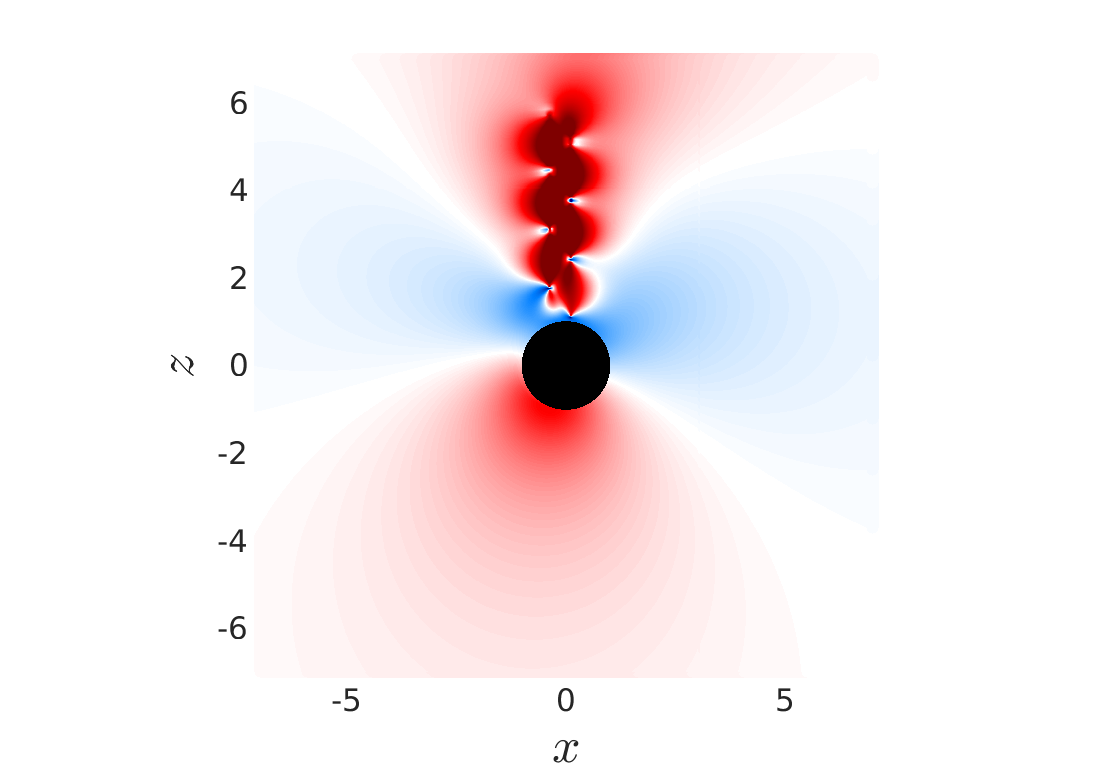}
	\includegraphics[scale=0.35]{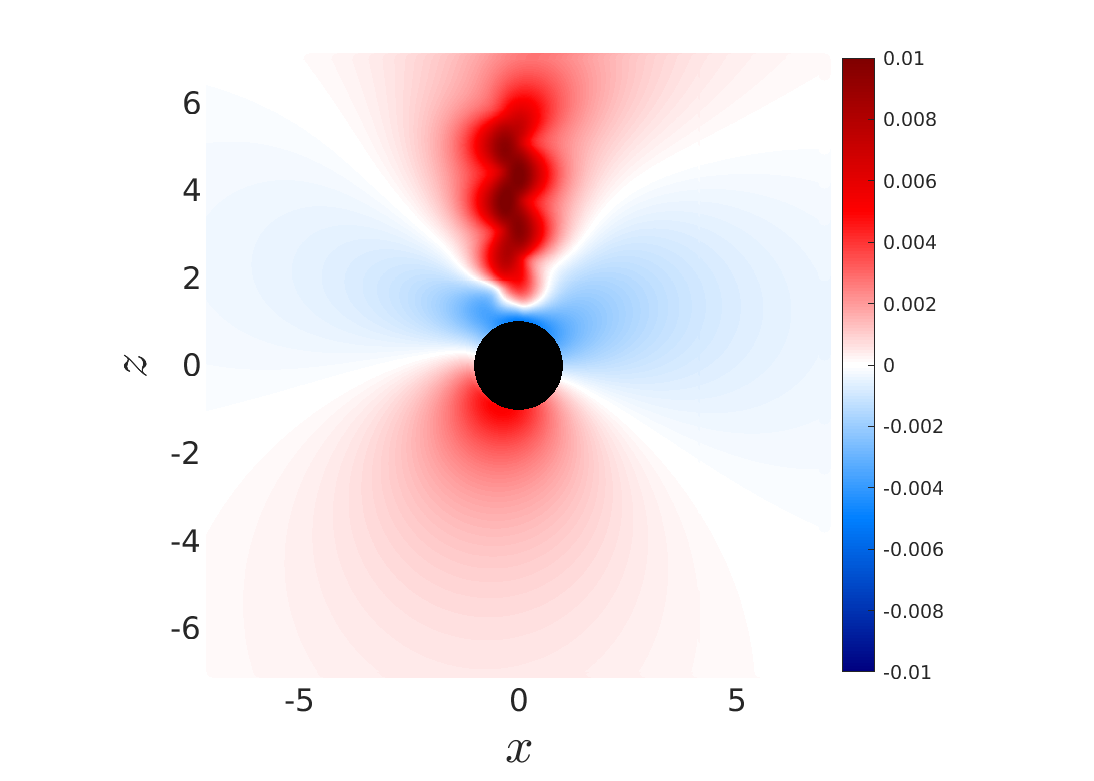}
	\caption{Radial velocity fields of the solvent and polymer fluids for (a),(b) $L_B/R_H = 10$ and (c),(d) $L_B/R_H = 0.25$. At small $L_B$, the polymer fluid experiences local forcing from the flagellar bundle, while at large $L_B$, the flow resembles that induced by a translating and rotating sphere, with no (direct) forcing from the bundle. The velocity is non-dimensional and is scaled by $U_c \sim T_M/\mu_s R_H^2$, where $T_M = C_1 = 1250 \,pN\,nm$ as shown in table \ref{tab:EcoliBergPoon}.}
	\label{fig:VelField2FluidNewt}
\end{figure}

Figure~\ref{fig:SwimmingVelocity2FluidNewt}b shows the swimming speed $U_S$, normalized by the speed in a pure solvent ($\lambda = 0$, viscosity $\mu_s$), as a function of $\lambda$ at fixed $L_B/R_H$. The speed decreases monotonically with increasing polymer viscosity. Figures~\ref{fig:SwimmingVelocity2FluidNewt}c and~\ref{fig:SwimmingVelocity2FluidNewt}d show the normalized angular velocities of the head ($\omega_H$) and flagellar bundle ($\omega_F$), respectively. Although $\omega_H$ is nearly independent of $L_B$ and agrees with the RFT predictions, $\omega_F$ increases with $L_B$ for fixed $\lambda$, highlighting the non-trivial coupling between the flagellar bundle and the microstructure of the medium. This is because, the head in this case always sees the two-fluid medium as a mixture (since no-slip condition is applied to both solvent and polymer), whereas the bundle gets decoupled from the polymer as $L_B$ increases. Thus for large $L_B$, the bundle essentially rotates in a solvent, and therefore experiences less resistance to rotation than when $L_B \rightarrow 0$. This can also be seen from figure \ref{fig:VelField2FluidNewt}, which plots the radial velocity in the solvent and polymer fluids due to the motion of the bacterium for two values of $L_B/R_H$. We see that the polymer field closely resembles the solvent field as $L_B \rightarrow 0$, whereas the polymer flow corresponds to that of the flow due to a translating and rotating sphere for $L_B/R_H \gg 1$. This decoupling explains the $L_B$ dependent increase in flagellar rotation seen in figure~\ref{fig:SwimmingVelocity2FluidNewt}d.

\begin{figure}
	\centering
	\includegraphics[scale=0.5]{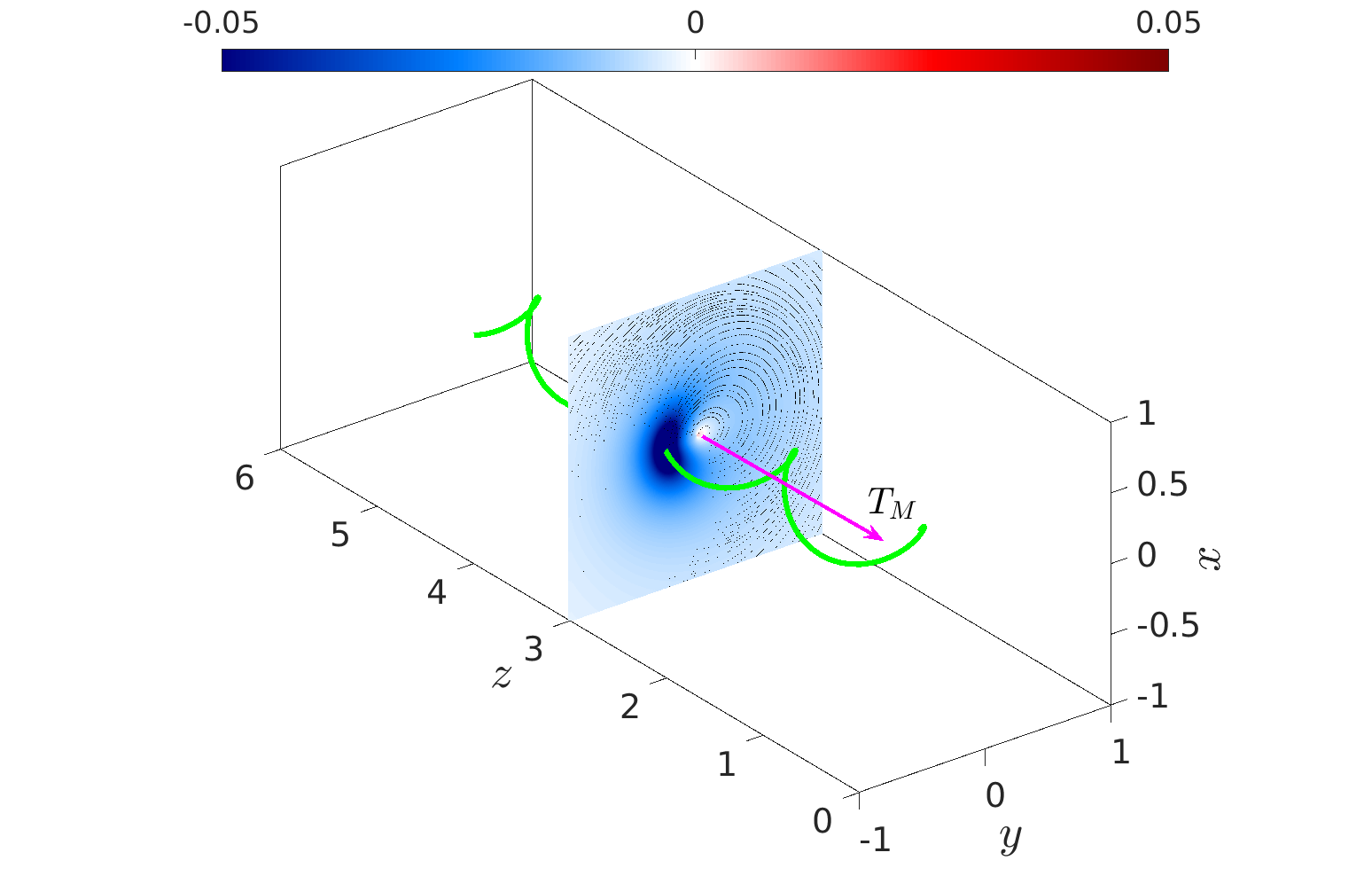}
	\includegraphics[scale=0.4]{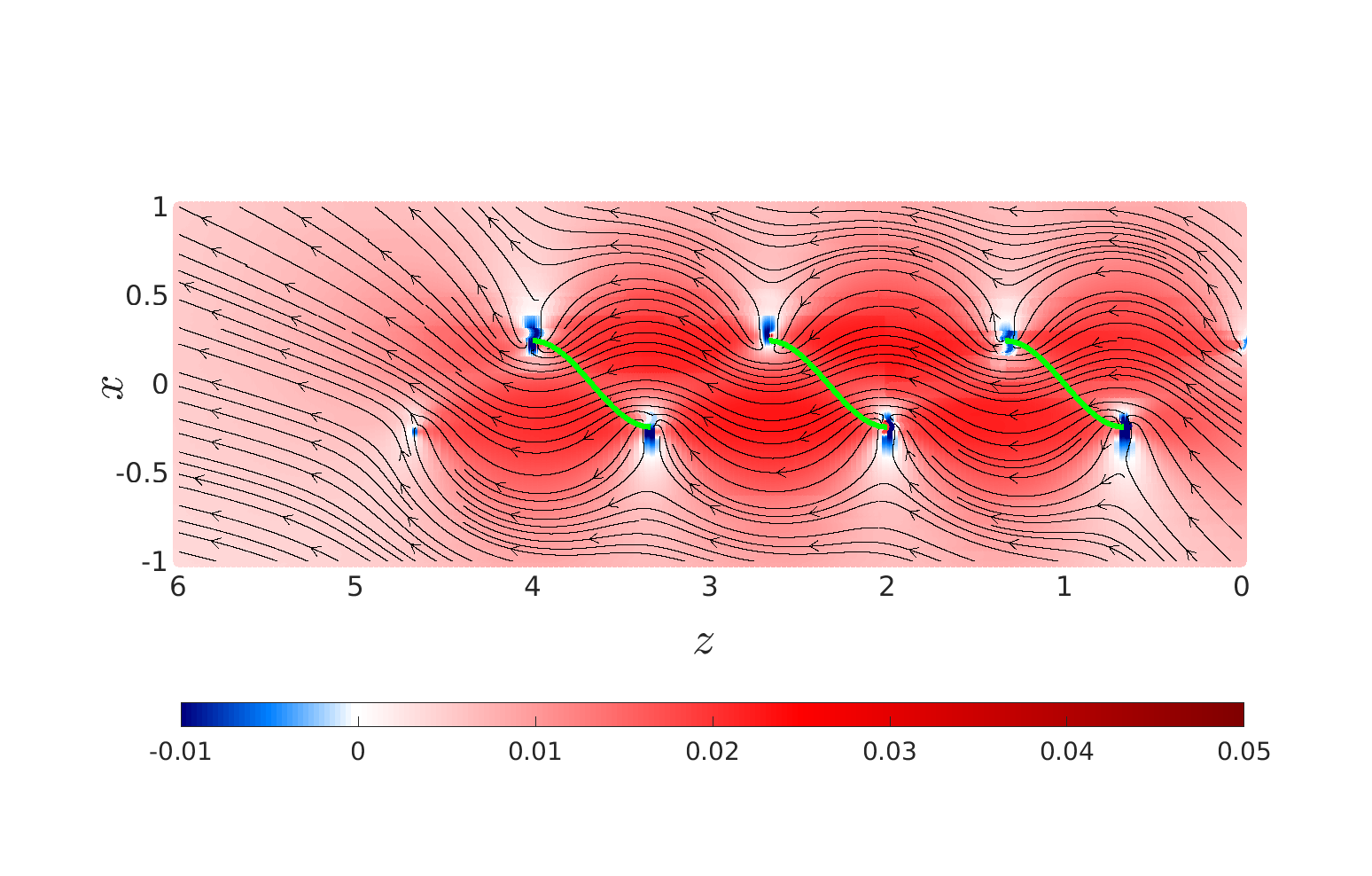}	
	\caption{(a) Azimuthal velocity field ($u_s^{\phi}$) and (b) axial velocity field ($u_s^z$) of the solvent for $L_B/R_H = 0.25$ due to a helix subject to an input torque $T_M$. The flow field is rotational in a plane perpendicular to the helical axis, whereas the rotation of helix also produces an axial flow, which generates the thrust needed for propulsion.}
	\label{fig:velfieldHelixTF}
\end{figure}

There are two major features in the swimming speed of a bacterium, that are caused by the microstructure of the swimming medium. First, one observes a non-monotonic variation of the swimming speed with $L_B$ (figure \ref{fig:SwimmingVelocity2FluidNewt}a), and second, the enhancement in velocity relative to mixture is larger for larger viscosity ratio $\lambda$. The physical reason for these two observations are provided below. In order to understand the non-monotonic behavior, we first plot the flow field produced by a rotating helix, when subject to an input torque, in figure~\ref{fig:velfieldHelixTF}. We see that the rotation of the helix produces an axial flow along the length of the helix, whereas very close to the surface of the helical bundle, the flow field is rotational (as can be seen by the streamlines in figure~\ref{fig:velfieldHelixTF}a). This implies that the thrust force produced by the rotating helix (for a given input torque) is proportional to both the angular speed of the helix and the resistance to the (axial) flow offered by the medium over a length scale of the helix. 

\begin{figure}
	\centering
	\includegraphics[scale=0.35]{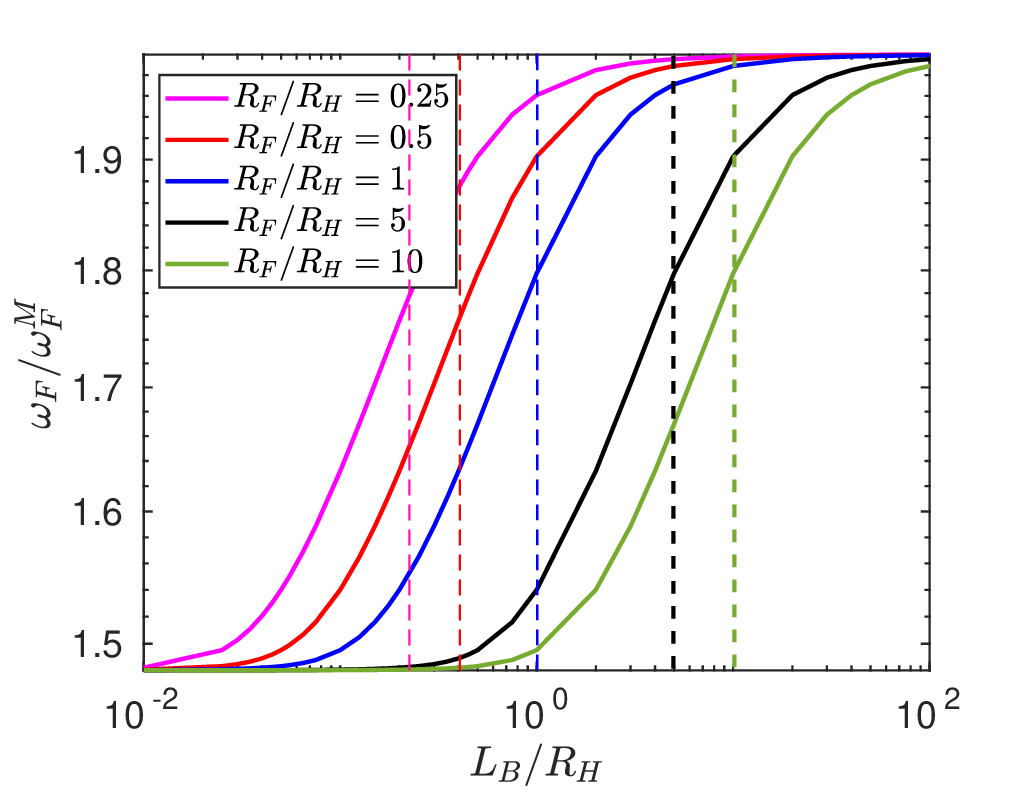}  
	\includegraphics[scale=0.35]{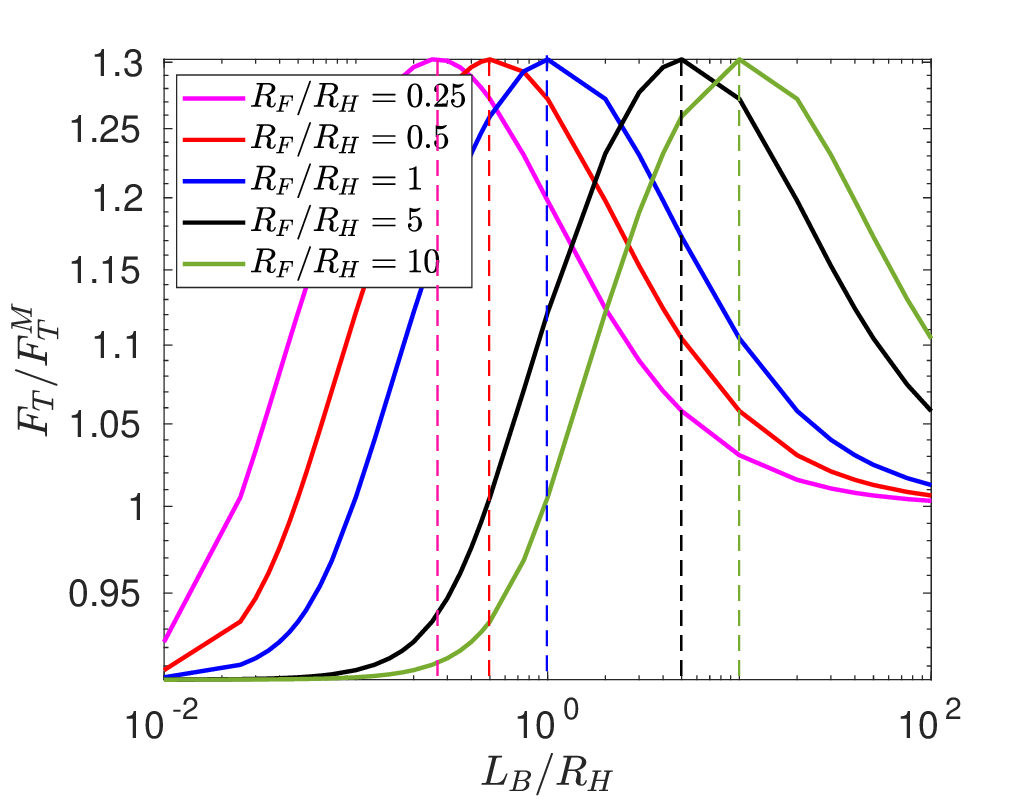}
	\caption{Plot of normalized angular velocity of the helix ($\omega_F$) and (b) the thrust force generated by a rotating helix, when subject to a given input torque ($T_M$) for different values of $R_F/R_H$, with the ratio of pitch to helix radius ($\beta/R_F$) kept constant. Note that the thrust is non-monotonic with the maximum always occurring at $L_B \approx R_F$. The results are obtained from slender body theory for a helix in two-fluid medium\,(eq.\eqref{SBTEQ}, \cite{Sabarish}).}
	\label{fig:SBTexplanation}
\end{figure}

In figure \ref{fig:SBTexplanation}a, we plot the angular speed of a helix rotated by an input torque as a function of $L_B/R_H$ for different values of $R_F/R_H$\,(note that in our bacterium simulations $R_F/R_H \approx 0.25$). These are calculated from the slender body theory for a helix in two-fluid medium (eq.\eqref{SBTEQ}) without the presence of bacterial head. Figure~\ref{fig:SBTexplanation}b plots the thrust produced by the same helix as a function of $L_B/R_H$ for different values of $R_F/R_H$. Both these values are normalized by their corresponding values in a mixture (viscosity $\mu_s + \mu_p$). We note from figure~\ref{fig:SBTexplanation}b, that the thrust produced is non-monotonic. This can be understood in conjunction with figure~\ref{fig:SBTexplanation}a, where we see that as $L_B/R_H$ increases from zero, the angular velocity of the helix increases. This is because the resistance to flow closer to the helix surface decreases with increasing $L_B$, where now locally, the helix starts seeing less of the polymers. But on larger length scales (of $O(L_F)$), the resistance to flow is still that due to the mixture. This increase in angular velocity, combined with the large length scale resistance corresponding to that of the mixture (maximum possible resistance to flow), leads to an increase in thrust. This increase continues until $L_B/R_F \approx 1$, when the (local) resistance to rotation of the helix has dropped entirely to that corresponding to solvent alone, and subsequent increase in $L_B/R_F$ now results in a drop in the large length scale resistance to flow (as the polymer becomes decoupled from helix over larger length scale). This results in the angular velocity plateauing off and the thrust decreasing for $L_B > R_F$. This results in the non-monotonic variation of the thrust, that also results in the non-monotonic variation in swimming speed. 

\begin{figure}
	\centering
	\includegraphics[scale=0.35]{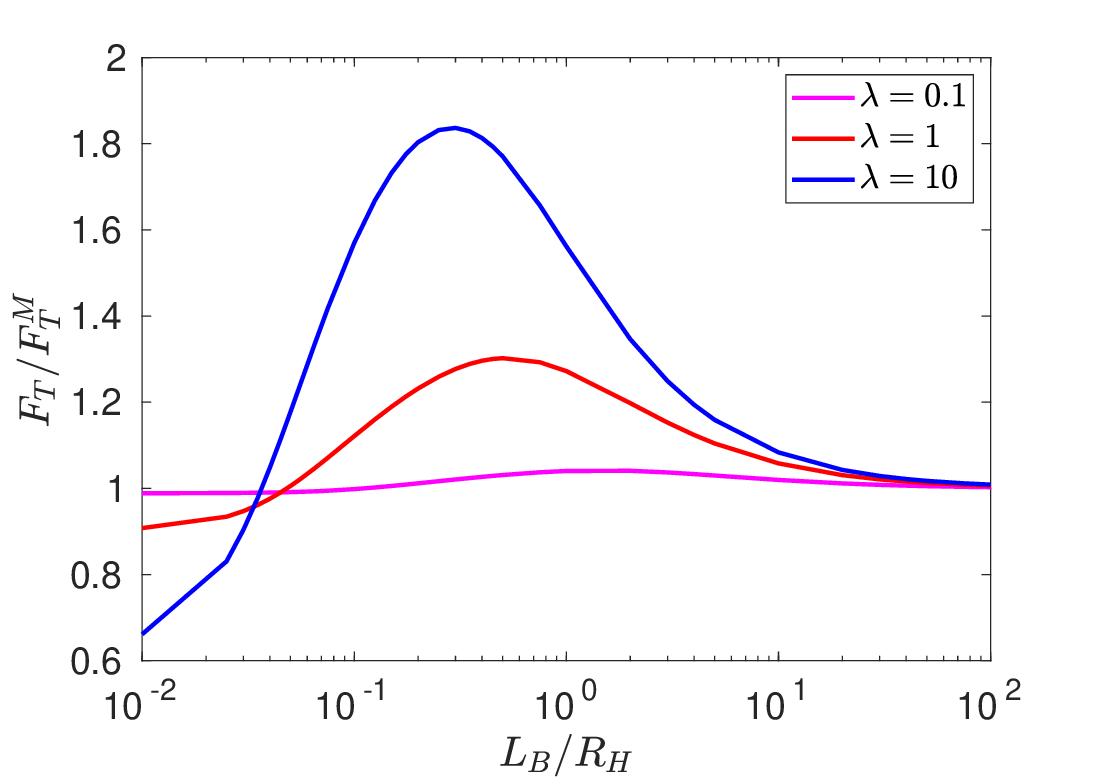}  
	\includegraphics[scale=0.35]{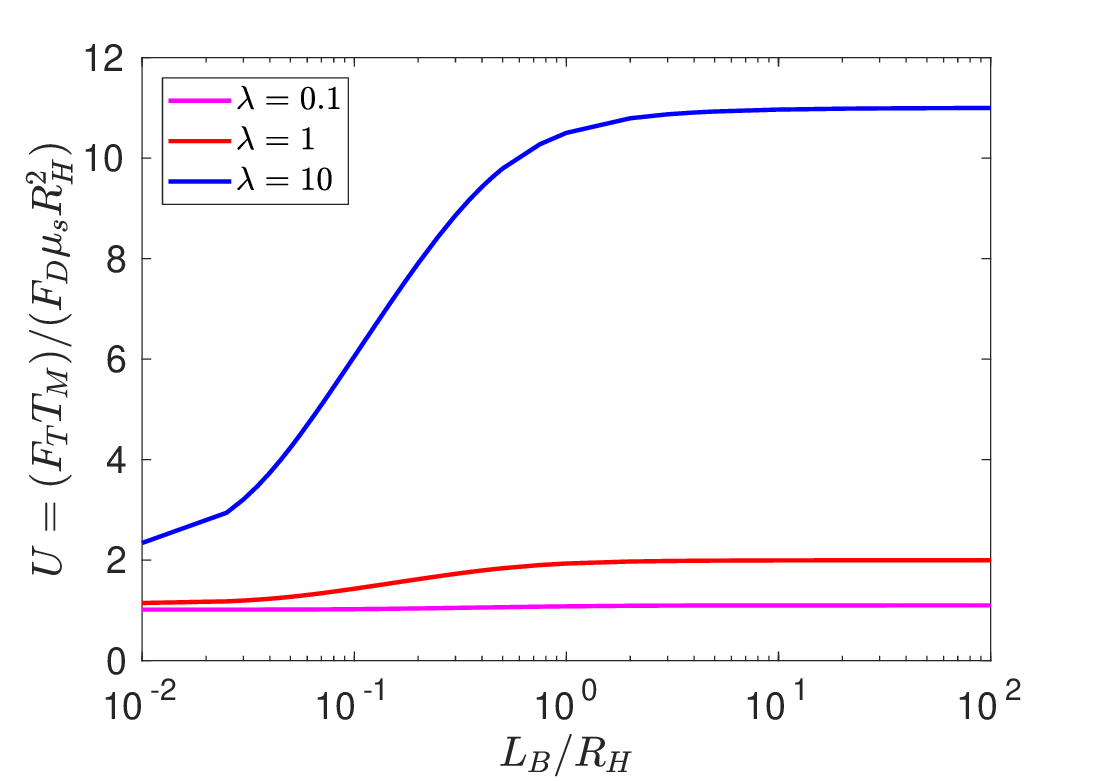}
	\caption{Plot of (a) normalized thrust force generated by a rotating helix, and (b) the ratio of thrust to drag (free-swimming velocity) when subject to a given input torque ($T_M$) for different values of $\lambda$. Note that for this calculation the radius of helix is fixed at $R_F/R_H = 0.5$. The results are obtained from slender body theory for a helix in two-fluid medium\,(eq.\eqref{SBTEQ}, \cite{Sabarish}).}
	\label{fig:SBTexplanation2}
\end{figure}

To understand the second feature, we plot the normalized thrust and the thrust-to-drag ratio for a force-free rotating helix, subject to a given input torque ($T_M$) for different $\lambda$ in figure \ref{fig:SBTexplanation2}. The thrust-to-drag ratio is obtained by dividing the thrust generated for a given torque to drag on a translating helix moving with a unit velocity $U$. This ratio therefore is equal to the force-free swimming velocity of a helix subject to a given input torque. From the plot, we see that this ratio is larger than the mixture value, with this ratio being larger for larger $\lambda$. This can be understood as follows, for both $L_B/R_H \rightarrow 0$ (mixture with viscosity $\mu_s + \mu_p$) and $L_B/R_H \rightarrow \infty$ (pure solvent with viscosity $\mu_s$), the thrust produced by a helix for a given torque is the same owing to Stokesian linearity and is independent of viscosity. However, the drag on the helix increases with increasing $\lambda$ in the limit $L_B/R_H \rightarrow 0$, and is independent of $\lambda$ in the limit $L_B/R_H \rightarrow \infty$ as the helix only moves in a pure solvent in this limit. Therefore for the same thrust, at large and finite $L_B/R_H$, the drag experienced by the helix drops, and therefore the force-free swimming velocity increases as $L_B/R_H$ increases, and the increase is larger for larger $\lambda$ as the drag on helix drops by a larger amount.
Together, these two features of the two-fluid slender body theory directly contribute to the same features observed in the plot of swimming velocity of a bacterium (figure \ref{fig:SwimmingVelocity2FluidNewt}a). 

\begin{figure}
	\centering
	\includegraphics[scale=0.3]{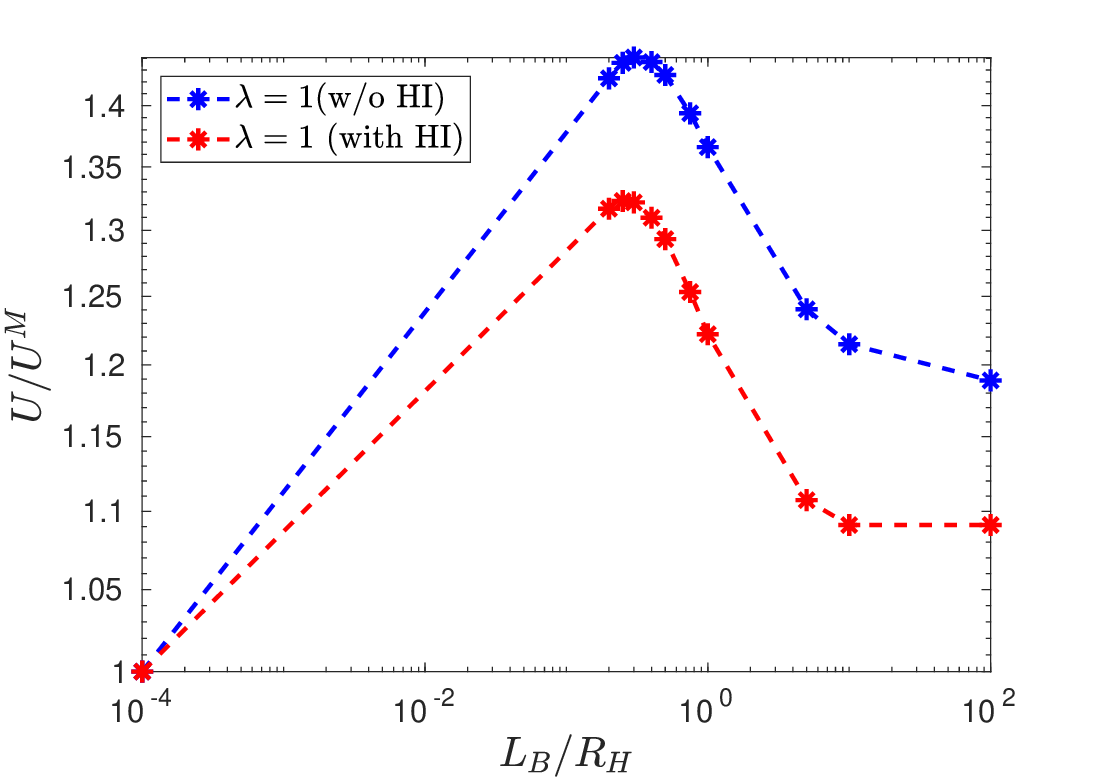}\quad
	\includegraphics[scale=0.3]{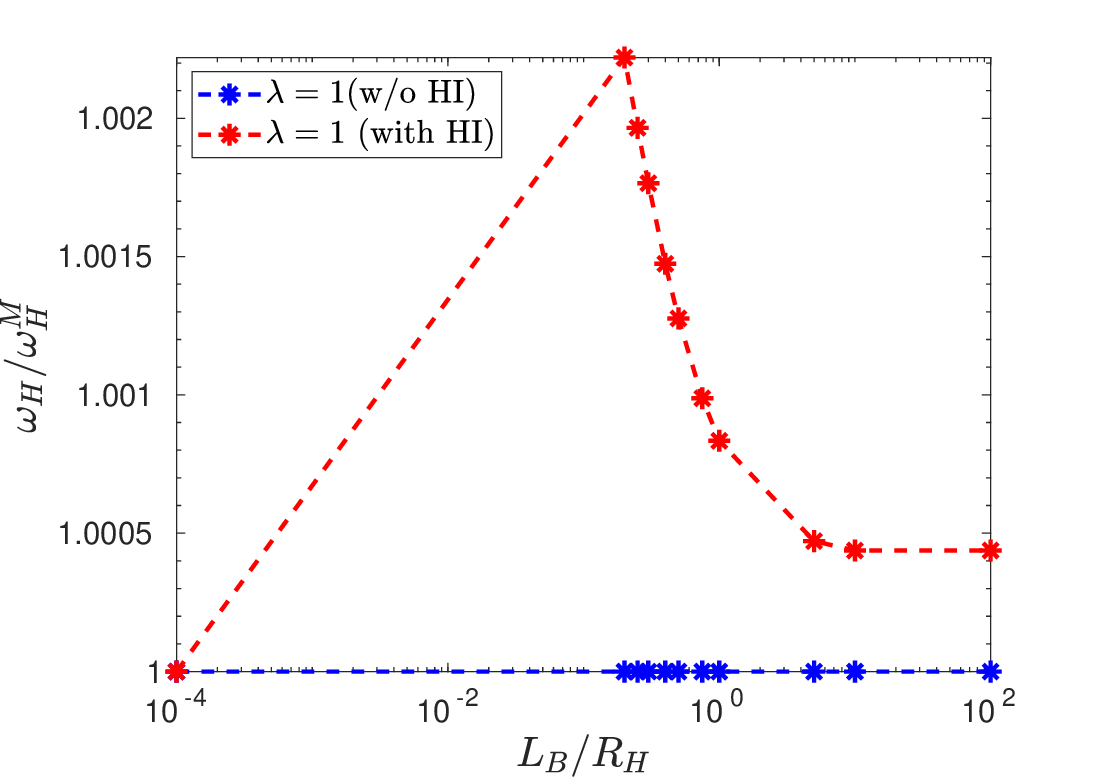}\quad
    \includegraphics[scale=0.3]{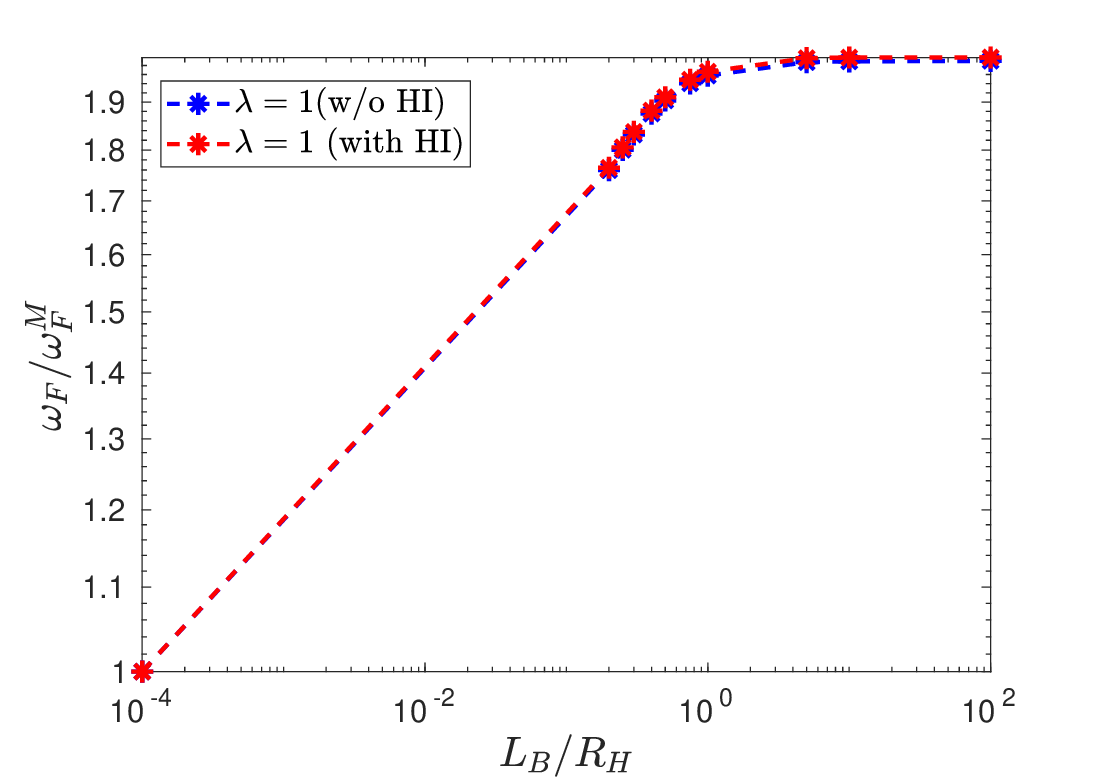}
	\caption{Plot of (a) normalized swimming velocity, (b) head angular velocity and (c) tail angular velocity for $\lambda = 1$. The velocities from the reduced system (eq.\eqref{ResisReduced} without HI between head and bundle) are compared from the prediction from the full calculation (eq.\eqref{Resis}}
	\label{fig:withwithoutHI}
\end{figure}

The fact that the trends obtained for a helix in a two-fluid medium are also observed for a swimming bacterium can be further justified by using a resistivity formulation for bacterium motion, accounting for the direction of swimming:
\begin{align}
	\begin{bmatrix}
		{F}_{\text{Head}} \\
		{L}_{\text{Head}} \\
		{F}_{\text{Flag}} \\
		{L}_{\text{Flag}}
	\end{bmatrix} = \begin{bmatrix}
		R_{11} &R_{12} &R_{13} &R_{14}\\
		R_{21} &R_{22} &R_{23} &R_{24}\\
		R_{31} &R_{32} &R_{33} &R_{34}\\
		R_{41} &R_{42} &R_{43} &R_{44}    
	\end{bmatrix} \begin{bmatrix}
		{U} \\
		{\omega}_H \\
		{U} \\
		{\omega}_F
	\end{bmatrix}.\label{Resis}
\end{align}
The off-diagonal terms represent the coupling between translational and rotational motion, which is absent in the flow around passive spherical or spheroidal particles. The coefficient $R_{11}$ represents the contribution to the force on the head due to translation of the head (with unit velocity), $R_{12}$ is the contribution due to rotation of the head (with unit angular velocity), and $R_{13}$ and $R_{14}$ represent the contributions to the head force due to translation and rotation of the flagellar bundle, respectively. Importantly, the coefficient $R_{34}$ represents the dominant contribution to the thrust due to the rotation of the bundle. Our numerical method offers us a unique advantage of selectively understanding each of these coefficients and their contribution to the overall bacterial motion, a feature that has so far been unaddressed in earlier studies on the same. To that end, we first solve a simpler system where the hydrodynamic interactions (HI) between the head and the bundle are turned off. Here, the resistivity formulation reduces to:
\begin{align}
	\begin{bmatrix}
		{F}_{\text{Head}} \\
		{L}_{\text{Head}} \\
		{F}_{\text{Flag}} \\
		{L}_{\text{Flag}}
	\end{bmatrix} = \begin{bmatrix}
		R_{11} &0 &0 &0\\
		0 &R_{22} &0 &0\\
		0 &0 &R_{33} &R_{34}\\
		0 &0 &R_{43} &R_{44}    
	\end{bmatrix} \begin{bmatrix}
		{U} \\
		{\omega}_H \\
		{U} \\
		{\omega}_F
	\end{bmatrix}.\label{ResisReduced}
\end{align}
Here, $R_11$ and $R_22$ are the Stokesian resistance to the motion of the bacterial head in the two-fluid medium, which for the case considered here, reduces to $6 \pi \mu_s (1+\lambda) R_H$ and $8 \pi \mu_s (1+\lambda) R_H^3$ for the spherical head. Similarly, the coefficients $R_{33}, R_{34}, R_{43}, R_{44}$ can be calculated using the two-fluid slender body theory\,(eq. \eqref{SBTEQ},\cite{Sabarish}), where unlike in RFT, the slender body theory accounts for the HI between different points on the helix. Using these coefficients, and by imposing force-free, torque-free and input torque constraints, one can calculate the swimming parameters for a bacterium, which is plotted in figure \ref{fig:withwithoutHI}, where it is compared with the swimming velocity from the full calculation (corresponding to eq. \eqref{Resis} with HI between head and the bundle included), and the RFT prediction\,\citep{Sabarish}. These results serve two crucial purposes. First, it shows that HI between different points on the helix affects the swimming motion more than HI between the head and the helix, inclusion of which only slightly changes the reduced system prediction. Second, it also justifies our earlier explanation, where characteristic features from the SBT calculation for a helix carried over to the swimming velocity of the bacterium. This however does not imply HI between the head and the bundle aren't important. While the results above correspond to purely Newtonian two-fluid medium, the presence of non-linear effects like viscoelasticity and inertia can significantly alter the contributions from HI. In these cases, all the coefficients in eq.\eqref{Resis} become time-dependent and they can be tracked with our numerical method. 

\begin{figure}
	\centering
	\includegraphics[scale=0.33]{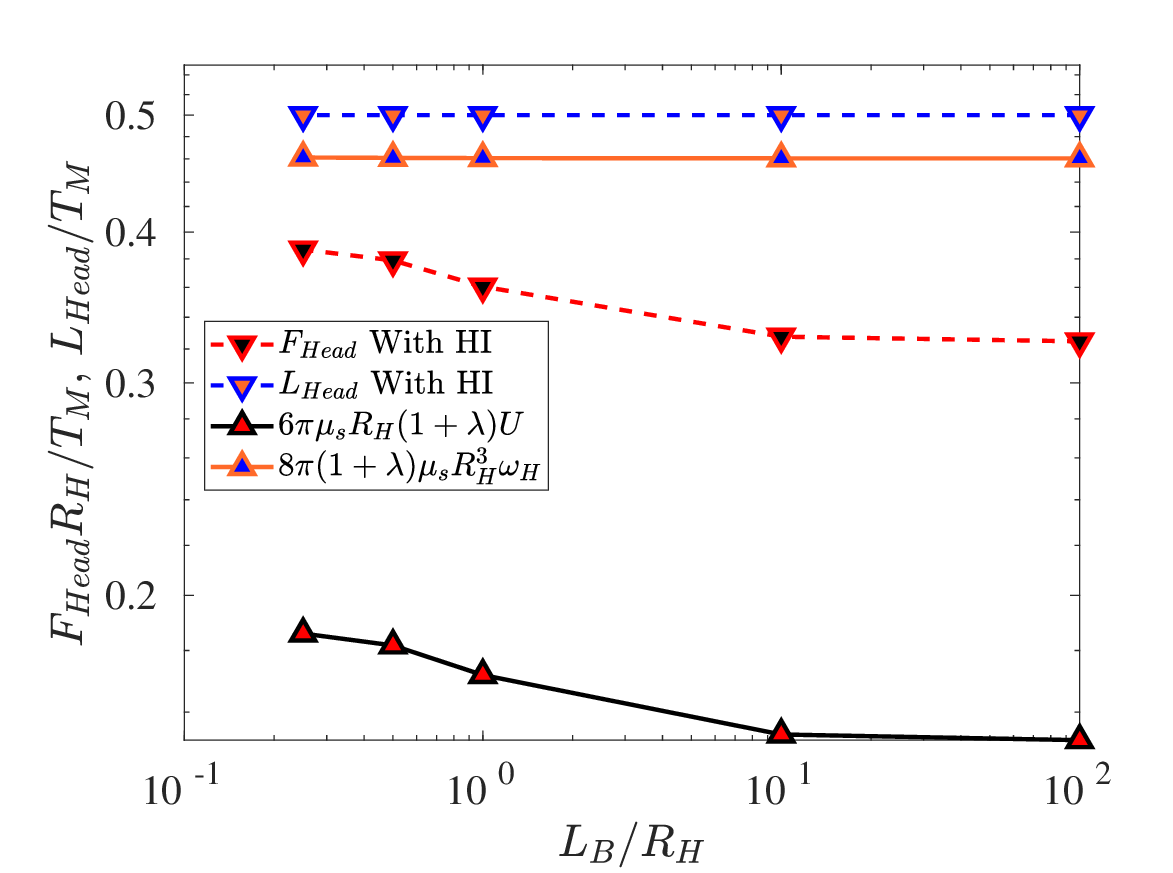}
	\includegraphics[scale=0.33]{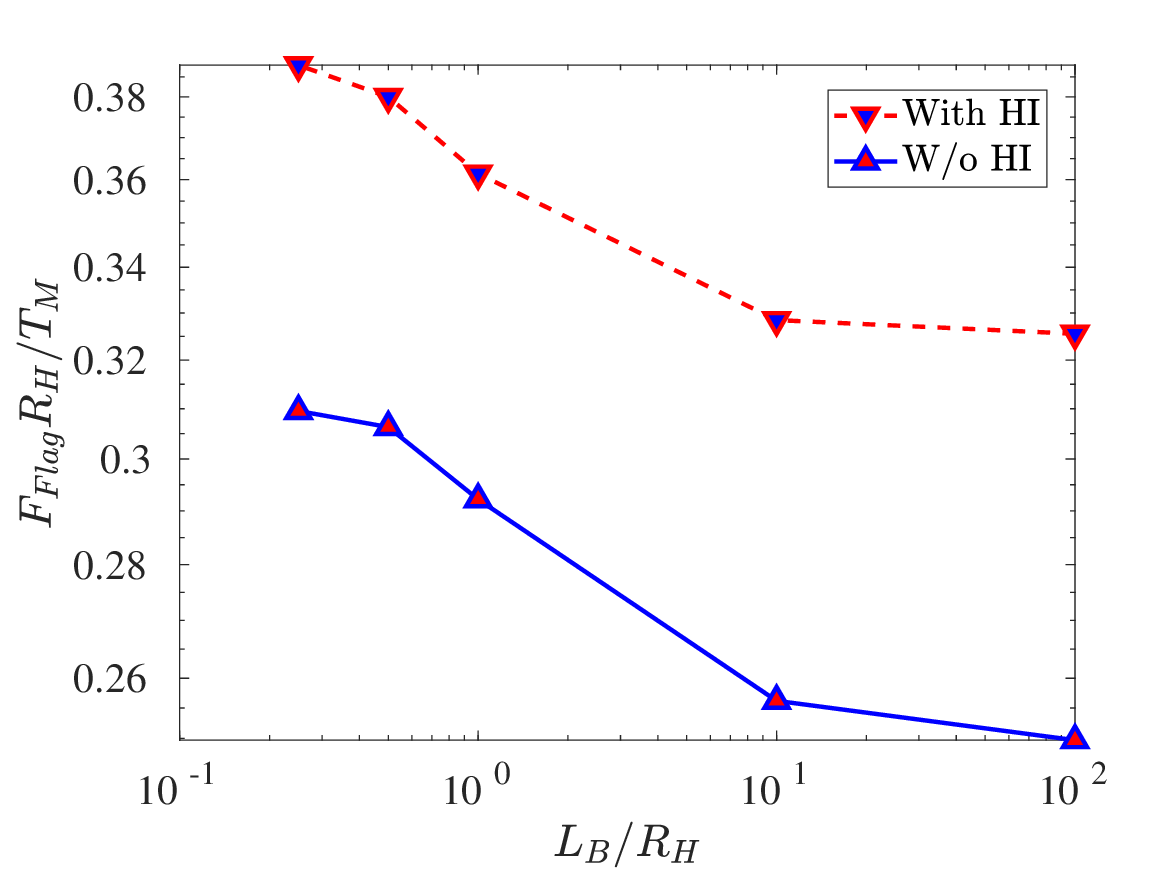}  
	\caption{(a) Drag and torque on the head as functions of $L_B/R_H$, for a bacterium swimming in the two-fluid medium with and without HI between the head and the bundle. (b) Net thrust produced by the flagellar bundle for the same. Note that HI between head and the bundle increase both drag and thrust with the former being significantly more than the latter.}
	\label{fig:SwimmingResistivities}
\end{figure}

While HI between the head and the bundle don't significantly affect the swimming parameters, relative to when these interactions aren't included, one can still understand how HI at different $L_B$ and $\lambda$ affect swimming motion in order to thoroughly explain bacterial motion. To this end, we plot the drag on the head and the (net) thrust generated by the flagellar bundle in figure~\ref{fig:SwimmingResistivities} and compare it with the drag and thrust without HI. In plotting the former, we essentially calculate $F_{Head} = (R_{11} + R{13}) U + R_{12} \omega_H +R_{14} \omega_F$, and similarly for the thrust $F_{Flag}$, after calculating $U$, $\omega_H$ and $\omega_F$ that satisfies the motion constraints. We see that HI between head and the bundle increases the (net) thrust generated by the bacterium, and also the drag on the head. However the swimming velocity is slightly smaller for the bacterium when HI is included between the head and the bundle because the increase in thrust is over-compensated by an increase in the drag. Similar comments can be made about the torque on the head and the bundle, which result in smaller angular velocities of the head and the bundle with HI. 

\begin{figure}
	\centering
	\includegraphics[scale=0.3]{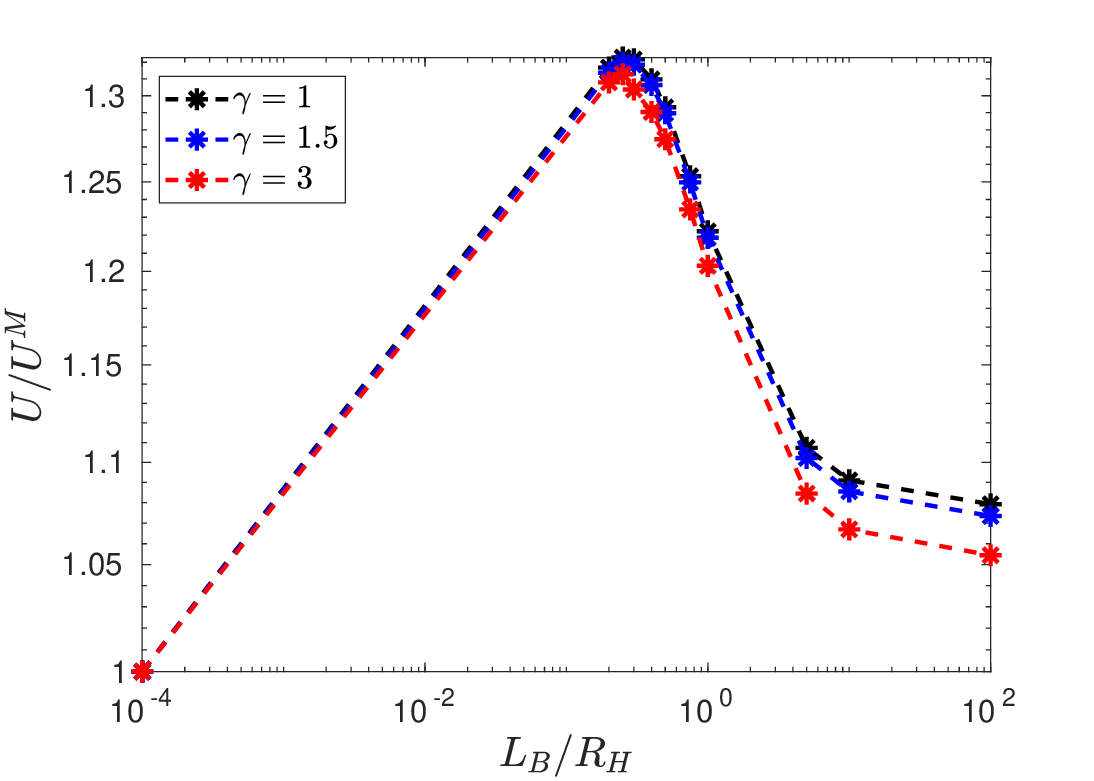}\quad
	\includegraphics[scale=0.3]{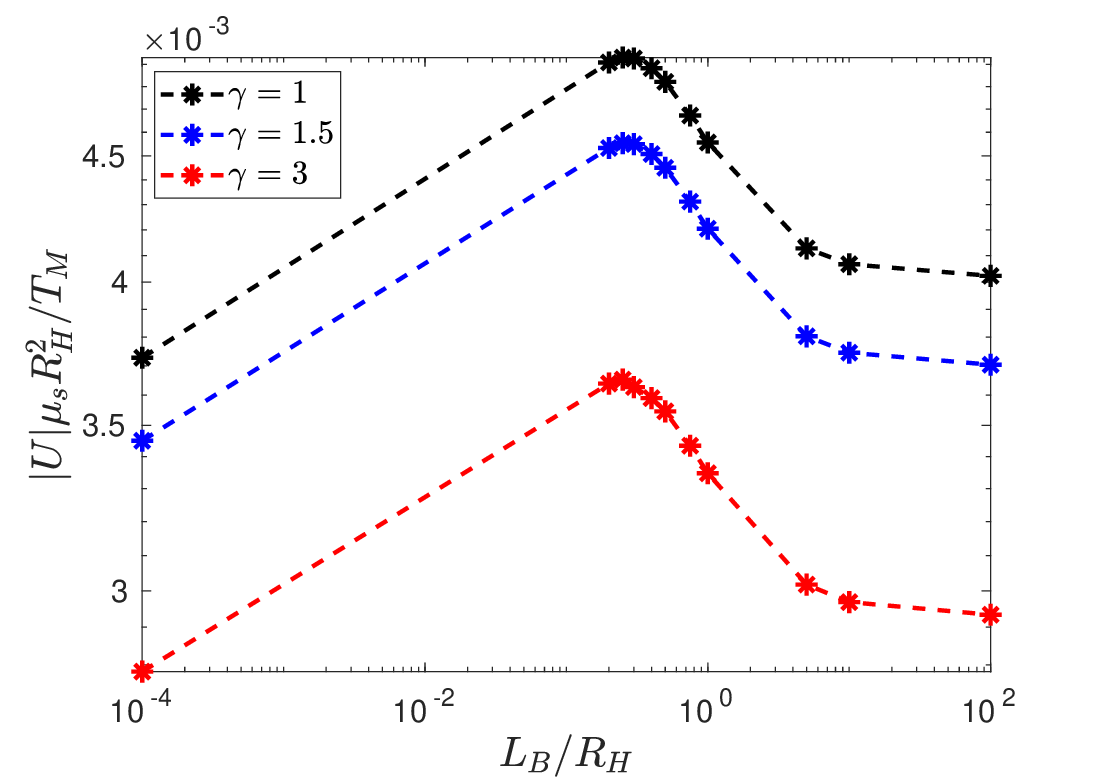}\quad
    \includegraphics[scale=0.3]{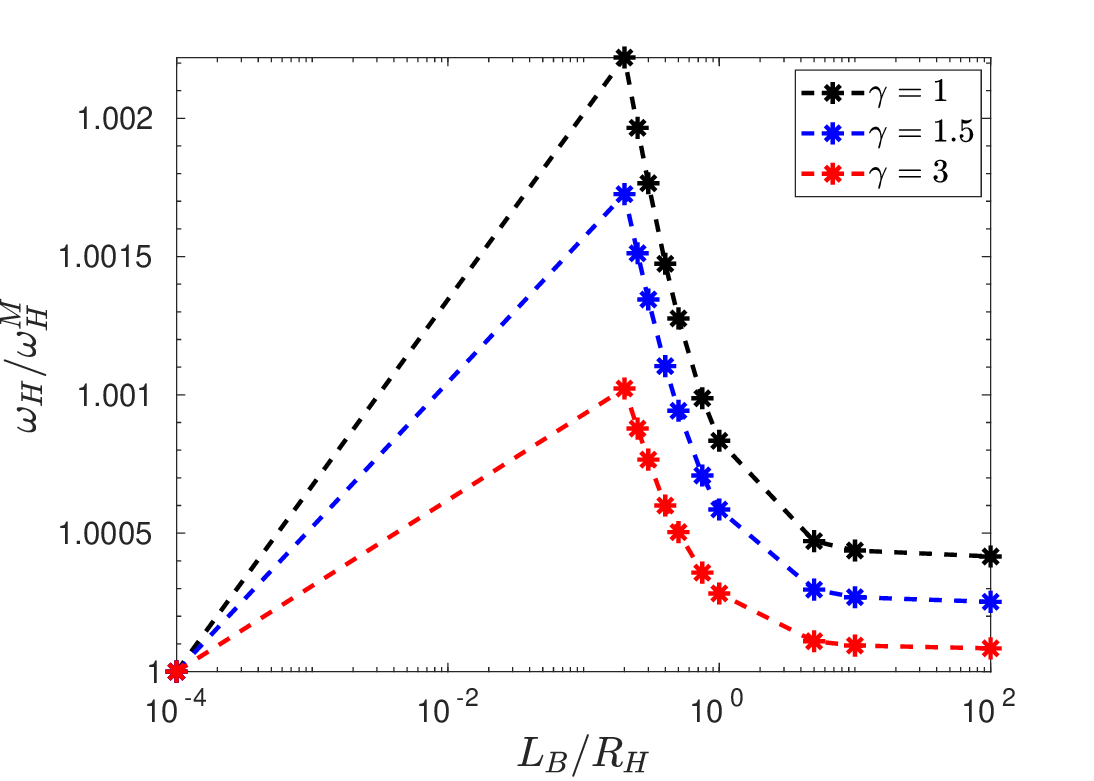}\quad
    \includegraphics[scale=0.3]{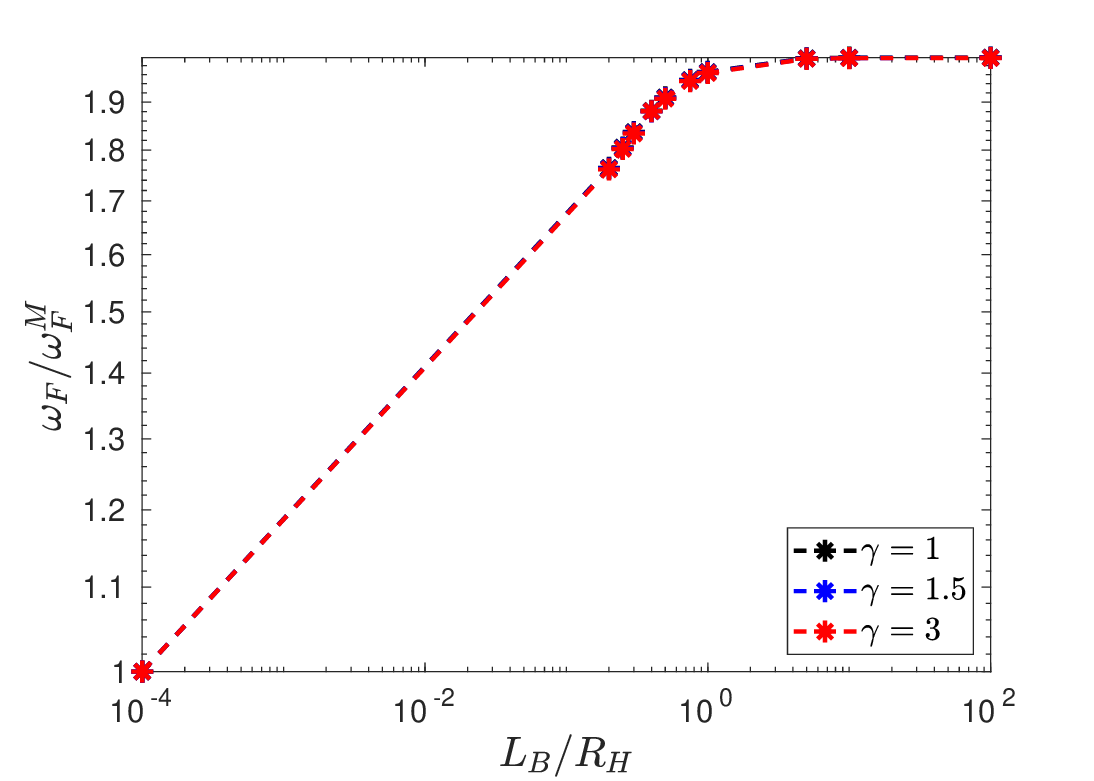}
	\caption{Plot of (a) normalized swimming velocity, (b) absolute dimensionless swimming velocity, (c) head angular velocity and (c) tail angular velocity for $\lambda = 1$, and three different head aspect ratio $\gamma$. The velocities follow similar trend as the spherical case, with higher head aspect ratio leading to smaller swimming and head angular velocity relative to the spherical case.}
	\label{fig:spheroidNoslip}
\end{figure}

While the results above pertained to bacterium with a spherical head, our method can in fact model a spheroidal head with aspect ratio greater than 1. In figure \ref{fig:spheroidNoslip}a,b we plot the swimming parameters for a bacterium with a spheroidal head with aspect ratio $\chi = 1.5$ and $\chi = 3$, which are roughly the range of aspect ratios of {\it E. coli} head measured in experiments. We see that here, the results follow similar trend as the case with spherical head ($\chi \approx 1$, but the absolute values of swimming velocity decreases slightly with increasing $\gamma$. This can be attributed to the higher drag on a spheroid relative to the drag on a sphere in Stokesian regime.

\subsection{Effects of microstructure at the length scale of the head}\label{sec:LargePolymers}
As mentioned earlier, when the polymer network exhibits larger pore size or when the polymer chains are large, they can slip past the bacterial head and are therefore modeled using the slip boundary conditions given in equations~\eqref{BC2B}–\eqref{BC2D}. In contrast to the previous case, where only the flagellar bundle experienced different resistances from the solvent and polymer fluids, here, the head does as well.

\begin{figure}
	\centering
	\includegraphics[scale=0.35]{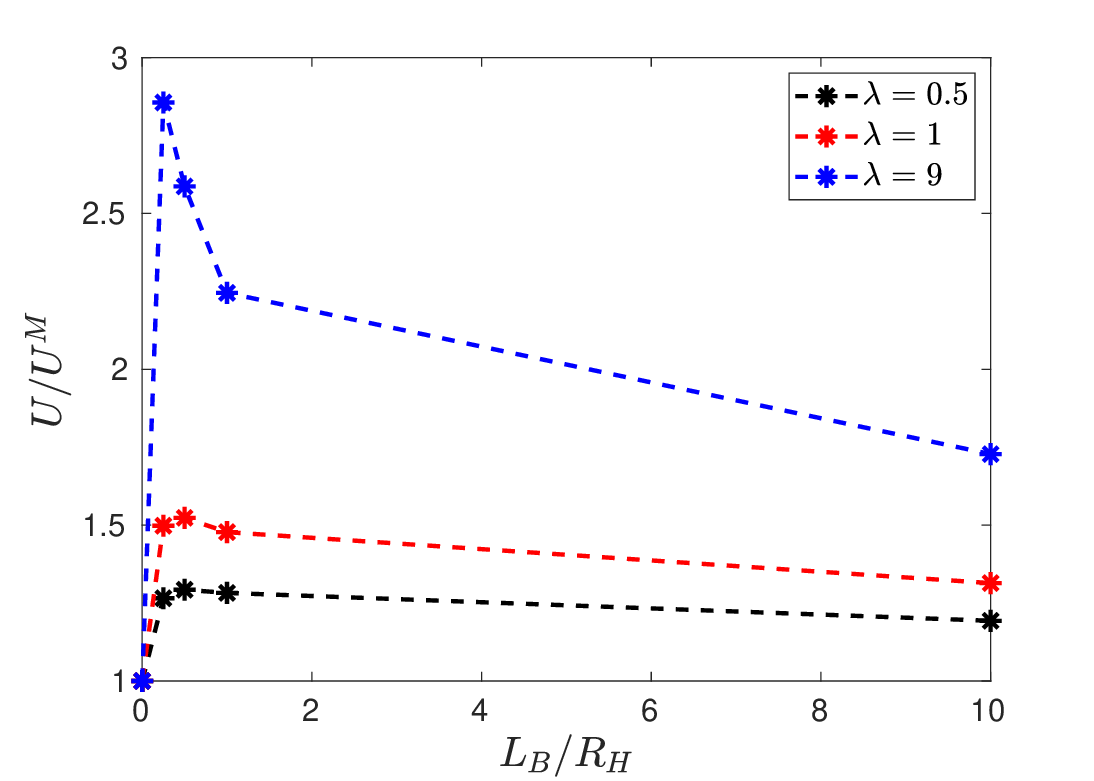}  
	\includegraphics[scale=0.35]{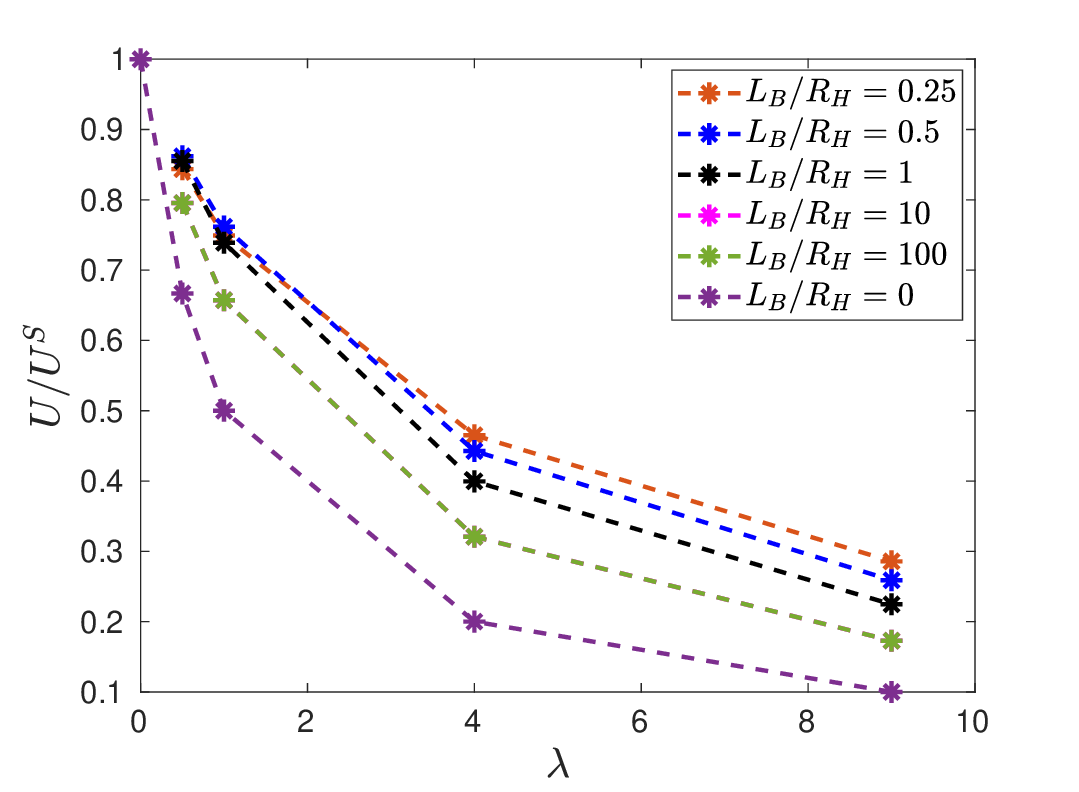}\\
	\includegraphics[scale=0.35]{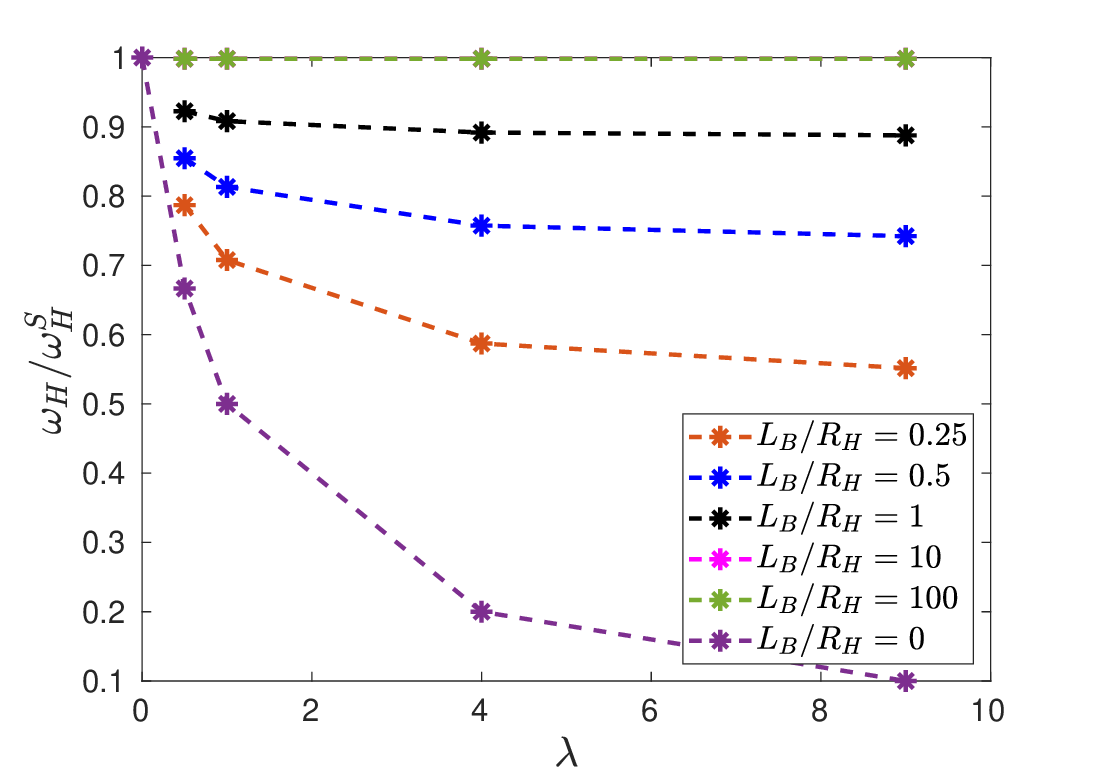}
	\includegraphics[scale=0.35]{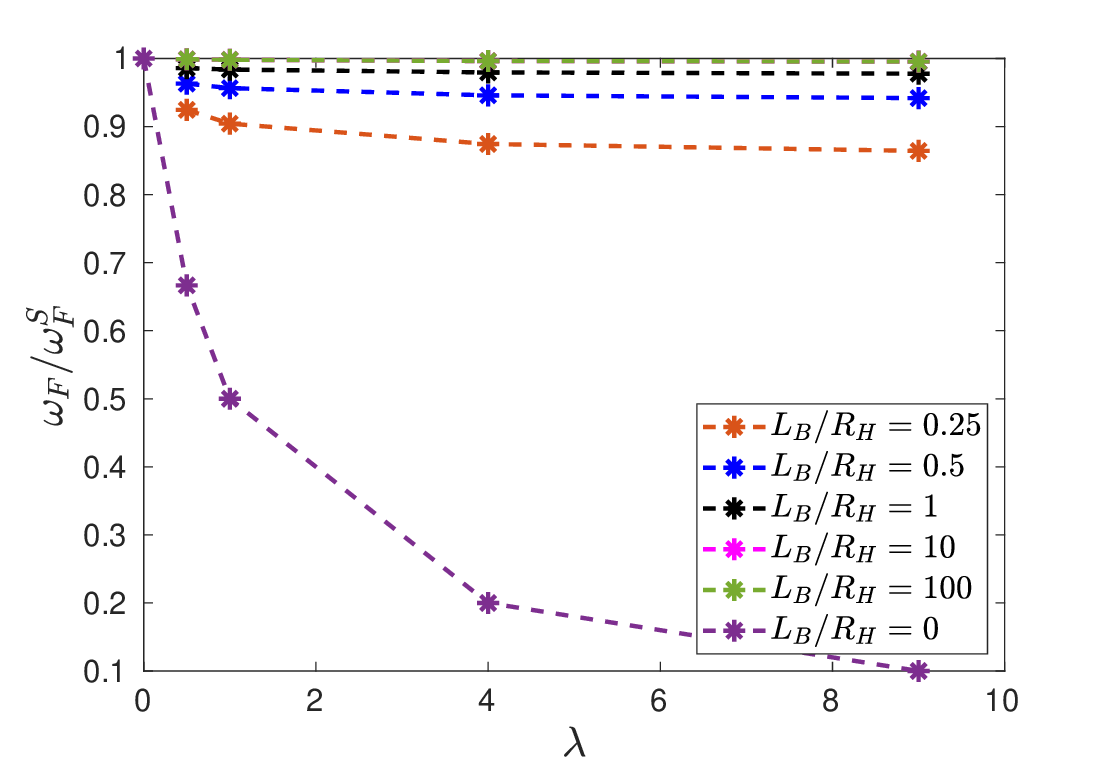}
	\caption{Same plots as in figure~\ref{fig:SwimmingVelocity2FluidNewt}, but for the case of slipping polymers.}
	\label{fig:SlippingPolySwim}
\end{figure}

Figure~\ref{fig:SlippingPolySwim} shows the same swimming parameters as in figure~\ref{fig:SwimmingVelocity2FluidNewt}, but for the case of slipping polymers. The overall trends are qualitatively similar to those observed for the no-slip case. However, the enhancement in swimming velocity at $L_B/R_H \approx 1$ is notably larger, reaching 2.9 times the speed in a single fluid mixture of viscosity $\mu_s(1+\lambda)$ for $\lambda = 9$, compared to 1.8 for no-slip polymers. This may be attributed to the reduced resistance from the polymer fluid when slip is allowed at the head. Another significant difference is the behavior of the head’s angular velocity, $\omega_H$, which now increases with $L_B$ (figure~\ref{fig:SlippingPolySwim}c). In contrast, for the no-slip case, $\omega_H$ was independent of $L_B$. This can once again be attributed to the reduced resistance to rotation when the polymer is allowed to slip. These results underscore the influence of polymer microstructure not just on the flagellar bundle but also on the bacterial head.

\section{Conclusion}\label{sec:Conclusions}
We have developed a numerical framework that simultaneously incorporates several important features of bacterial swimming and the surrounding biological fluid that are often neglected in computational modeling. These include the non-Newtonian elastoviscoplastic (EVP) rheology of the bio-fluid, the length-scale dependent spatial heterogeneity of the bio-fluid (polymer microstructure), and the length-scale dependent differential response of the bio-fluid. To strike a balance between capturing key physical features and maintaining computational feasibility, we do not resolve individual polymer-flagellar bundle interactions. Instead, we adopt a two-fluid model consisting of a polymer fluid and a Newtonian solvent fluid. In this model, the flagellar bundle directly drives the solvent, and momentum is exchanged between the two fluids via a drag force proportional to their relative velocity and inversely proportional to a characteristic screening length, $L_B$, which is a measure of the microstructural length scale. We model two types of polymer-bacterial head interactions: no-slip and perfectly slipping polymers that generate no tangential stress on the bacterial head. These boundary conditions mimic the behavior of short and long polymer chains, respectively. The flagellar forcing on the solvent is modeled using slender-body theory(SBT) as formulated by \citet{Sabarish}. In addition to microstructural interactions, the polymer also generates a polymer stress, $\bm{\Pi}$, which captures the EVP response of the medium. As illustrated in figure~\ref{fig:ModelSchematic}, the computational framework couples three main components: the two-fluid mass and momentum equations ($\hat{\mathcal{P}}_\text{MFP}$), slender-body theory for flagellar forcing ($\hat{\mathcal{P}}_\text{Flag}$ and $\mathcal{P}_\text{SBT}$), and constitutive equations for polymer stress.

In the absence of polymer stress $\bm{\Pi}$, the entire problem is linear in the unknown variables: the translational and angular velocities of the bacterium, the background flow, and the force distribution along the flagellar bundle. Recognizing this allows us to decompose both the flow field around the swimmer and the hydrodynamic moments (net force and torque) into three contributions: those induced by rigid-body kinematics, by flagellar forcing, and by $\bm{\Pi}$. The flows and moments due to kinematics and flagellar forcing can themselves be further decomposed into elementary components that are pre-computed for a given bacterial head shape and flagellar geometry. Once $\bm{\Pi}$ is introduced, the total flow and moment are computed by adding the contributions from these pre-computed components and the polymer stress. This linear decomposition enables a resistivity formulation in which the swimming velocities and the flagellar force distribution are obtained by solving a linear system involving a resistivity matrix. This matrix is assembled from pre-computed flow fields, and the resulting formulation satisfies force- and torque-free conditions, as well as slender-body theory constraints, in a non-iterative manner. This not only simplifies the algorithm but also makes it significantly more efficient.

When $\bm{\Pi} = 0$, the problem is quasi-steady. However, the viscoelastic stress $\bm{\Pi}$ evolves according to a time-dependent constitutive law, making the full EVP problem transient. At each time step, only two components must be updated: the stress $\bm{\Pi}$ (via time integration of the constitutive equations) and the flow induced by $\bm{\Pi}$ (obtained by inverting one component of the mass–momentum equation). Both steps are performed using a two-fluid extension of the finite-difference solver developed by \citet{Sharma23}, written in the particle reference frame and prolate spheroidal coordinates. The computational domain consists of the bacterium centered within a nearly spherical outer boundary. Pre-computed flows due to kinematic and flagellar forcing are also obtained by inverting the mass–momentum equations using this finite-difference technique. The discretization of the flagellar bundle in the slender-body formulation follows the method of \citet{Swinney}. This modular strategy combines operator decomposition, pre-computed basis flows, and a flexible two-fluid solver into a versatile numerical framework for simulating bacterial swimming in an elastoviscoplastic, microstructured media. Arbitrary linear or uniform background flows can also be imposed at the outer boundary, enabling exploration of a wide range of swimming behavior such as rheotaxis, upstream swimming etc. Additionally, the framework can also be extended to compute suspension-averaged stress in dilute bacterial suspensions using ensemble-averaging techniques \citep{koch2016stress}, under the assumption of negligible swimmer–swimmer interactions.

In this work, we have used the numerical framework to study the effect of polymer microstructure on bacterial swimming in a quiescent fluid by turning off the EVP response. Focusing on a bacterium with a spherical head, we systematically varied the screening length $L_B$ to study its influence on motility. We found that the improvement of swimming speed near $L_B \approx R_F$, originally predicted by \citet{Sabarish} using resistive force theory (RFT), is significantly amplified when hydrodynamic interactions (HI) are included between the head and the flagellar bundle. Using a linear resistivity framework, we demonstrate that these interactions enhance drag, thrust, and redistribute the force density along the flagellar bundle. Interestingly, while the angular velocity of the head remains largely unaffected by $L_B$, the flagellar rotation rate shows a strong dependence on the microstructural length scale. At $L_B$ values near the peak enhancement, the polymer velocity field is substantially disturbed by flagellar forcing, whereas at large $L_B$, the polymer response is largely due to the head and the swimming speed gain diminishes. This enhancement is even more pronounced for long polymers that can slip past the bacterial head, compared to the no-slip case. Contrary to the no-slip scenario, polymer slipping also increases the rotation speed of the bacterial head. These findings underscore the critical role of both microstructure and polymer–surface boundary conditions in modulating bacterial propulsion in complex fluids.

As the solver incorporates multiple key features of bacterial swimming in biological fluids, it can now be used to explore the interplay between different non-Newtonian properties, such as yield stress, viscoelasticity, and microstructural length scale ($L_B$), and their impact on swimming efficiency in quiescent media. By quantifying how polymer microstructure alters bacterial translation and rotation, the solver offers predictive insight that can inform future laboratory experiments and biomedical applications. Recent work by \citet{torres2024enhancement} has shown that shear-thinning viscoelasticity enhances upstream swimming of bacteria near walls by inducing a reorientation torque that aligns the swimmer against the flow. This behavior has important implications for wastewater contamination and the design of medical devices to prevent bacterial invasion, such as upstream migration in the urinary tract. Since the flow near walls approximates simple shear, and previous studies have shown that viscoelasticity can produce a wide range of orientation dynamics for spheroids in shear flow \citep{d2014bistability,sharma2023rotation}, our framework provides a natural extension to investigate how EVP rheology and microstructure influence swimmer alignment and navigation in such environments. Another interesting phenomena that can be addressed with our framework is the effect of bacterial wobbling on the swimming characteristics. Past experiments by \cite{Allison15, Xiang22} etc. have suggested swimming enhancement can result from a reduction in bacterial wobbling and the method can also be retrofitted to simulate a wobbling bacterium, for instance by using the model proposed by \cite{Xiang22}, in order to understand the relative importance of the different physical features (wobbling, shear-thinning, yield stresses etc.) that govern the swimming motion.

Together, these capabilities position the framework as a powerful tool for exploring how the interplay of microstructure, viscoelasticity, and boundary interactions shapes microbial locomotion in complex biological fluids, with implications for both fundamental science and biomedical applications.

\section*{Author contributions} AS and SVN contributed equally to the work.

\section*{Competing Interests} The authors declare no conflicts of interest.

\section*{Funding Sources}
AS was supported by the U.S. Department of Energy, Office of Science, Advanced Scientific Computing Research (ASCR) Early Career Research Program. This paper describes objective technical results and analysis. Any subjective views or opinions that might be expressed in the paper do not necessarily represent the views of the U.S. Department of Energy or the United States Government. Sandia National Laboratories is a multimission laboratory managed and operated by National Technology and Engineering Solutions of Sandia, LLC, a wholly owned subsidiary of Honeywell International Inc., for the U.S. Department of Energy’s National Nuclear Security Administration under contract DE-NA0003525.\\

This work was supported by the NSF CBET: Fluid Dynamics Award 2135617

\appendix

\section{Evaluating flow fields at points along the flagellar bundle using boundary integral formula} \label{sec:SphdCoordToFlagella}
\begin{figure}
	\centering
	\includegraphics[scale = 0.4]{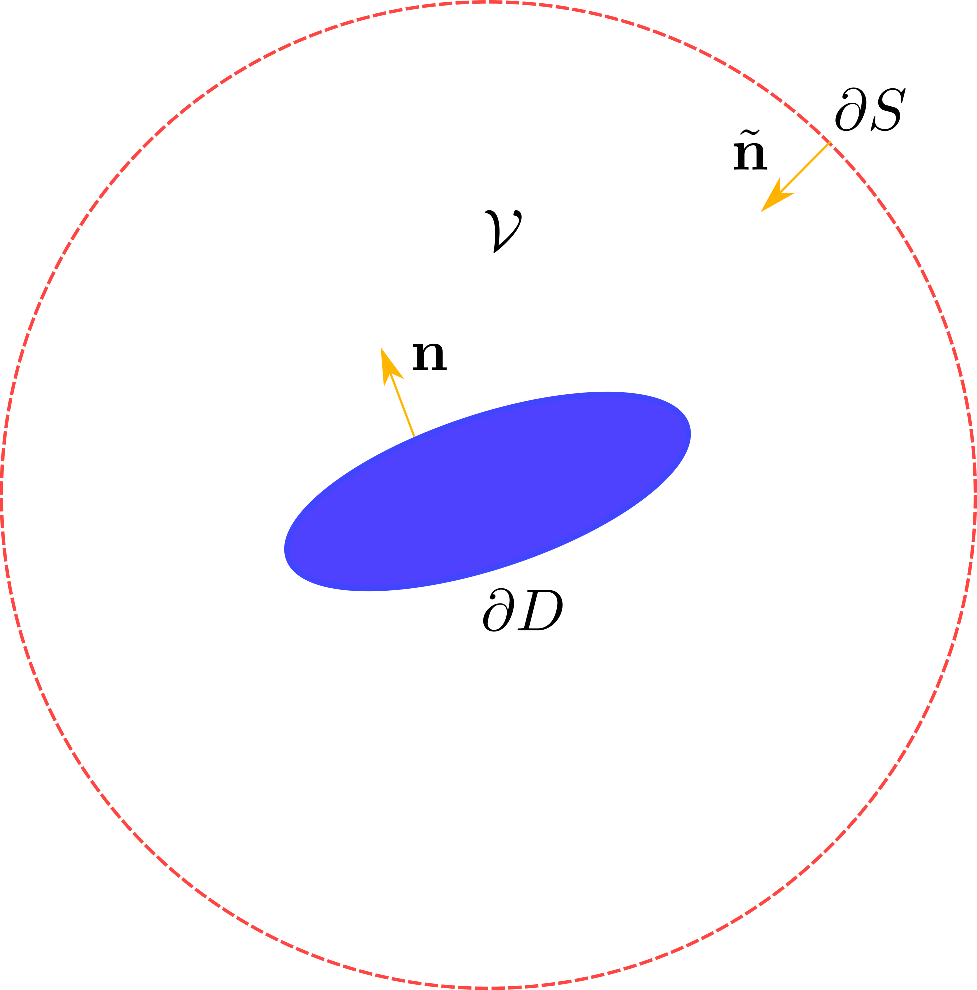}
	\caption{The domain $\mathcal{V}$ bounded by the head surface $\partial D$ and the fictitious outer surface $\partial S$, with normal vectors pointing into the fluid domain.}
	\label{fig:AppB1}
\end{figure}

As mentioned in section~\ref{sec:NumericalMethod}, to obtain the required flow velocities in the SBT equation~\eqref{SBTEQ} we use a boundary integral formulation (BIF) technique rather than interpolating the numerically evaluated fields from the prolate spheroidal grid to the helical flagellar bundle. We describe the BIF in this section. For notational convenience, we set the solvent viscosity $\mu_s = 1$ and the polymer viscosity $\mu_p = \lambda$. The governing equations for the two-fluid Newtonian medium in the absence of flagellar forcing are given by:
\begin{align}
	&\nabla \cdot \bm{u}_s = 0, \quad \nabla \cdot \bm{u}_p = 0, \label{B1} \\
	&\nabla^2 \bm{u}_s - \frac{1}{L_B^2}(\bm{u}_s - \bm{u}_p) - \nabla p_s = 0, \label{B2} \\
	&\lambda \nabla^2 \bm{u}_p + \frac{1}{L_B^2}(\bm{u}_s - \bm{u}_p) - \nabla p_p = 0. \label{B3}
\end{align}
We define the mixture and difference fields as $\bm{u}_m = \bm{u}_s + \lambda \bm{u}_p$ and $\bm{u}_d = \bm{u}_p - \bm{u}_s$, which yields:
\begin{align}
	&\nabla \cdot \bm{u}_m = 0, \quad \nabla^2 \bm{u}_m - \nabla p_m = 0, \label{B4} \\
	&\nabla \cdot \bm{u}_d = 0, \quad \nabla^2 \bm{u}_d - \nabla p_d - \frac{1 + \lambda}{\lambda L_B^2} \bm{u}_d = 0. \label{B5}
\end{align}
Note that equation~\eqref{B4} corresponds to the Stokes equations and equation~\eqref{B5} to the Brinkman equations. Thus, the solution can be expressed using a suitable combination of the boundary integral formulas for Stokes and Brinkman flows. We consider a domain $\mathcal{V}$ bounded by the head surface $\partial D$ and a fictitious surface $\partial S$ that will eventually be pushed to infinity (see figure~\ref{fig:AppB1}). Applying the boundary integral representation for Stokes flow in such a domain gives:
\begin{equation}
	\begin{split}
		\bm{u}_m (\bm{r}) = - \int_{\partial D} dA_{\bm{\xi}} \bm{G}_{St}(\bm{r} - \bm{\xi}) \cdot \bm{\sigma}_m(\bm{\xi}) \cdot \bm{n}(\bm{\xi}) + \int_{\partial D} dA_{\bm{\xi}} \bm{T}_{St}(\bm{r} - \bm{\xi}) \cdot \bm{u}_m(\bm{\xi}) \cdot \bm{n}(\bm{\xi}).
	\end{split} \label{B9}
\end{equation}
Similarly, the boundary integral expression for the difference field is:
\begin{equation}
	\begin{split}
		\bm{u}_d (\bm{r}) = \lambda \int_{\partial D} dA_{\bm{\xi}} \bm{G}_{Br}(\bm{r} - \bm{\xi}) \cdot \bm{\sigma}_d(\bm{\xi}) \cdot \bm{n}(\bm{\xi}) - \lambda \int_{\partial D} dA_{\bm{\xi}} \bm{T}_{Br}(\bm{r} - \bm{\xi}) \cdot \bm{u}_d(\bm{\xi}) \cdot \bm{n}(\bm{\xi}).
	\end{split} \label{B10}
\end{equation}
Finally, the solvent and polymer velocities in equations~\eqref{B1}–\eqref{B3} can be recovered from the BIF-computed fields $\bm{u}_m$ and $\bm{u}_d$ as:
\begin{align}
	\bm{u}_s = \frac{\bm{u}_m - \lambda \bm{u}_d}{1 + \lambda},\quad \bm{u}_p = \frac{\bm{u}_m + \bm{u}_d}{1 + \lambda}. \label{B8}
\end{align}
That can be expressed in a condensed form:
\begin{align}
	\bm{u}(\bm{r}) = - \int_{\partial D} dA_{\bm{\xi}} \bm{G}(\bm{r} - \bm{\xi}) \cdot \bm{\sigma}(\bm{\xi}) \cdot \bm{n}(\bm{\xi}) + \int_{\partial D} dA_{\bm{\xi}} \bm{T}(\bm{r} - \bm{\xi}) \cdot \bm{u}(\bm{\xi}) \cdot \bm{n}(\bm{\xi}). \label{TFBIF}
\end{align}
where,
\begin{center}
	$\bm{u}=\begin{bmatrix}\bm{u}_s\\\bm{u}_p\end{bmatrix}$, $\bm{\sigma}=\begin{bmatrix}\bm{\sigma}_s\\\bm{\sigma}_p\end{bmatrix}$, 
	$\bm{G}=\begin{bmatrix}\bm{G}_\text{SS}& \lambda \bm{G}_\text{SP}\\\bm{G}_\text{PS}& \lambda \bm{G}_\text{PP}\end{bmatrix}$, $\bm{T}=\begin{bmatrix}\bm{T}_\text{SS}& \lambda \bm{T}_\text{SP}\\\bm{T}_\text{PS}& \lambda \bm{T}_\text{PP}\end{bmatrix}$.
\end{center}
Note that the tensor $\bm{G}$ is the two-fluid Green's function whose components are given by\,\citep{Sabarish}:
\begin{align}
	\bm{G}_\text{SS} = \frac{\bm{G}_{St} + \lambda \bm{G}_{Br}}{1 + \lambda},\quad
	\bm{G}_\text{SP} = \frac{\bm{G}_{St} - \bm{G}_{Br}}{1 + \lambda} ,\quad
	\bm{G}_\text{PS} = \frac{\bm{G}_{St} - \bm{G}_{Br}}{1 + \lambda} ,\quad
	\bm{G}_\text{PP} = \frac{\lambda \bm{G}_{St} + \bm{G}_{Br}}{1 + \lambda}.
\end{align}
The numerical inputs for these integrals are the fluid stress $\bm{\sigma}$ and velocity $\bm{u}$ at the bacterial head. While $\bm{u} = 0$ at the surface for the main flow component corresponding to $\hat{\mathcal{P}}_\text{MFP}^{(M)}$, the flagellar flow component, $\hat{\mathcal{P}}_\text{MFP}^{(\text{Flag})}$, is obtained as a correction to the singularity-driven flow defined by equations~\eqref{eq:UtildeEqn}–\eqref{eq:UtildeBC2c}. This leads to a finite $\bm{u}$ at the surface, justifying the inclusion of the second term in equation~\eqref{TFBIF}.

Due to computational cost, we do not evaluate these integrals to obtain velocities corresponding to $\hat{\mathcal{P}}_\text{MFP}^{(M)}$ and $\hat{\mathcal{P}}_\text{MFP}^{(\text{Flag})}$ at every point in the domain. Instead, we compute them only at the $N_S$ discrete points along the flagellar bundle, which is computationally inexpensive and avoids interpolation errors. A similar boundary integral formulation can be derived for the velocity induced by the polymeric stresses in equation~\eqref{eq:IntMFPPoly}. In that case, however, the volumetric forcing term $\nabla \cdot \bm{\Pi}$ introduces an additional volume integral alongside the surface integrals. We compared the interpolated velocities due to $\bm{\Pi}$ at the helix points using such a boundary integral formula against a Lagranian polynomial interpolation scheme and found the former to be less accurate than the latter. Therefore we use a Lagrange polynomial interpolation of a higher order in the solver. 

\bibliographystyle{jfm}
\bibliography{CompBacteria}

\end{document}